\documentclass[]{elsarticle}
\usepackage{a4wide}
\usepackage{csquotes}

\usepackage[dvipsnames]{xcolor}
\usepackage{graphicx,color,booktabs,listings,caption,subcaption,tikz,amsmath,amssymb,placeins,subfiles,tabto, parskip}
\usetikzlibrary{fit, positioning, backgrounds}

\usepackage[titletoc]{appendix}
\usepackage{algpseudocode}
\usepackage{algorithm}
\usepackage{esvect}
\usepackage{gensymb}
\usepackage{inconsolata}
\usepackage{hyperref}
\usepackage{natbib}
\hypersetup{
    colorlinks=true,
    linkcolor=blue,
    filecolor=magenta,      
    urlcolor=cyan,
    pdftitle={Reduced Subgrid-Scale Terms for Turbulence Simulations with the Lattice Boltzmann Method},
    }
\usepackage{pdfpages}
\usetikzlibrary{math}
\usepackage{arydshln}
\newcommand{\D}{\mathrm{d}}

\DeclareMathOperator*{\argmin}{arg\,min}

\title{Reduced Subgrid-Scale Terms for Turbulence Simulations with the Lattice Boltzmann Method}
\date{September 26, 2026}
\journal{arXiv}

\begin{document}
\begin{frontmatter}
\author[inst1]{Rik Hoekstra \corref{cor1}}
\author[inst2]{Xiao Xue}
\author[inst2,inst3]{Peter V. Coveney}
\author[inst1,inst4]{Wouter Edeling}
\affiliation[inst1]{organization={Centrum Wiskunde \& Informatica, Scientific Computing group},%Department and Organization 
            city={Amsterdam},
            country={the Netherlands}}
\affiliation[inst2]{organization={University College London, Centre for Computational Science},%Department and Organization
            city={London},
            country={UK}}
\affiliation[inst3]{organization={University College London, Centre for Advanced Research Computing},%Department and Organization
            city={London},
            country={UK}}
\affiliation[inst4]{organization={University of Twente, RAM laboratory, Faculty of
Electrical Engineering, Mathematics \& Computer Science},%Department and Organization 
            city={Enschede},
            country={the Netherlands}}

\cortext[cor1]{Corresponding author, e-mail address: rik.hoekstra@cwi.nl}

\begin{abstract}   
Turbulent flows span a wide range of interacting scales, making direct numerical simulation (DNS) prohibitively expensive at high Reynolds numbers. Large eddy simulation (LES) offers a pragmatic alternative by only resolving the large-scale motions, but its accuracy hinges on the subgrid-scale (SGS) model. We introduce the Tau-orthogonal (TO) framework to the lattice Boltzmann method (LBM) using a macroscopic-velocity forcing. The TO method shifts the modeling target from a high-dimensional SGS field to the time dynamics of a small set of spatially integrated quantities of interest (QoIs). The SGS forcing is decomposed into analytically derived spatial patterns, so that each QoI is controlled to first order by a single scalar coefficient. Training data are generated by a predictor-corrector tracking procedure that nudges coarse simulations toward reference QoI trajectories without requiring a differentiable solver. Lightweight linear regression models with stochastic, multivariate Gaussian residuals (LRS) then drive the LES autonomously. Compact convolution-based kernels replace the sharp Fourier filters of the original formulation, lifting its restriction to rectangular periodic domains.
    
We first evaluate the closure on two-dimensional Kolmogorov flow at $Re=10\,000$, where TO-LRS yields stable, statistically accurate online LES and outperforms a calibrated Smagorinsky baseline on the long-term QoI distributions. However, bounded high-frequency oscillations are observed in the solution fields. The decisive test is three-dimensional turbulent channel flow at $Re_\tau=180$, simulated on a grid five times coarser in every direction than the high-fidelity reference, with the first fluid node at $y^+\approx 3.8$. Here the plain BGK solver is unstable, and even calibrated Smagorinsky and WALE models over-dissipate the near-wall flow, missing both the mean streamwise velocity profile and the long-term QoI distributions. TO-LRS, applied on top of an eddy-viscosity substrate without coefficient calibration and with the QoI set augmented by mean-profile QoIs, reproduces the reference distributions of all twelve QoIs and holds the DNS mean velocity profile across the full channel, at $40\times$ lower online cost per simulated time unit than the high-fidelity simulation. These results establish reduced, QoI-based closures as an interpretable and inexpensive alternative to high-dimensional neural SGS models for LBM, including wall-bounded flows on severely under-resolved grids.
\end{abstract}

\end{frontmatter}

\section{Introduction}\label{sec:introduction}
Turbulent flows involve a vast range of interacting scales: energy injected at the largest scales cascades through an inertial range down to the dissipation scales, where viscosity converts kinetic energy into heat. A direct numerical simulation (DNS) that resolves all of these scales demands a very fine grid whose number of points grows as $Re^{9/4}$ in three dimensions, where $Re$ is the Reynolds number. This scaling makes DNS prohibitively expensive for most engineering applications, where Reynolds numbers are typically high. Large eddy simulation (LES) strikes a pragmatic middle ground: it resolves the large scales on a coarse grid while representing the effect of the unresolved small scales through a subgrid-scale (SGS) model. The fidelity of an LES therefore hinges on the quality of this SGS closure.

SGS modeling has a long history in the context of classical Navier--Stokes solvers, where the central challenge is to balance accuracy, stability, and computational cost. The Smagorinsky model \cite{smagorinsky_general_1963} represents the unresolved stresses through an eddy viscosity proportional to the local strain rate. Later dynamic variants introduce spatial or scale-dependent variation of this coefficient \cite{germano_dynamic_1991}, while approximate-deconvolution closures use the resolved field itself to reconstruct subgrid information \cite{stolz_approximate_1999}. More recent data-driven closures instead learn the SGS contribution directly from high-fidelity reference data, often through neural networks trained end-to-end with the solver \cite{kochkov_machine_2021}, requiring backpropagation through the solver.

Alongside these advances in SGS modeling, the lattice Boltzmann method (LBM) has established itself as a competitive solution technique for the Navier--Stokes equations. Rooted in kinetic theory, LBM is a velocity-discretized form of the Boltzmann equation that evolves single-particle distribution functions on a lattice through a simple two-step cycle: a local collision that relaxes the distributions toward a discrete equilibrium, followed by a streaming step that propagates them along the lattice links~\cite{he1997theory,succi2001lattice,kruger_lattice_2017,lallemand2021lattice}.

Crucially, this kinetic representation carries more information than the macroscopic fields it reproduces, giving access to hydrodynamic regimes beyond Navier--Stokes~\cite{grad1949kinetic,grad1958principles,shan2006kinetic,xue2026fast}. The locality of the collision and the regularity of the streaming pattern give LBM near-perfect parallel scalability and a straightforward treatment of complex, evolving boundaries. Together, these properties have driven the adoption of LBM well beyond canonical single-phase hydrodynamics: multiphase and multicomponent flows~\cite{shan1993lattice,kupershtokh_equations_2009}, micro- and nanoscale flows with thermal fluctuations~\cite{ladd1993short,xue2018effects}, fully developed turbulence~\cite{chikatamarla2013entropic}, biomedical simulations~\cite{mazzeo2008hemelb,lo2025multi,lo2024uncertainty}, plasma and warm-fluid kinetics relevant to wakefield acceleration~\cite{simeoni2024lattice}.

Performing LES within the LBM framework, however, is not a straightforward adaptation of macroscopic Navier--Stokes closures. Because the dynamic variables are particle distributions rather than velocity and pressure fields, the SGS contribution must ultimately be re-injected at the kinetic level. This naturally suggests two routes: absorbing the unresolved dissipation into an effective turbulent relaxation time so that the standard collision operator carries the closure~\cite{malaspinas_consistent_2012,xue2022synthetic}, or modifying the collision step (or an external forcing term) directly so that distribution-level moments match those of the resolved flow~\cite{khan_physics-constrained_2026}. 

Classical eddy-viscosity closures, most notably the Smagorinsky model and its dynamic variant, have been ported to LBM by folding the turbulent viscosity into the relaxation time~\cite{premnath_dynamic_2009}, although this route might not in general reproduce the correct filtered macroscopic equations~\cite{malaspinas_consistent_2012}. A separate, LBM-specific line of work stabilizes under-resolved simulations through modified collision operators or selective filtering, without explicitly modeling the subgrid physics~\cite{nathen_adaptive_2018}. More recently, data-driven approaches have appeared. One line of work trains a multi-layer perceptron as a correction to the BGK collision operator for three-dimensional turbulence, enforcing conservation of mass and momentum through hard constraints~\cite{ortali_kinetic_2025}; because the network is trained end-to-end through the solver, this approach requires a differentiable (and therefore already stable) coarse LBM simulation. A related effort trains a compact neural network for the SGS stress tensor with explicit physics constraints on energy transfer and rotational equivariance~\cite{khan_physics-constrained_2026}. To circumvent the differentiability and stability prerequisites, multi-agent reinforcement learning has been used to treat the SGS model as a policy rewarded for producing accurate coarse-grained statistics~\cite{fischer_optimal_2025}. These data-driven LBM closures all target the bulk SGS term; closer to wall-bounded applications, a data-driven near-wall stress model has been learned for lattice Boltzmann LES~\cite{xue_physics_2024}.

In this paper, we adapt the Tau-orthogonal (TO) method~\cite{edeling_reducing_2020, hoekstra2024Reduced_data-driven, hoekstra_reduced_2026} to the LBM setting for the first time. Rather than predicting a high-dimensional SGS field directly, the TO method shifts the modeling task to the time dynamics of a small set of spatially integrated quantities of interest (QoIs); here kinetic energy and enstrophy across different spatial scales, augmented by mean-profile QoIs in the channel-flow case. The SGS forcing is built from a set of spatial patterns, constructed such that each influences only one QoI, so that the full field is fixed up to a handful of scalar time-series coefficients. These coefficients are learned by nudging the coarse LBM simulation toward reference QoI trajectories from a high-fidelity run, using a predictor-corrector procedure that does not require a differentiable solver. Casting the closure as nudging toward reference statistics places it in the broader family of data-assimilation-based LES closures~\cite{ephrati_continuous_2025}, here reduced to a small set of QoIs. We then fit lightweight linear regression models with stochastic, multivariate-Gaussian residuals (LRS), yielding an autonomous closure, which we refer to as TO-LRS. This model has far fewer than typical neural-network closures that predict the full SGS field. In contrast to closures that learn such a high-dimensional field, whether by end-to-end training through a differentiable solver \cite{ortali_kinetic_2025}, by physics-constrained supervised learning \cite{khan_physics-constrained_2026}, or by reinforcement learning \cite{fischer_optimal_2025}, the learning problem here reduces to a linear least-squares fit, trained in one shot from a single tracking run.

Compared to our earlier work \cite{hoekstra2024Reduced_data-driven, hoekstra_reduced_2026}, this paper contributes three new ingredients: (i) the adaptation of the TO framework to the kinetic LBM solver via macroscopic-velocity forcing; (ii) a reformulation of the scale-aware QoIs around compact convolution-based filters in place of the sharp Fourier filters used previously in the TO method, which frees the QoI construction from the rectangular periodic domains those filters require; and (iii) a new set of mean-profile QoIs that pin the mean streamwise velocity profile, introduced when extending the closure to three-dimensional turbulent channel flow.

The paper is organized as follows. Section \ref{sec:lattice-boltzmann-method} summarizes the LBM formulation. Section \ref{sec:coarse-graining-lattice-boltzmann-equation} introduces the coarse-grid LES setting. Section \ref{sec:tau-orthogonal-method-reduced-sgs-modeling} presents the TO-based SGS framework: scale-aware QoI definitions, spatial pattern construction, tracking, and the autonomous TO-LRS closure, in both two and three dimensions. Section \ref{sec:smagorinsky-model} reviews the Smagorinsky and WALE models. Sections \ref{sec:kolmogorov-flow} and \ref{sec:3d-channel} report the two test cases: a two-dimensional Kolmogorov flow and a three-dimensional channel flow. Section \ref{sec:conclusion} concludes. The appendices collect supporting material, including a repetition of the two-dimensional study with sharp Fourier filters (\ref{app:fourier-filters}).

\section{The lattice Boltzmann method}\label{sec:lattice-boltzmann-method}
The lattice Boltzmann method (LBM) is a class of fluid-flow solvers rooted in the mesoscopic Boltzmann equation; for a comprehensive introduction we refer to \cite{kruger_lattice_2017}. Unlike traditional Navier--Stokes solvers that discretize the macroscopic conservation laws, LBM describes the evolution of the particle distribution function $f(\boldsymbol{x}, \boldsymbol{\xi}, t)$, representing the density of particles with velocity $\boldsymbol{\xi}$ at position $\boldsymbol{x}$ and time $t$. Discretizing this equation in velocity space, physical space and time yields the lattice Boltzmann equation
\begin{equation} \label{eq:lattice-boltzmann-equation}
    f_i(\boldsymbol{x}+\boldsymbol{c}_i \Delta t, t+\Delta t)-f_i(\boldsymbol{x}, t)=\Omega_i(\boldsymbol{f}(\boldsymbol{x}, t)) + f^{\text{ext}}_i.
\end{equation}
At each time step, particles $f_i(\boldsymbol{x}, t)$ with discrete velocity $\boldsymbol{c}_i$ undergo a local collision governed by $\Omega_i$, an optional external forcing $f^{\text{ext}}_i$, and then stream to the neighboring lattice site $\boldsymbol{x}+\boldsymbol{c}_i \Delta t$. The velocity space is discretized into a small set of discrete velocities $\{\boldsymbol{c}_i\}$; in this work we use the D2Q9 stencil, which comprises nine velocities, for 2D simulations, and the D3Q19 stencil, consisting of 19 velocities, for 3D simulations. Standard lattice Boltzmann simulations are performed in lattice units, where the system is scaled such that $\Delta x = \Delta t = 1$, so that streaming is straightforward. Mass and momentum are recovered from the discrete distributions as:
\begin{equation}
    \rho = \sum_i f_i, \quad \rho \boldsymbol{u} = \sum_i f_i \boldsymbol{c}_i.
\end{equation}

We adopt the Bhatnagar--Gross--Krook (BGK) single-relaxation-time collision model
\begin{equation}
    \Omega_i(f)=-\frac{f_i-f_i^{\text{eq}}}{\tau} \Delta t,
\end{equation}
which relaxes the populations toward the local equilibrium $f^\text{eq}_i(\rho, \boldsymbol{u})$\footnote{The equilibrium is $f_i^{\text{eq}}(\rho, \boldsymbol{u})=w_i \rho\left(1+\frac{\boldsymbol{u} \cdot \boldsymbol{c}_i}{c_{\text{s}}^2}+\frac{\left(\boldsymbol{u} \cdot \boldsymbol{c}_i\right)^2}{2 c_{\text{s}}^4}-\frac{\boldsymbol{u} \cdot \boldsymbol{u}}{2 c_{\text{s}}^2}\right)$, with lattice specific weights $w_i$ and lattice speed of sound $c_s$.} at a rate set by the relaxation time $\tau$.
Through the Chapman--Enskog expansion, the BGK lattice Boltzmann equation recovers the incompressible Navier--Stokes equations to second-order accuracy in the low-Mach-number limit, with kinematic viscosity
\begin{equation}
    \nu = c_s^2\left( \tau - \frac{\Delta t}{2} \right),
\end{equation}
where $c_s$ is the lattice speed of sound. The BGK model is neither the most accurate nor the most stable collision model, but it is widely used for its simplicity; its instability on coarse grids also makes it a stringent test case for a low-fidelity closure.

\paragraph{Collision, forcing and streaming}
Typically, the lattice Boltzmann equation is split into a local collision step followed by a streaming step that propagates the resulting distributions along the lattice. Incorporating an external force through the exact-difference method \cite{kupershtokh_equations_2009} introduces an intermediate forcing substep, so that \eqref{eq:lattice-boltzmann-equation} becomes
\begin{align}
    & f_i^{\text{a}}(\boldsymbol{x}, t)=\left(1-\frac{\Delta t}{\tau}\right) f_i(\boldsymbol{x}, t)+ \frac{\Delta t}{\tau} f_i^{\text{eq}}(\rho, \boldsymbol{u}), \\
    &f_i^{\text{p}}(\boldsymbol{x}, t) = f_i^{\text{a}}(\boldsymbol{x}, t)
    + \bigg( f_i^{\text{eq}}(\rho, \boldsymbol{u} + \Delta \boldsymbol{u}_F) - f_i^{\text{eq}}(\rho, \boldsymbol{u})\bigg), \label{eq:collide-and-stream} \\
    & f_i\left(\boldsymbol{x}+\boldsymbol{c}_i \Delta t, t+\Delta t\right)=f_i^{\text{p}}(\boldsymbol{x}, t),
\end{align}
where $\boldsymbol{f}^\text{p}$ is the post-collision state\footnote{Forcing is typically seen as part of the collision step.}, $\boldsymbol{f}^\text{a}$ is the intermediate post-relaxation, pre-forcing state, and $\Delta \boldsymbol{u}_F$ denotes the local velocity change due to the forcing, calculated from the force density $\boldsymbol{F}$ as $\Delta \boldsymbol{u}_F = \boldsymbol{F}\Delta t/\rho$.

\section{Coarse graining the lattice Boltzmann equation}\label{sec:coarse-graining-lattice-boltzmann-equation}
We now formulate LES in the LBM setting. Following \cite{ortali_kinetic_2025, fischer_optimal_2025}, we obtain a low-fidelity (LF) field by projecting the high-fidelity (HF) field directly onto the coarser lattice. Denoting projected variables by $\overline{\left(\cdot\right)}$, and the projection onto an $m$-times coarser grid by $\phi_m$, we write
\begin{equation}
    \overline{\boldsymbol{f}} = \phi_m \boldsymbol{f}, \quad
    \overline{\boldsymbol{u}} = \phi_m \boldsymbol{u}, \quad
    \overline{\rho} = \phi_m \rho.
\end{equation}

We use two concrete projections in this work. For the doubly-periodic Kolmogorov test case of Section~\ref{sec:kolmogorov-flow}, we take $\phi_m$ to be bicubic interpolation, as in the experiments of \cite{fischer_optimal_2025}. For the channel of Section~\ref{sec:3d-channel}, bicubic interpolation would blend mid-channel velocities into the near-wall LF cells. We therefore replace it by a stride-$m$ subsampling operator.

On the coarse grid we run an LES with a coarse lattice Boltzmann solver augmented by an SGS forcing $\Delta \boldsymbol{v}_{SGS}$:
\begin{align}
     &\tilde{f}_i^{\text{a}}(\boldsymbol{x}, t)=\left(1-\frac{\Delta t}{\tilde{\tau}}\right) \tilde{f}_i(\boldsymbol{x}, t)+
    \frac{\Delta t}{\tilde{\tau}} \tilde{f}_i^{\text{eq}}(\tilde{\rho}, \boldsymbol{v}), \label{eq:LES-1}\\
    &\tilde{f}_i^{\text{p}}(\boldsymbol{x}, t) = \tilde{f}_i^{\text{a}}(\boldsymbol{x}, t) + \bigg( \tilde{f}_i^{\text{eq}}(\tilde{\rho}, \boldsymbol{v} + \Delta \boldsymbol{v}_F + \Delta \boldsymbol{v}_{SGS}) - \tilde{f}_i^{\text{eq}}(\tilde{\rho}, \boldsymbol{v})\bigg), \label{eq:LES-2}\\
    &\tilde{f}_i\left(\boldsymbol{x}+\boldsymbol{c}_i \Delta t, t+\Delta t\right)=\tilde{f}_i^{\text{p}}(\boldsymbol{x}, t), \label{eq:LES-3}
\end{align}
where $\tilde{f}$ and $\boldsymbol{v}$ denote the mesoscopic and macroscopic LES solution fields, respectively, and $\Delta t$ is $m$ times larger than in the DNS. In practice both simulations are performed in their own set of lattice units.

We add the subgrid contribution as a macroscopic velocity forcing $\Delta \boldsymbol{v}_{SGS}$ rather than as a modification of the collision operator. This choice (also used in \cite{khan_physics-constrained_2026}) keeps the SGS modeling task in macroscopic-velocity space, where our quantities of interest are naturally defined. Let $\delta\tilde f_i$ denote the change in the post-collision populations caused by $\Delta\boldsymbol{v}_{SGS}$. The equilibrium moment identities give
\[
    \sum_i \delta\tilde f_i = 0, \qquad
    \sum_i \boldsymbol{c}_i\,\delta\tilde f_i
    = \tilde\rho\,\Delta\boldsymbol{v}_{SGS}.
\]
The correction therefore preserves local mass and supplies the prescribed momentum increment.

For clarity, we use the following notation throughout the paper: $\boldsymbol{u}$ is the DNS velocity, $\overline{\boldsymbol{u}}=\phi_m\boldsymbol{u}$ is the projected DNS velocity on the coarse grid, and $\boldsymbol{v}$ is the LES velocity computed on that grid. Within the $n$-th solver step we distinguish three states: the post-streaming state $\boldsymbol{v}^n$, the intermediate post-relaxation, pre-forcing state $\boldsymbol{v}^{\text{a},n}$, and the post-collision state $\boldsymbol{v}^{\text{p},n}$.

Quantities of interest (QoIs) can be evaluated on any of these states. With the population-moment velocity defined above, BGK relaxation conserves local momentum, so $\boldsymbol{v}^{\text{a},n}=\boldsymbol{v}^{n}$; the subsequent forcing changes the velocity by $\Delta\boldsymbol{v}_F+\Delta\boldsymbol{v}_{SGS}$. The results in Sections~\ref{sec:kolmogorov-flow} and~\ref{sec:3d-channel} are reported on the post-streaming state.

The objective of an SGS model is to make the LES velocity approximate the projected DNS solution, i.e.\ $\boldsymbol{v}(\boldsymbol{x},t) \approx \overline{\boldsymbol{u}}(\boldsymbol{x}, t)$. Because turbulent flows are chaotic, pointwise agreement over long time horizons is unattainable. More realistic targets are correct decorrelation times when both simulations share the same initial conditions, or accurate reproduction of statistical quantities such as the energy spectrum. In this work we pursue the latter perspective and focus on accurately reproducing the long-term distributions of a small set of spatially integrated QoIs.

\subsection{Evaluation metric: the KS-distance}\label{sec:ks-distance}
To quantify similarity between long-term QoI distributions (scalar time series), we use the Kolmogorov--Smirnov (KS) distance, which compares two distributions through the maximum difference between their cumulative distribution functions $F(x)$ and $G(x)$:
\begin{equation}
    \mathrm{KS}\Big(F(x),G(x)\Big) = \max_x \Big\lvert F(x)-G(x) \Big\rvert.
    \label{eq:KS}
\end{equation}
The KS-distance is dimensionless, bounded by one, and invariant under monotone rescalings of its argument, which makes it directly comparable across QoIs with different magnitudes and units. We therefore summarize discrepancies across a set of QoI distributions using the summed KS-distance.

\section{The Tau-orthogonal method for reduced SGS modeling}\label{sec:tau-orthogonal-method-reduced-sgs-modeling}
The SGS forcing field $\Delta \boldsymbol{v}_{SGS}(\boldsymbol{x},t)$ has as many degrees of freedom as the velocity field itself, which, despite the projection to an LF grid, makes it a high-dimensional modeling target. The Tau-orthogonal (TO) method \cite{edeling_reducing_2020} reduces this complexity by modeling only the time dynamics of a small set of spatially integrated QoIs:
\begin{equation} \label{eq:qoi-def}
    Q = \int_{\Omega} q(\boldsymbol{v}, \boldsymbol{x}) \D \boldsymbol{x}.
\end{equation}
The SGS forcing is then approximated as a weighted sum of spatial patterns, one per QoI:
\begin{equation} \label{eq:TO-ansatz}
    \Delta \boldsymbol{v}_{SGS} = \sum_{i=1}^{N_Q} \gamma_i(t) \boldsymbol{O}_i(\boldsymbol{v}, \boldsymbol{x}),
\end{equation}
where $N_Q$ is the number of QoIs, the scalar coefficients $\gamma_i(t)$ represent the SGS contribution to the $i$-th QoI, and the spatial patterns $\boldsymbol{O}_i(\boldsymbol{v}, \boldsymbol{x})$ are constructed such that each QoI is controlled independently to first order in the velocity increment. To avoid confusion with the LBM relaxation time $\tau$, we denote these coefficients by $\gamma_i$, instead of $\tau_i$ as in previous work \cite{hoekstra2024Reduced_data-driven}. The method nonetheless keeps its established \emph{Tau-orthogonal} name. This independence is enforced through the orthogonality conditions
\begin{equation} \label{eq:orthogonality}
    \int_\Omega \frac{\delta Q_i}{\delta \boldsymbol{v}} \cdot \boldsymbol{O}_j(\boldsymbol{v}, \boldsymbol{x}) \D \boldsymbol{x} = 0 \quad \text{for } i \neq j,
\end{equation}
where $\delta Q_i / \delta \boldsymbol{v}$ is the functional derivative \footnote{It is defined as the field representing the directional derivative of $Q_i$ through the $L^2$ inner product, $\left.\frac{\D}{\D \epsilon} Q_i(\boldsymbol{v} + \epsilon \boldsymbol{\varphi})\right|_{\epsilon=0} = \int_\Omega \frac{\delta Q_i}{\delta \boldsymbol{v}} \cdot \boldsymbol{\varphi} \, \D \boldsymbol{x}$ for all $\boldsymbol{\varphi} \in L^2(\Omega)^d$. Such a field is unique whenever it exists, since the $L^2$ inner product is non-degenerate; we derive it in closed form for each QoI in \ref{app:derivatives of QoIs}.} (or $L^2$-gradient) of the QoI functional $Q_i$ with respect to the velocity field $\boldsymbol{v}$. These conditions guarantee that each spatial pattern $\boldsymbol{O}_i$ has zero first-order influence on the other QoIs. The ansatz, the orthogonality conditions, and the time-series closure developed below hold in any spatial dimension; only the QoI sensitivities, the compact filters, and the treatment of solid walls depend on the dimension or the boundary geometry. For a more detailed description of the TO method we refer to \cite{edeling_reducing_2020, hoekstra2024Reduced_data-driven, hoekstra_reduced_2026}.

Throughout this section we evaluate QoIs on \emph{post-collision} states $\boldsymbol{v}^{\text{p},n}$ rather than on post-streaming states. Since the SGS forcing in \eqref{eq:LES-2} acts before streaming, this choice gives a direct predictor-corrector relation between the imposed velocity increment and the QoI correction.

To see how the coefficients $\gamma_i$ enter the dynamics, consider the $i$-th QoI over a single LES step between successive post-collision states $\boldsymbol{v}^{\text{p},n-1}$ and $\boldsymbol{v}^{\text{p},n}$. Advancing the solver \eqref{eq:LES-3}, \eqref{eq:LES-1}, \eqref{eq:LES-2} without SGS forcing produces a predictor $\boldsymbol{v}^{\text{p},n^*}$; adding the forcing then yields the corrected state
\begin{equation}
    Q_i(\boldsymbol{v}^{\text{p},n}) = Q_i\!\left(\boldsymbol{v}^{\text{p},n^*}+\Delta \boldsymbol{v}_{SGS}^n\right).
\end{equation}
A first-order Taylor expansion gives
\begin{equation}
    Q_i(\boldsymbol{v}^{\text{p},n}) \approx Q_i(\boldsymbol{v}^{\text{p},n^*})+ \int_\Omega \frac{\delta Q_i}{\delta \boldsymbol{v}}\bigg|_{\boldsymbol{v}^{\text{p},n^*}} \cdot \Delta \boldsymbol{v}_{SGS}^{n} \ \D \boldsymbol{x},
\end{equation}
and inserting the TO ansatz \eqref{eq:TO-ansatz} together with the orthogonality conditions \eqref{eq:orthogonality} reduces this to
\begin{equation}\label{eq:influence-SGS-term-on-LF-QoI-trajectories}
    Q_i(\boldsymbol{v}^{\text{p},n}) - Q_i(\boldsymbol{v}^{\text{p},n^*}) \approx
    \gamma_i(t^n) \int_\Omega \frac{\delta Q_i}{\delta \boldsymbol{v}}\bigg|_{\boldsymbol{v}^{\text{p},n^*}} \cdot \boldsymbol{O}_i(\boldsymbol{v}^{\text{p},n^*}, \boldsymbol{x}) \ \D \boldsymbol{x}.
\end{equation}
Each $\gamma_i$ thus controls the $i$-th QoI through a single scalar relation. The burden of subgrid modeling shifts from high-dimensional spatial fields to a set of time series.

\subsection{Quantities of interest}\label{sec:scale-aware-qois}
The TO closure builds on a small set of QoIs, which we introduce here. We use two groups: scale-aware energies and enstrophies, and for wall-bounded flows mean-profile QoIs, which constrain the mean streamwise velocity profile.

The scale-aware group is built from two physically motivated quantities, kinetic energy and enstrophy, which characterize the spectral distribution of turbulent fluctuations and their transfer across scales. On the computational domain $\Omega$ they are defined as
\begin{align}
    E = \frac{1}{2} \int_\Omega \|{\boldsymbol{v}}\|^2 \, \D \boldsymbol{x}, \quad \quad
    Z = \frac{1}{2} \int_\Omega \|{\boldsymbol{\omega}}\|^2 \, \D\boldsymbol{x} = \frac{1}{2} \int_\Omega  \|\text{curl}( {\boldsymbol{v}})\|^2 \, \D\boldsymbol{x}.
\end{align}
We always evaluate QoIs on coarse(-grained) velocity fields.

To resolve interactions across scales, we make these QoIs scale-aware by evaluating them after a scale-selective filter. In previous work \cite{hoekstra2024Reduced_data-driven, hoekstra_reduced_2026} we used sharp Fourier filters that partition the spectrum into wavenumber bins $[l, m]$:
\begin{equation} \label{eq:scale-aware-energy-fourier}
    E_{R_{[l, m]}}
    =  \frac{1}{2} \int_\Omega \left\|R_{[l, m]} {\boldsymbol{v}}\right\|^2 \ \D\boldsymbol{x}, \quad Z_{R_{[l, m]}}
    =  \frac{1}{2} \int_\Omega  \left\|R_{[l, m]} {\boldsymbol{\omega}}\right\|^2 \ \D\boldsymbol{x},
\end{equation}
where $R_{[l,m]}$ is the sharp Fourier filter
\begin{align}
    R_{[l,m]} &= \mathcal{F}^{-1} \hat{R}_{[l,m]} \mathcal{F}, \nonumber\\
    \hat{R}_{[l, m]} \,\hat{\boldsymbol{y}}_{\bf k} &=
    \begin{cases}
        \hat{\boldsymbol{y}}_{\mathbf{k}} & \text{if} \quad l-\frac{1}{2} \leq \lVert{\bf k}\rVert_2 < m + \frac{1}{2} \\
        0 & \text{otherwise}
    \end{cases}.
\end{align}
Here $\hat{\boldsymbol{y}} = \mathcal{F}{\boldsymbol{y}}$ is the discrete Fourier transform of the field and $\mathbf{k}$ are the wavenumber vectors. Such filters are well defined only on rectangular periodic domains, which limits their use beyond academic test cases.

To remove this geometric restriction, we replace the Fourier filters with convolutions against compact kernels in physical space. We use two complementary kernels: a Gaussian low-pass filter $G_s$ with standard deviation $s$, and a Laplacian-based high-pass filter $L_s$ that highlights fluctuations on scale $s$ \cite{nathen_adaptive_2018}. Writing the integer lattice offset as $\boldsymbol{r}$, with $d$ the spatial dimension, the Gaussian low-pass kernel is given by
\begin{equation}\label{eq:gaussian}
    G_s(\boldsymbol{r}) = \frac{\exp\!\left( -\|\boldsymbol{r}\|^2 / (2 s^2) \right)}{\sum_{\boldsymbol{r} \in [-\alpha s, \alpha s]^d}\exp\!\left( -\|\boldsymbol{r}\|^2 / (2 s^2) \right)},
\end{equation}
truncated to $\boldsymbol{r} \in [-\alpha s, \alpha s]^d$, where we used $\alpha = 4$ in 2D and $\alpha = 2$ in 3D. The Laplacian-based high-pass kernel is built from the square of the discrete Laplacian, the discrete biharmonic stencil $\ell = [1,-4,6,-4,1]$. At unit scale, the $d$-dimensional kernel places this stencil along each coordinate axis through the origin,
\begin{equation}\label{eq:laplace}
    L_1(\boldsymbol{r}) = \sum_{a=1}^{d} \ell \left( \lvert r_a \rvert+2 \right) \prod_{b \neq a} \delta_{r_b, 0},
\end{equation}
where $\delta$ is the Kronecker delta, so that $L_1$ is nonzero only on the axes, resulting in a $5\times5$ cross-stencil in two dimensions. The scale-$s$ kernel $L_s$ is obtained by dilating $L_1$: each entry is replaced by an $s\times\cdots\times s$ block of equal value (a Kronecker product with the all-ones tensor), widening the stencil to scale $s$ while preserving its shape. As $s$ increases, the response peak shifts toward smaller wavenumbers. We normalize $L_s$ to $\sum_{\boldsymbol{r}} |L_s| = 2$. Like the Gaussian, this kernel is self-adjoint. On non-periodic, wall-bounded axes the filters require a boundary extension of the field, for which we use zero padding.

\begin{figure}
    \centering
    \includegraphics[width=0.8\textwidth]{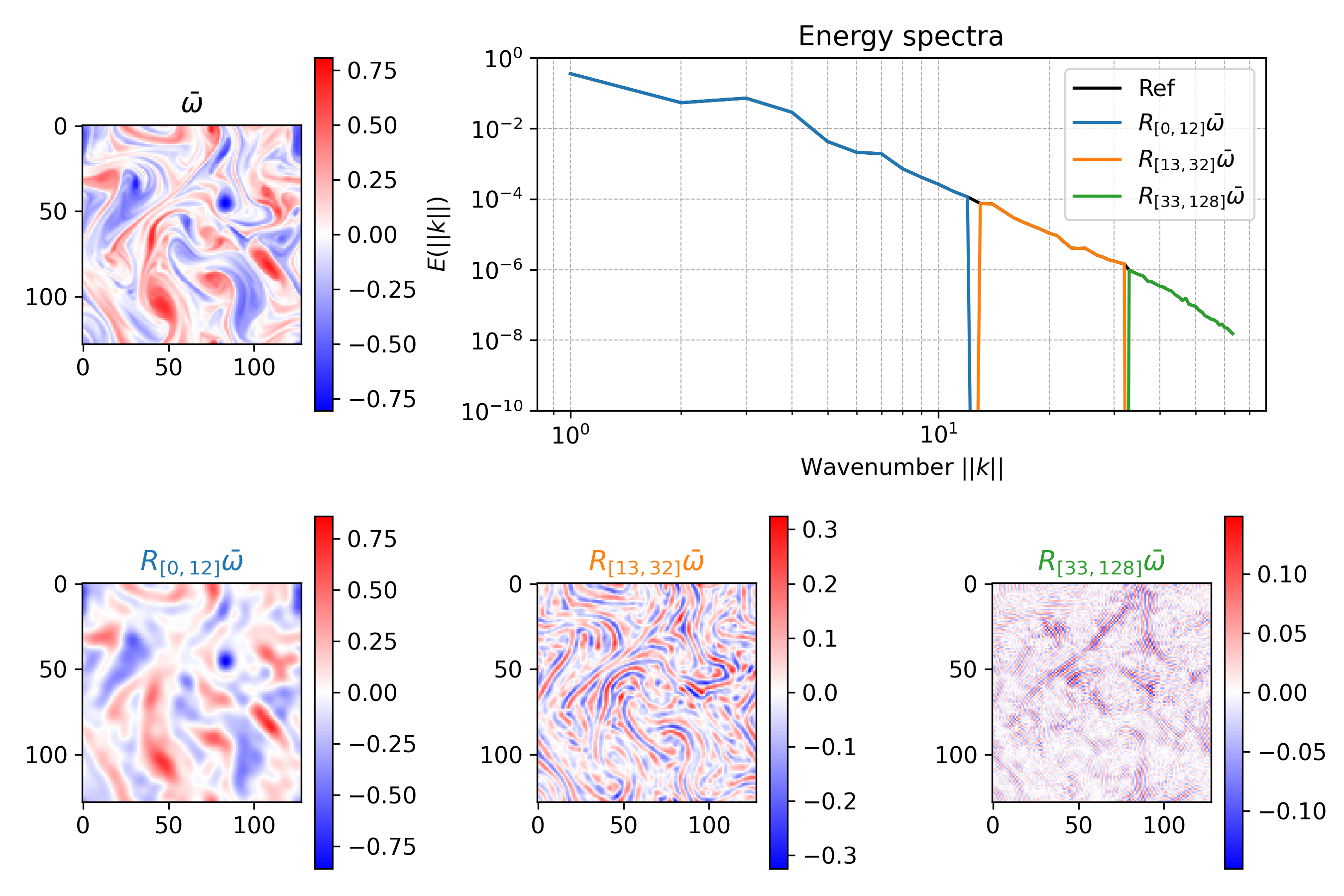}
    \caption{Effect of the sharp Fourier filters $\{R_{[0,12]}, R_{[13,32]}, R_{[33,128]}\}$ on a fully developed turbulent vorticity field from a two-dimensional, $Re=10\,000$ Kolmogorov flow.}
    \label{fig:fourier-filters}
\end{figure}

\begin{figure}
    \centering
    \includegraphics[width=0.8\textwidth]{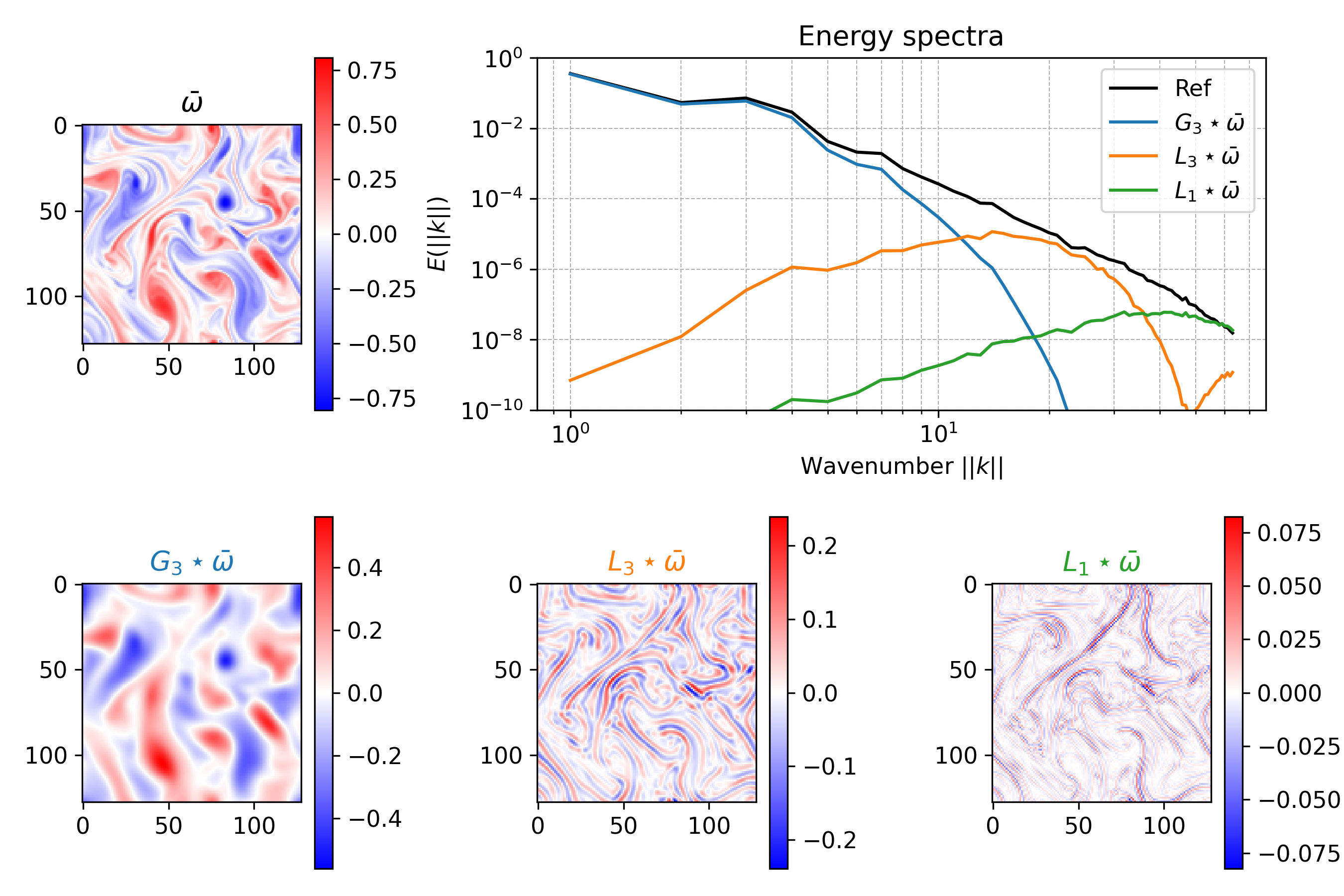}
    \caption{Effect of the compact kernel filters $\{G_3, L_3, L_1\}$ on a fully developed turbulent vorticity field from a two-dimensional, $Re=10\,000$ Kolmogorov flow.}
    \label{fig:kernel-filters}
\end{figure}

Figures~\ref{fig:fourier-filters} and \ref{fig:kernel-filters} compare the two filter families on a fully developed two-dimensional turbulent flow field. Figure~\ref{fig:fourier-filters} shows the effect of the Fourier filters $\{R_{[0,12]}, R_{[13,32]}, R_{[33,128]}\}$, while Figure~\ref{fig:kernel-filters} shows the effect of the kernel filters $\{G_3, L_3, L_1\}$, applied with periodic padding. The two families produce a comparable separation of large, intermediate, and small scales, even though the kernel-based filters do not sharply cut off in Fourier space.


On wall-bounded geometries we add a second group of QoIs that act on the mean streamwise velocity profile directly. These mean-profile (mp) QoIs are linear functionals that weight the streamwise velocity over a wall-normal band,
\begin{equation}\label{eq:mean-profile-qoi}
    Q_j^{\text{mp}}(\boldsymbol{v}) = \int_{\Omega} w_j(y)\, {v}_x(\boldsymbol{x})\, \D \boldsymbol{x},
\end{equation}
where the band weights $w_j(y)$ are smooth. Adding the $Q_j^{\text{mp}}$ to the QoI set pins the mean streamwise velocity profile; the specific bands used for the channel test case are given in Section~\ref{sec:3d-channel-qois}.

\subsection{Constructing spatial patterns}

The spatial patterns $\boldsymbol{O}_i$ must satisfy the orthogonality constraints \eqref{eq:orthogonality}. We construct them from the functional derivatives (sensitivities) $\delta Q_i /\delta {\boldsymbol{v}}$ of the QoIs, which admit closed forms derived in \ref{app:derivatives of QoIs}.

We construct each spatial pattern as a linear combination:
\begin{equation}
    \boldsymbol{O}_i = \sum_{j=1}^{N_Q} c_{ij} \boldsymbol{V}_j(\boldsymbol{v}, \boldsymbol{x}), \quad \textrm{for}\;\;\; 1 \leq i \leq N_Q,
    \label{eq:O-from-V}
\end{equation}
where the basis fields $\boldsymbol{V}_j$ are set equal to the QoI sensitivities, $\boldsymbol{V}_j = \delta Q_j/\delta\boldsymbol{v}$. The coefficients $c_{ij}$ are determined from the orthogonality constraints in \eqref{eq:orthogonality}, additionally requiring $c_{ii} = 1$. This choice of basis functions allows the spatial patterns to be written in closed form. The resulting linear system is solvable provided the sensitivities $\boldsymbol{V}_j$ are linearly independent.

\subsection{Tracking}
Equation~\eqref{eq:influence-SGS-term-on-LF-QoI-trajectories} provides a direct mechanism for nudging an LES toward prescribed QoI trajectories. Given reference values $Q_i^{ref}(t^n)$, we adopt a predictor-corrector procedure: first advance the solver without SGS forcing to obtain $Q_i(\boldsymbol{v}^{\text{p},n^*})$, then compute the coefficient $\gamma_i(t^n)$ from the discrepancy with the reference:
\begin{align}
    dQ_i^{n} = \left(Q_i^{ref}(t^{n}) - Q_i(\boldsymbol{v}^{\text{p},n^*}) \right), \label{eq:dQ-from-ref} \\
    \gamma_i(t^n) =  dQ_i^{n} / \left( \int_\Omega \frac{\delta Q_i}{\delta \boldsymbol{v}}\bigg|_{\boldsymbol{v}^{\text{p},n^*}} \cdot \, \boldsymbol{O}_i(\boldsymbol{v}^{\text{p},n^*}, \boldsymbol{x}) \ \D \boldsymbol{x} \right). \label{eq:tau-from-dQ-for-tracking}
\end{align}
Here $dQ_i^{n}$ is the SGS correction to the $i$-th QoI at
 the end of the $n$-th time step.

\begin{figure}
    \centering
    \resizebox{\textwidth}{!}{
    \begin{tikzpicture}[
        boxnode/.style={draw, align=center, minimum width=2cm, minimum height=1cm},
        arrow/.style={->, thick},
        yellowarrow/.style={->, thick, Sepia},
        redarrow/.style={->, thick, red},
        labelstyle/.style={midway, font=\small, right}
        ]
        \tikzmath{\h = 3; \w = 3.3;}

        \normalsize
        \node (a) at (0, \h) {$\boldsymbol{f}^p(t_{n-1})$};
        \node (b) at (2.9, \h) {$\boldsymbol{f}(t_{n-1}+\Delta t_{})$};
        \node (c) at (6.1, \h) {$\boldsymbol{f}^a(t_{n-1}+\Delta t_{})$};
        \node (d) at (3*\w, \h) {$\boldsymbol{f}^p(t_{n-1}+\Delta t_{})$};
        \node (e) at (13, \h) {$\boldsymbol{f}^p(t_{n-1}+m \Delta t_{})$};

        \scriptsize
        \draw[arrow] (a) -- node[midway, above] {stream} (b);
        \draw[arrow] (b) -- node[midway, above] (highest hf) {collide} (c);
        \draw[arrow] (c) -- node[midway, above] {force $\Delta\boldsymbol{u}_F$} (d);
        \draw[arrow] (d) edge[dashed] (e);

        \normalsize
        \node (f) at (0, 0) {$\tilde{\boldsymbol{f}}^p(t_{n-1})$};
        \node (g) at (\w, 0) {$\tilde{\boldsymbol{f}}(t_n)$};
        \node (h) at (2*\w, 0) {$\tilde{\boldsymbol{f}}^{a}(t_n)$};
        \node (i) at (10, 0) {$\tilde{\boldsymbol{f}}^{p^*}(t_n)$};
        \node (j) at (13.76, 0) {$\tilde{\boldsymbol{f}}^p(t_n)$};

        \node (l) at (13, 1) {$Q(\boldsymbol{v}^{p,n^*})$};

        \node (m) at (13, 2.2) {$Q(\bar{\boldsymbol{u}}^{p,n})$};
        
        \scriptsize
        \draw[arrow] (f) -- node[midway, above] {stream} (g);
        \draw[arrow] (g) -- node[midway, above] {collide} (h);
        \draw[arrow] (h) -- node[midway, above] {force $\Delta\boldsymbol{v}_F$} (i);
        \draw[arrow] (i) -- node[midway, above] (q) {$\Delta\boldsymbol{v}_{SGS}(dQ^{n})$} (j);
        \node[] at (q.north) {force}; 
        
        \draw[arrow] (i) |- node[above right] (highest lf) {} (l);
        \draw[arrow] (e) -- (m);
        
        \normalsize
        \draw[<->, thick] (l) -- node[right] {$dQ^{n}$} (m);

        \begin{scope}[on background layer]
            \node[draw=blue, fill=blue!20, thick, inner sep=1pt, fit=(a) (e) (m) (highest hf)] (HF) {};
        \end{scope}
        \begin{scope}[on background layer]
            \node[draw=green, fill=green!20, thick, inner sep=1pt, fit=(f) (l) (j)] (LF) {};
        \end{scope}
        \node[above right, blue!50!black] at (HF.south west) {HF};
        \node[below right, green!50!black] at (LF.north west) {LF};

\end{tikzpicture}
    }
    \caption{Predictor-corrector tracking of a HF reference simulation by a LF LBM solver. Here $\Delta t$ is the time step of the HF solver and $t_{n-1}$, $t_n$ are the LF time steps, with $t_n = t_{n-1}+m\Delta t$. The SGS correction $dQ$ is determined from the difference between the post-collision QoI of the LF predictor and the post-collision QoI of the HF reference projected onto the coarse grid.}
    \label{fig:tracking-diagram}
\end{figure}
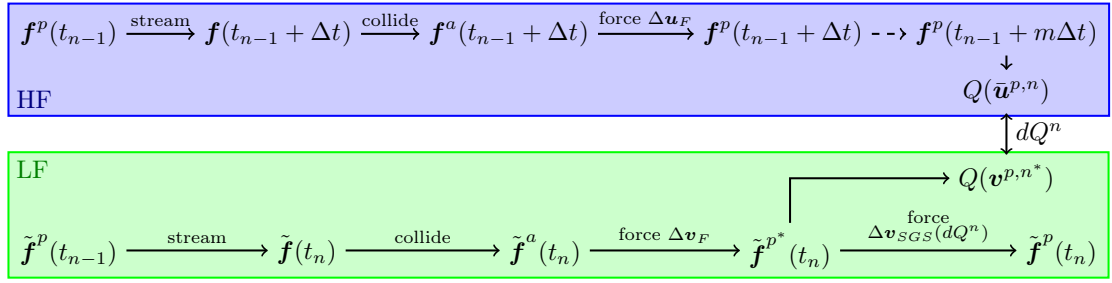

Figure \ref{fig:tracking-diagram} illustrates how this setup can be used to track QoI reference trajectories from a high-fidelity simulation. While this tracking procedure is useful for understanding and validating the SGS dynamics, it requires access to reference trajectories from a high-fidelity simulation at every time step during the LES.

\subsection{Time-series models for SGS correction} \label{section:Time-Series Models for SGS Correction}

For practical applications, the SGS model must predict corrections without access to reference trajectories. We therefore train a multivariate time-series model on the tracking data, learning the mapping from past QoI states to SGS corrections.

\begin{figure}
    \centering
    \begin{tikzpicture}[
        boxnode/.style={draw, align=center, minimum width=2cm, minimum height=1cm},
        arrow/.style={->, dashed, thick},
        yellowarrow/.style={->, thick},
        redarrow/.style={->, thick, red},
        labelstyle/.style={midway, font=\small, right}
        ]

        % Nodes
        \node[boxnode] (qstar_n2) at (0,2) {$\boldsymbol{q}^{n-2^*}$};
        \node[boxnode] (q_n2) at (0,0) {$\boldsymbol{q}^{n-2}$};
        \node[boxnode] (qstar_n1) at (3,2) {$\boldsymbol{q}^{n-1^*}$};
        \node[boxnode] (q_n1) at (3,0) {$\boldsymbol{q}^{n-1}$};
        \node[boxnode] (qstar_n) at (6,2) {$\boldsymbol{q}^{n^*}$};
        \node[boxnode] (q_n) at (9,0) {{$\boldsymbol{q}^{n}$}};

        % Arrows (Standard transitions with labels on them)
        \draw[arrow] (qstar_n2) -- (q_n2) node[labelstyle] {$dQ^{n-2}$} ;
        \draw[arrow] (qstar_n1) -- node[labelstyle] {$dQ^{n-1}$} (q_n1);
        \draw[arrow] (qstar_n) -- node[midway, right] {$\;\; dQ^{n}$} (q_n);

        % Yellow arrows (LF solver)
        \draw[yellowarrow] (q_n2) -- (qstar_n1);
        \draw[yellowarrow] (q_n1) -- (qstar_n) node[midway, right]{\, LF solver};

        % Model hist box (Dotted black outline)
        \node[draw, dashed, thick, inner sep=10pt, fit=(qstar_n2) (q_n1)] (modelhist) {};
        \node[above right] at (modelhist.south east) {Model hist};

        % Model input box (Blue with low opacity, covering both rows)
        \begin{scope}[on background layer]
            \node[draw=blue, fill=blue!20, thick, inner sep=15pt, fit=(q_n2) (qstar_n)] (modelinput) {};
        \end{scope}
        \node[above right, blue] at (modelinput.north west) {LinReg input};
    \end{tikzpicture}

    \caption{Setup of the time-series model. Solid arrows are LF LES steps without SGS forcing (predictor). During tracking the corrections (dashed arrows) are computed from \eqref{eq:dQ-from-ref}; at inference time the time-series model predicts $\boldsymbol{q}^n$ from the historical inputs in the blue box, and $\gamma_i$ then follows from \eqref{eq:tau-from-dQ-for-tracking}.}
    \label{fig:pred-cor-sgs-model-setup}
\end{figure}
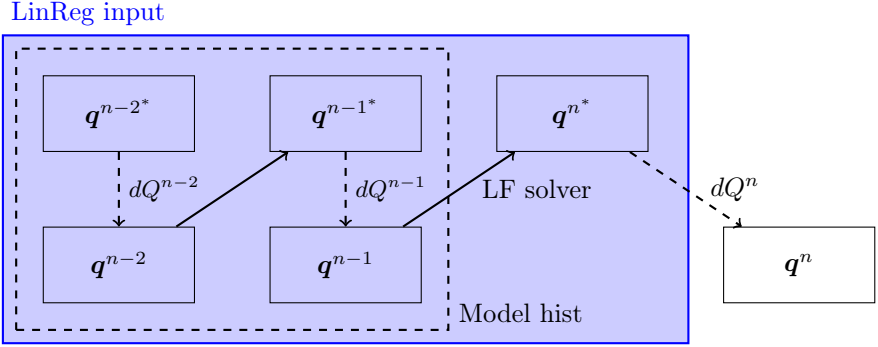

The overall setup is summarized in Figure~\ref{fig:pred-cor-sgs-model-setup}. We denote the QoI vectors at time step $n$ by $\boldsymbol{q}^n = Q(\boldsymbol{v}^{\text{p},n})$ for the corrected state and $\boldsymbol{q}^{n^*} = Q(\boldsymbol{v}^{\text{p},n^*})$ for the predictor.

To model the unresolved time-series dynamics, we use a linear regression with stochastic residuals (LRS) \cite{hoekstra_reduced_2026}. The model predicts the corrected QoI vector $\boldsymbol{q}^{n}$ from a history of length $h$ of previous states:
\begin{align}
    \mathbf{LRS}(\underline{\boldsymbol{q}}_h) = \underline{\boldsymbol{q}}_h C + \boldsymbol{\eta}, \quad \boldsymbol{\eta} \sim \mathcal{N}(\boldsymbol{\mu}, \Sigma), \nonumber\\
    \underline{\boldsymbol{q}}_h = [\boldsymbol{q}^{n-1}, \dots, \boldsymbol{q}^{n-h}, \boldsymbol{q}^{n^*}, \boldsymbol{q}^{n-1^*}, \dots, \boldsymbol{q}^{n-h^*}, 1]. \label{eq:Linear-Regression-with-Stochastic-Residuals}
\end{align}
where $\underline{\boldsymbol{q}}_h$ contains all model inputs for history length $h$, $C \in \mathbb{R}^{ \mathrm{len}(\underline{\boldsymbol{q}}_h) \times N_Q}$ is a learnable regression matrix, and $\mathcal{N}$ is an $N_Q$-dimensional multivariate Gaussian distribution. The final entry ``1'' provides a bias term. We fit the model to predict corrected QoIs $\boldsymbol{q}^{n}$. Before fitting, the inputs and the prediction targets are scaled by the standard deviation of the reference trajectories; no mean removal is applied. 

We fit $C$ by regularized least squares with a Frobenius penalty, which improves long-term stability of the coupled LES--SGS system:
\begin{equation}
    C = \argmin_{X \in \mathbb{R}^{ \mathrm{len}(\underline{\boldsymbol{q}}_h) \times N_Q}} \frac{1}{2} \lVert \mathbf{Q}_h X - \mathbf{Q} \rVert_F^2 + \frac{\lambda}{2} \lVert X \rVert_F^2,
\end{equation}
where the $n$-th rows of $\mathbf{Q}_h$ and $\mathbf{Q}$ are $\underline{\boldsymbol{q}}_h^n$ and $\boldsymbol{q}^{n}$ from the training data, and $\lambda \geq 0$ is the regularization parameter \footnote{Even after rescaling the inputs and targets, the $dQ$ corrections vary strongly in magnitude across QoIs, so we rescale the penalty per QoI: the $i$-th column of $C$ uses $\lambda_i = \lambda \, (\sigma_i / \sigma_1)^2$, with $\sigma_i$ the standard deviation of the scaled corrections. Quoted $\lambda$ values are the base $\lambda$.}. The residual statistics $(\boldsymbol{\mu}, \Sigma)$ are estimated from the training residuals $\mathbf{Q} - \mathbf{Q}_h C$.

\subsection{Final integration into LES}
Replacing the reference-dependent tracking with autonomous predictions from the trained time-series model yields a fully independent SGS closure; we refer to LES driven by the trained model as \emph{online} simulations, as opposed to the reference-driven \emph{tracking} simulations. The resulting TO-LRS model is remarkably parsimonious: it has only two hyperparameters, the history length $h$ and the regularization strength $\lambda$, and $\mathcal{O}(h \, {N_Q}^2)$ learnable parameters ($10^4$--$10^5$ for the cases studied here). Because the model predicts corrections to physically meaningful QoIs rather than to high-dimensional fields, its outputs remain directly interpretable.

\section{Eddy-viscosity baselines}\label{sec:smagorinsky-model}
We compare the TO-LRS model against two eddy-viscosity models: the classical Smagorinsky model and, for the wall-bounded channel flow, the Wall-Adapting Local Eddy-viscosity (WALE) model. These models serve a dual role: as standalone baselines against which TO-LRS is judged, and in the channel case as a substrate on which the TO correction operates. 

Both model the unresolved stresses through an eddy viscosity $\nu_t$ that augments the molecular viscosity in the collision operator, resulting in an effective relaxation time:
\begin{equation}\label{eq:eff-relaxation}
    \tau_{\text{eff}} = \frac{\nu + \nu_t}{c_s^2} + \frac{\Delta t}{2}.
\end{equation}
We follow the LBM-specific implementation of \cite{gkoudesnes_implementation_nodate} for both. The two models differ only in how $\nu_t$ is computed from the resolved field.

\subsection{The Smagorinsky model}
The Smagorinsky model takes an eddy viscosity proportional to the local strain rate \cite{smagorinsky_general_1963},
\begin{equation}\label{eq:smag}
    \nu_t = C_{smag}^2 \Delta x^2 \sqrt{2 S_{ij} S_{ij}},
\end{equation}
where $S_{ij} = \frac{1}{2}\left(\partial_i v_j + \partial_j v_i\right)$ is the resolved strain-rate tensor, $C_{smag}$ the tunable Smagorinsky constant, and $\Delta x$ denotes the lattice spacing. In the lattice Boltzmann method the strain rate can be obtained locally from the non-equilibrium moments of the distribution functions,
\begin{equation} \label{eq:strain-from-non-equilibrium}
    S_{ij} = -\frac{1}{2 \rho c_s^2 \tau_{\text{eff}}} \Pi^{\text{neq}}_{ij}, \quad \Pi^{\text{neq}}_{ij} = \sum_k c_{ki} c_{kj} (f_k - f_k^{\text{eq}}).
\end{equation}
Because $S_{ij}$ itself depends on $\tau_{\text{eff}}$, inserting $\nu_t$ from \eqref{eq:smag} into \eqref{eq:eff-relaxation} gives an implicit equation for $\tau_{\text{eff}}$; solving yields the closed form
\begin{equation}
    \tau_{\text{eff}} = \frac{1}{2} \left( \tau + \sqrt{\tau^2 + \frac{18 C_{smag}^2 \lvert \Pi^{\text{neq}} \rvert}{\rho}}\right), \quad \lvert \Pi^{\text{neq}} \rvert = \sqrt{2\, \Pi^{\text{neq}}_{ij} \Pi^{\text{neq}}_{ij}},
\end{equation}
where we used $\Delta x = \Delta t = 1$ and $c_s^2 = 1/3$.

\subsection{The WALE model}\label{sec:wale-model}
For the channel flow we additionally consider the WALE model \cite{nicoud_ducros_1999}, which---unlike the Smagorinsky model---recovers the correct near-wall asymptotic scaling of the eddy viscosity:
\begin{equation}
    \nu_t = C_w^2 \Delta x^2 \frac{\left(S^d_{ij} S^d_{ij}\right)^{3/2}}{\left(S_{ij} S_{ij}\right)^{5/2} + \left(S^d_{ij} S^d_{ij}\right)^{5/4}},
\end{equation}
where $C_w$ is the WALE constant, $S_{ij}$ the resolved strain-rate tensor as above, and $S^d_{ij}$ the traceless symmetric part of the square of the velocity-gradient tensor $g_{ij} = \partial_j v_i$,
\begin{equation}
    S^d_{ij} = \tfrac{1}{2}\left(g_{ik}g_{kj} + g_{jk}g_{ki}\right) - \tfrac{1}{3}\delta_{ij}\, g_{lk}g_{kl}.
\end{equation}
In contrast to the Smagorinsky strain rate, $g_{ij}$ is computed by second-order central differences in lattice units.

\section{Two-dimensional Kolmogorov flow}\label{sec:kolmogorov-flow}
Two-dimensional Kolmogorov flow is a standard test case for data-driven subgrid-scale modeling \cite{fischer_optimal_2025, kochkov_machine_2021, hoekstra2024Reduced_data-driven}. A sinusoidal forcing along one axis generates a shear flow that subsequently breaks up into homogeneous turbulence.

\subsection{Governing equations} \label{sec:governing eqations kolmogorov flow}
We consider two-dimensional Kolmogorov flow on a doubly-periodic square domain $[0, 2\pi] \times [0, 2\pi]$:
\begin{align}
    \frac{\partial \boldsymbol{u}}{\partial t} + \boldsymbol{u} \cdot \nabla \boldsymbol{u} &= \frac{1}{Re} \nabla^2 \boldsymbol{u} - \frac{1}{\rho} \nabla p + \boldsymbol{f}, \\
    \nabla \cdot \boldsymbol{u} &= 0,
\end{align}
where $\boldsymbol{u}$ denotes the velocity field, $Re$ the Reynolds number, $p$ the pressure field and $\rho$ the fluid density, which we set to 1. The forcing term is given by
\begin{equation}
    \boldsymbol{f} = \sin(4y) \boldsymbol{e}_x - 0.1 \boldsymbol{u}.
\end{equation}
Here $\boldsymbol{e}_x$ denotes the unit vector in the x-direction. This setup matches that of \cite{fischer_optimal_2025}.

\paragraph{Lattice units}
Lattice Boltzmann simulations are performed in lattice units, denoted in this section by a star, e.g.\ $\boldsymbol{u}^\star$ and $t^\star$. We choose conversion factors from physical to lattice units such that $\Delta t^\star = \Delta x^\star = \rho^\star = 1$ and ${c_s^\star}^2 = 1/3$. We take $U = 4$ as the characteristic velocity of the flow, and we fix the velocity conversion such that this characteristic velocity corresponds to $0.1\, c_s^\star$ in lattice units, well below the lattice speed of sound. The complete set of conversion factors for length, velocity, and density for an $N \times N$ LBM simulation is then
\begin{equation}
    C_l = \frac{2\pi}{N}, \quad C_u=\frac{U}{0.1 c_s^{\star}} = \frac{4}{0.1 c_s^{\star}}, \quad C_\rho = \rho = 1.
\end{equation}
Note that the conversion factor for velocity is independent of the grid resolution. This allows us to coarse-grain our solution fields without rescaling. Conversion for all other parameters and variables follows from these conversion factors and the choice of Reynolds number.

\subsection{Test setup}
We evaluate SGS models in LES on an LF grid of size $128 \times 128$ against a reference DNS at $Re=10\,000$ on a $2048 \times 2048$ grid.

We first run a DNS burn-in phase to obtain a fully developed turbulent field. Starting from this field, we then run DNS for 125 time units (TU) to generate the reference QoI trajectories. The first 14, 28, or 57 TU of these trajectories are used to train the TO SGS models, and the full 125 TU trajectory serves as the reference against which the long-term QoI distributions of the LF simulations with trained models are evaluated. The DNS uses $2.3 \cdot 10^4$ steps per TU, which is reduced to $1.4 \cdot 10^3$ steps per TU in the LF simulations.

\subsection{Calibrating the Smagorinsky constant}
\begin{figure}[htbp]
    \centering
    \includegraphics[width=0.4\linewidth]{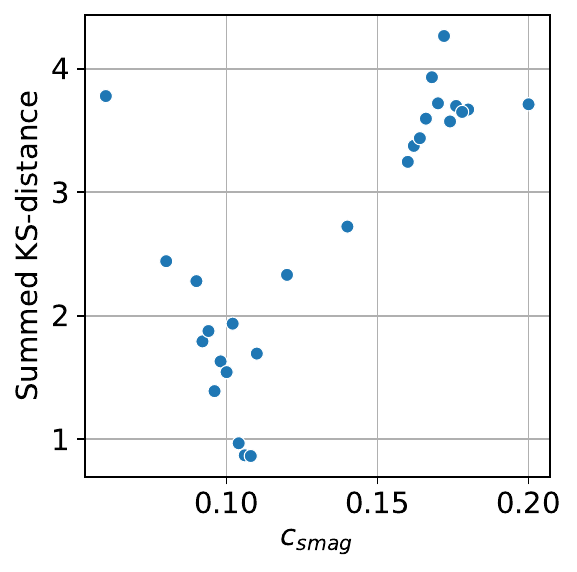}
    \caption{Optimization of the Smagorinsky constant for the Kolmogorov flow with respect to the summed KS-distance of kernel-based QoIs: $\{E_{G_3}, Z_{G_3}, E_{L_3}, Z_{L_3}, E_{L_1}, Z_{L_1}\}$. The QoI distributions are compared over 57 TU.}
    \label{fig:optimize-smag-summed-KS}
\end{figure}

We calibrate the Smagorinsky model by running a set of LF simulations with different values for the model constant. Each LF simulation is run for 57 TU, matching the largest training window used for the data-driven models in later sections, and the resulting QoI distributions are compared with those of the corresponding part of the reference simulation. Figure \ref{fig:optimize-smag-summed-KS} reports the summed KS-distance for the different model constant values. The minimum summed KS-distance is obtained at $C_{smag}=0.106$. KS-distances for the individual QoIs are reported in \ref{appendix:KS-distances for individual QOIs}.    

\subsection{Training the TO method}
Training consists of tracking the reference QoI trajectories with an LF simulation, which yields a time series of corrections per QoI to which we fit the LRS model.

Nudging at the post-collision state is highly effective, yielding only minor deviations from the reference trajectories (see~\ref{app:tracking}). On the corresponding post-streaming states, the small-scale energy $E_{L_1}$ is overestimated because the grid-scale energy removed by the SGS forcing in each step is partly regenerated by streaming. The nudged LF simulations remain stable, whereas the same solver without SGS forcing becomes unstable within 10 TU.

\begin{figure}[t]
    \centering
    \includegraphics[width=0.9\linewidth]{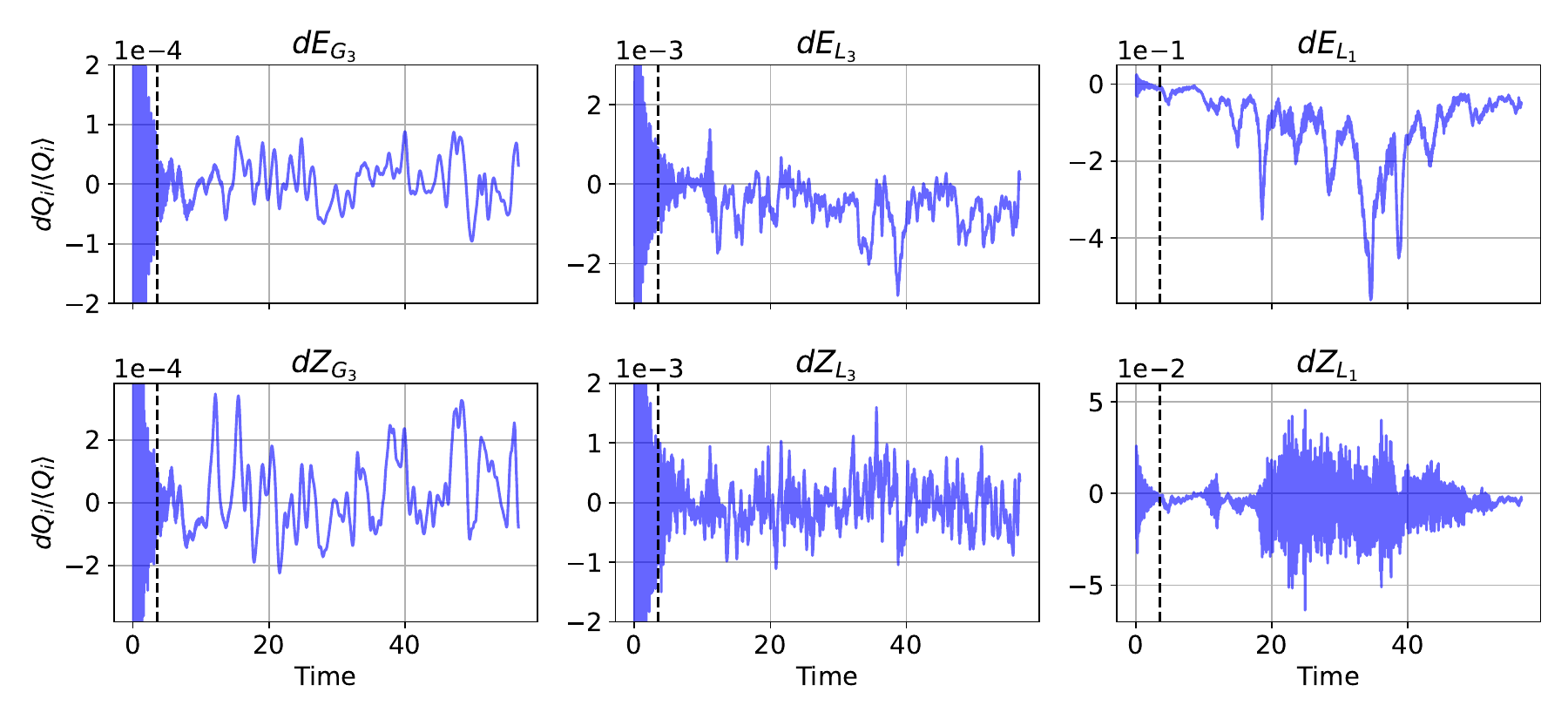}
    \caption{SGS corrections $dQ_i^n$ applied during tracking of the Kolmogorov flow, normalized by the time-averaged value of the corresponding QoI. The dashed vertical line marks the end of the spin-up window; only the trajectories to the right are used for training.}
    \label{fig:dQ-in-tracking}
\end{figure}

Figure \ref{fig:dQ-in-tracking} shows the corrections applied during tracking, normalized by each QoI's time-averaged value. They are generally small. The largest corrections are on small-scale energy, and predominantly remove energy from these scales, as physically expected. Early in tracking the corrections are large but they decay rapidly. We treat this as a spin-up transient from the abrupt change in solver resolution and exclude the first 5000 time steps (3.5 TU) from the training data.

\subsection{Online results}\label{sec:results}
We now report the online performance of the TO-LRS model. All LF simulations were run for 125 TU. Because TO-LRS injects stochastic forcing, each configuration was evaluated in an ensemble of five replicas. We first examine the effect of the two model hyperparameters, history length $h$ and regularization strength $\lambda$, together with the size of the training window. We then examine the turbulence statistics of the best models.

\subsubsection*{TO-LRS hyperparameters}
\begin{figure}[htbp]
    \centering
    \includegraphics[width=0.8\linewidth]{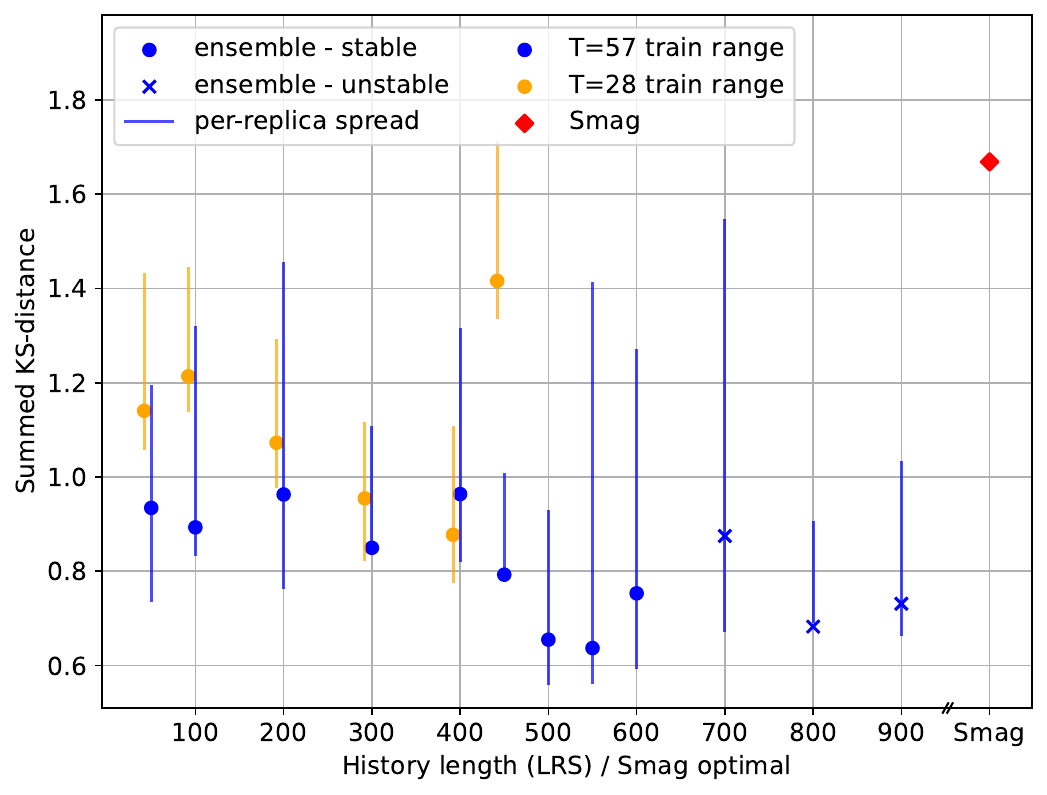}
    \caption{Online performance over 125 TU for TO-LRS models with different history lengths, trained on 57 or 28 TU without regularization, on the Kolmogorov flow. Results are based on five-replica ensembles; dots show the pooled summed KS-distance over stable replicas, vertical bars show the replica spread, and an `X' marks ensembles in which at least one replica was unstable. The Smagorinsky baseline, calibrated on 57 TU of training data, is included for reference.}
    \label{fig:TO-LRS-online-performance-lambda0}
\end{figure}

We start from the unregularized case ($\lambda = 0$) and vary the history length and the training window. Figure~\ref{fig:TO-LRS-online-performance-lambda0} reports the summed KS-distance for models trained on the 28- and 57-TU training windows; the 14-TU trained models have no stable simulations at $\lambda = 0$ and are omitted.

With the largest training window (57 TU), history lengths between 50 and 600 time steps produce stable simulations, while longer histories begin to destabilize some replicas. The best performance is obtained around $h=500$, and every stable history length outperforms the Smagorinsky baseline in summed KS-distance. Shrinking the training window to 28 TU slightly degrades the accuracy and makes histories above $h=450$ unstable.

\begin{figure}[htbp]
    \centering
    \includegraphics[width=0.8\linewidth]{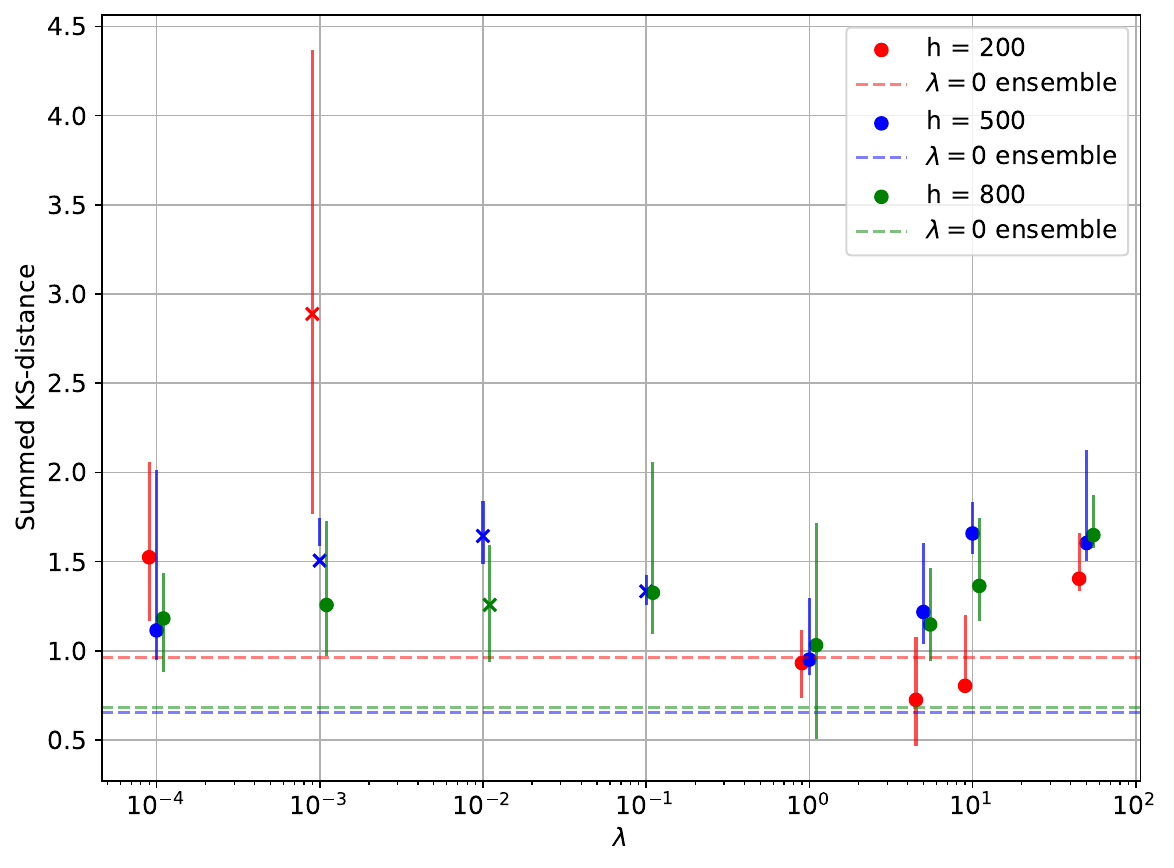}
    \caption{Influence of regularization on TO-LRS models trained on 57 TU, on the Kolmogorov flow. Results are based on five-replica ensembles over 125 TU. Dots show the pooled summed KS-distance over stable replicas; an `X' marks ensembles in which at least one replica became unstable. Vertical bars show the replica spread, and dashed lines show the pooled summed KS-distance at $\lambda = 0$.}
    \label{fig:TO-LRS-online-performance-against-regularization-T57}
\end{figure}

We next add regularization. Figure~\ref{fig:TO-LRS-online-performance-against-regularization-T57} shows the effect on the largest training set (57 TU) for three history lengths, with dashed lines marking the corresponding $\lambda=0$ baseline. The picture is mixed: accuracy generally decreases with $\lambda$, although for $h=200$ a few regularized ensembles outperform the unregularized model. Strong regularization stabilizes long histories: $h=800$, which had unstable replicas at $\lambda=0$, becomes consistently stable. Weak regularization, on the other hand, can be destabilizing: for $h=200$, all replicas at $\lambda \in \{0.01, 0.1\}$ are unstable. Among the stable regularized models, the best summed KS-distance is typically obtained for $\lambda \in [1,10]$.

\begin{figure}[htbp]
    \begin{subfigure}[b]{0.49\textwidth}
        \centering
        \includegraphics[width = \linewidth]{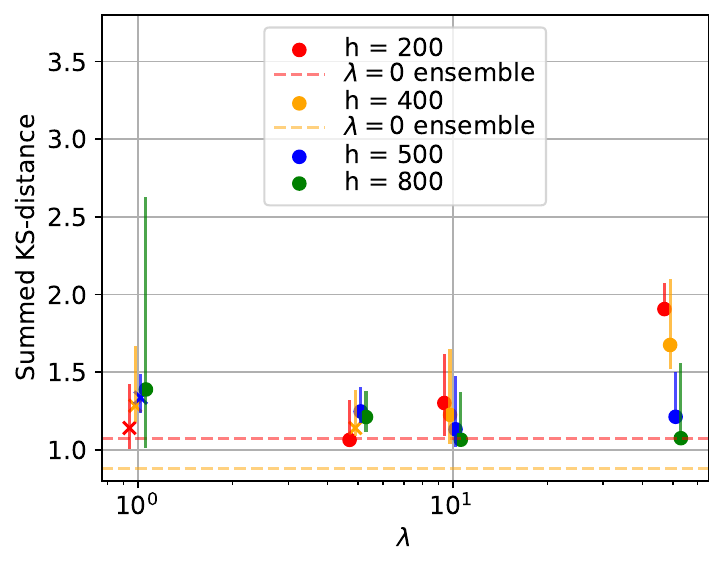}
        \caption{Training data up to $T=28$.}
    \end{subfigure}
    \begin{subfigure}[b]{0.49\textwidth}
        \centering
        \includegraphics[width = \linewidth]{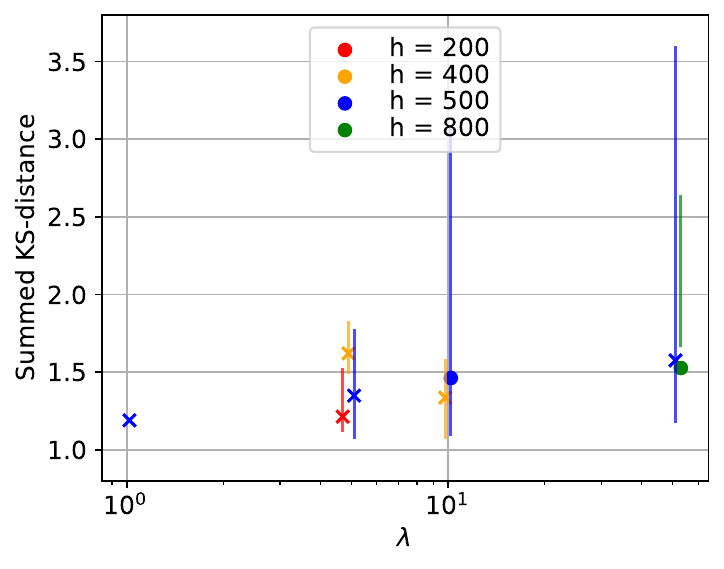}
        \caption{Training data up to $T=14$.}
    \end{subfigure}
    \caption{Influence of regularization on TO-LRS models trained on smaller datasets, on the Kolmogorov flow. Results are based on five-replica ensembles over 125 TU. Dots show the pooled summed KS-distance over stable replicas; an `X' marks ensembles in which at least one replica became unstable. Vertical bars show the replica spread, and dashed lines show the pooled summed KS-distance at $\lambda = 0$ when those ensembles were stable.}
    \label{fig:TO-LRS-online-performance-against-regularization-few-data}
\end{figure}

Figure~\ref{fig:TO-LRS-online-performance-against-regularization-few-data} shows the stabilizing effect of regularization on the smaller training sets. With 28 TU of data, regularization rescues several long-history ensembles that were unstable at $\lambda=0$, although the best accuracy of the unregularized model at the optimal history length is not recovered. With only 14 TU, where no unregularized configuration was stable, regularization produces several stable runs, but accuracy remains noticeably worse than with the larger training sets.

\newpage
\subsubsection*{Analysis of best models}
We now look more closely at the best-performing TO-LRS configurations on the three training windows: $h=500$ trained on 57 TU without regularization, $h=400$ trained on 28 TU without regularization, and $h=400$ trained on 14 TU with $\lambda=10$.

\begin{figure}[tbp]
    \centering
    \includegraphics[width=\linewidth]{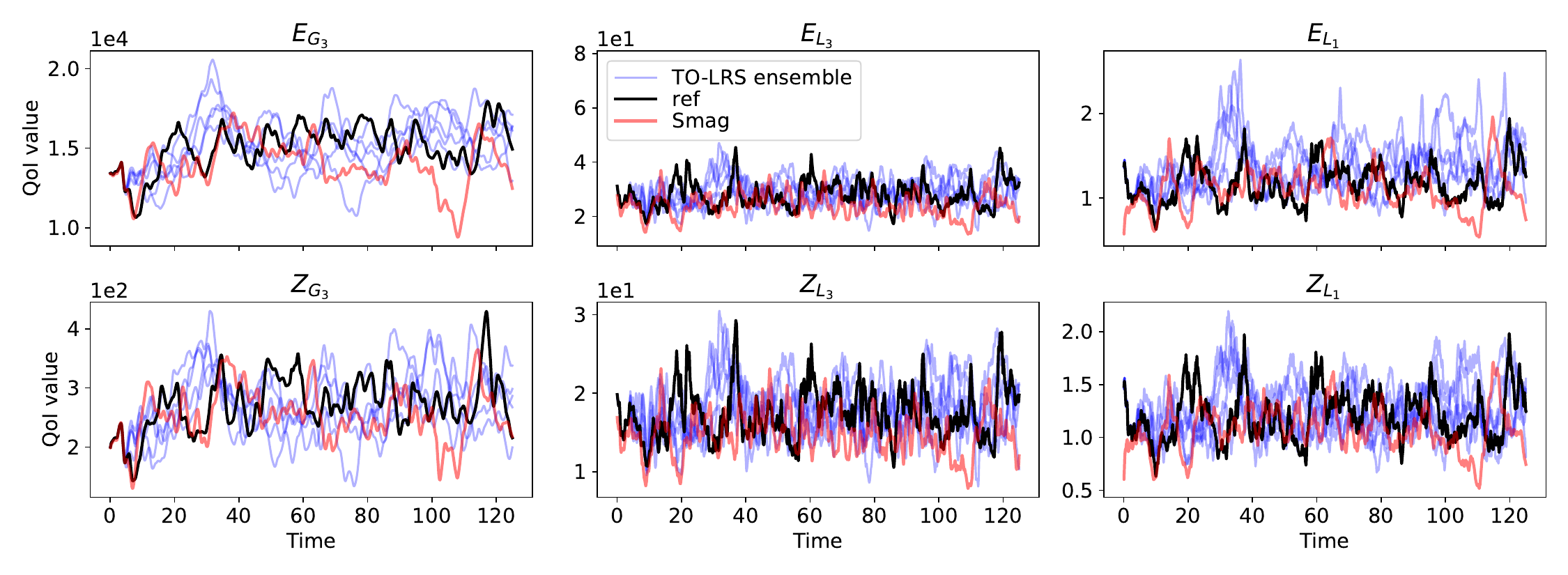}
    \caption{QoI trajectories from the online TO-LRS ensemble with history length $h=500$, $\lambda = 0$, and training data up to $T=57$, on the Kolmogorov flow. The reference DNS trajectory is shown together with the five replica trajectories of the ensemble and the calibrated Smagorinsky baseline.}
    \label{fig:qoi-trajectories-h500-T57}
\end{figure}

Figure~\ref{fig:qoi-trajectories-h500-T57} shows the QoI trajectories for the best model on the largest training set. During the first 5000 time steps (3.5 TU) of each online simulation, the solver is nudged to the reference QoI trajectories exactly as in tracking, while the history buffer fills; this also skips the resolution-change transient. Beyond that point all corrections come from the trained model. As expected for a chaotic system, individual ensemble trajectories diverge from the reference within a few TU but remain statistically close.

The trajectories for the shorter training sets and the corresponding predicted SGS corrections are shown in~\ref{app:online-best-models}. The model trained on 28 TU still tracks the reference distribution well. The model trained on only 14 TU, in contrast, fails to reproduce the reference dynamics: its QoI trajectories are smoother and have larger amplitudes than the reference, and its predicted $dQ$ series is dominated by Gaussian noise rather than a structured signal---a clear sign that the linear regression has not learned a useful conditional mean.

\begin{figure}[htbp]
    \centering
    \includegraphics[width=0.9\linewidth]{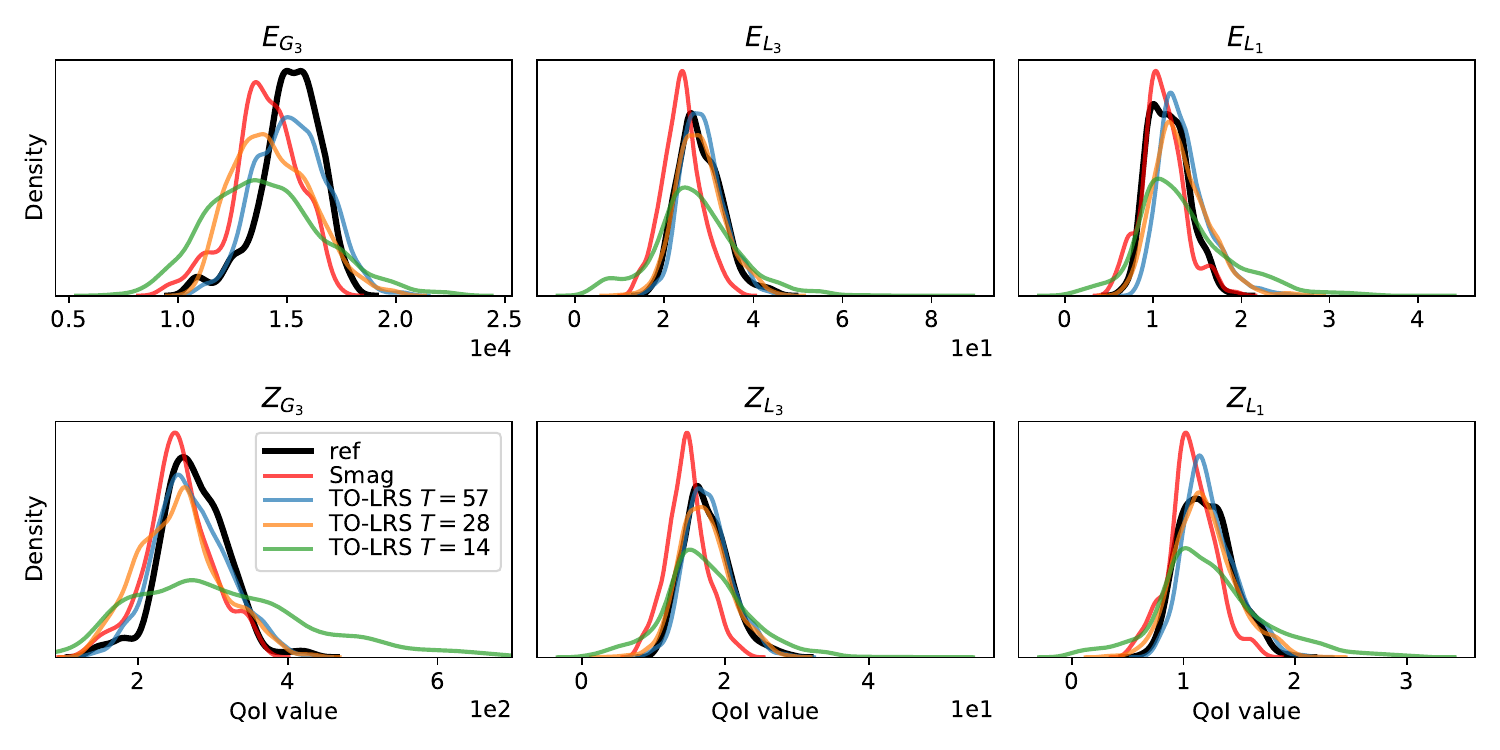}
    \caption{Long-term distributions of the six scale-aware QoIs for the best-performing TO-LRS models on three training-set sizes ($T \in \{14, 28, 57\}$ TU), compared with the calibrated Smagorinsky baseline and with the filtered-DNS reference, on the Kolmogorov flow. Each TO-LRS distribution is pooled over the stable replicas of the ensemble.}
    \label{fig:QoI-distributions}
\end{figure}
Figure~\ref{fig:QoI-distributions} shows the long-term QoI distributions for the same models. For TO-LRS, the distributions are pooled over all stable replicas in the ensemble. The models trained on the two larger training sets reproduce the reference distributions well, with mid-scale energy and enstrophy in particularly good agreement. Small-scale energy is slightly overpredicted, consistent with the behavior already observed in the tracking simulation. This is the only QoI on which Smagorinsky outperforms TO-LRS. Large-scale energy turns out to be the most difficult QoI to capture. Here, the model trained on 57 TU performs best.

\begin{figure}[htbp]
    \centering
    \includegraphics[width=0.9\linewidth]{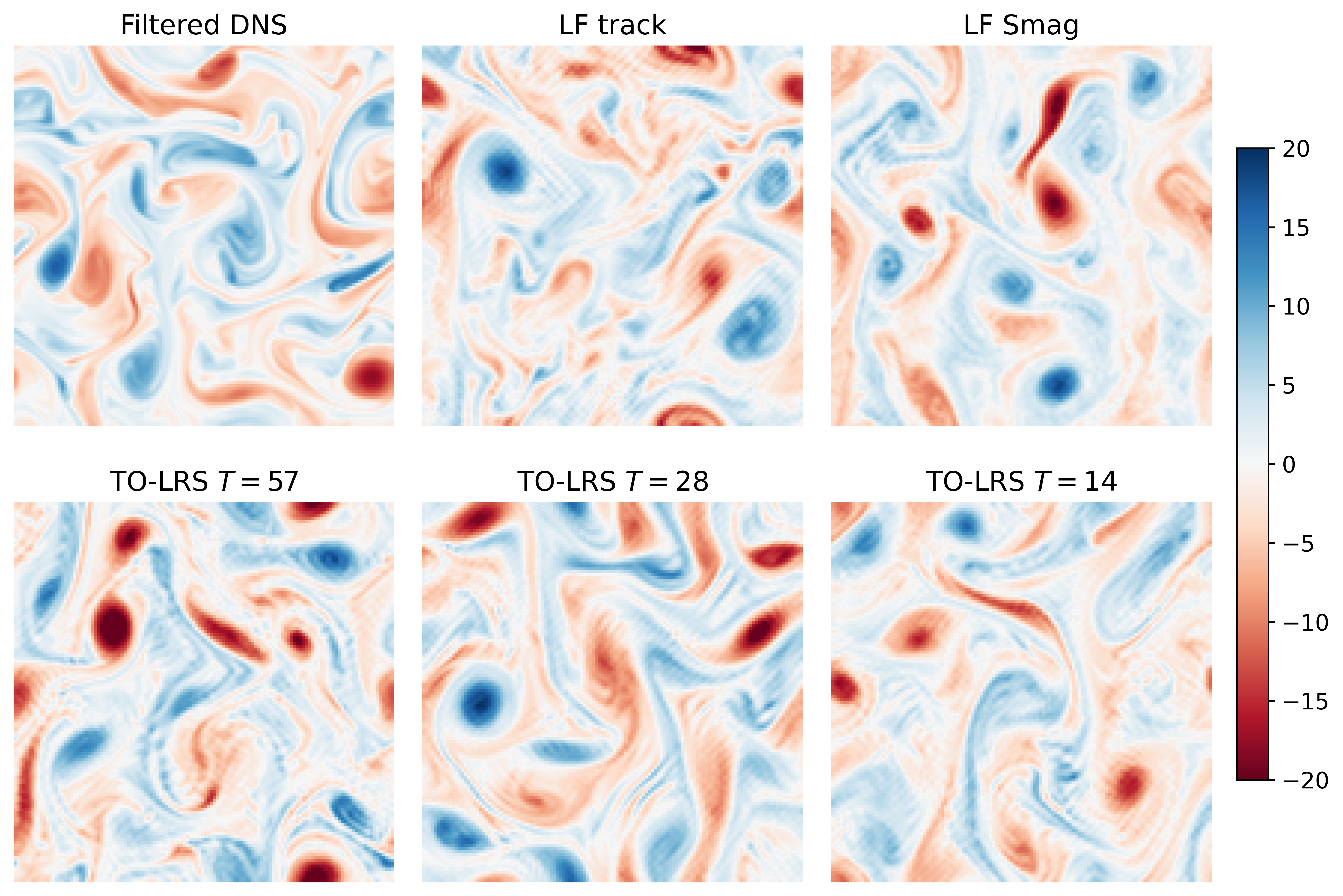}
    \caption{Vorticity fields at the end of the long-term Kolmogorov flow simulations ($t=125$ TU) for the filtered DNS, the low-fidelity tracking simulation, the calibrated Smagorinsky baseline, and the three best-performing TO-LRS models. For each TO-LRS ensemble the field from the first replica is shown.}
    \label{fig:final-fields}
\end{figure}

Figure~\ref{fig:final-fields} shows the vorticity field at the end of the simulations. All fields exhibit similar large-scale structures to the filtered DNS, but the TO-LRS fields display weak diagonal high-frequency oscillations that are absent from both DNS and Smagorinsky. The same pattern appears in the tracking simulation and in TO simulations driven by sharp Fourier filters (\ref{app:fourier-filters}). Filter discretization is therefore unlikely to be the cause; we attribute the pattern instead to the direction-dependent dispersion and dissipation of the D2Q9 stencil at wavenumbers approaching the lattice scale \cite{lallemand_theory_2000}. Because TO-LRS controls only band-integrated QoIs, it does not constrain the spectral distribution within a band and therefore allows this lattice imprint to express itself. The Smagorinsky model damps high-$k$ content directly and masks it. However we want to note that the simulations remain stable despite these oscillations, which develop already in the early part of the simulation.

\begin{figure}[htbp]
    \centering
    \includegraphics[width=0.7\linewidth]{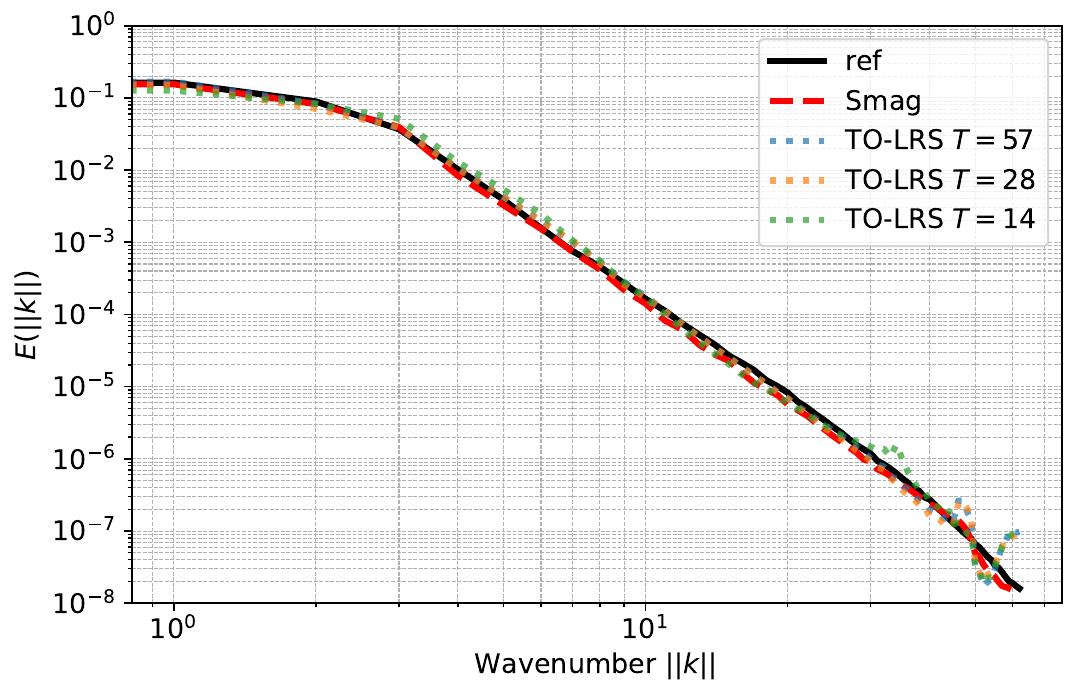}
    \caption{Energy spectra of the best-performing TO-LRS models, the calibrated Smagorinsky baseline, and the filtered-DNS reference, on the Kolmogorov flow. Each spectrum is averaged over 11 snapshots in $t \in [62.5, 125]$ and over the stable ensemble members.}
    \label{fig:energy-spectra}
\end{figure}

Figure~\ref{fig:energy-spectra} shows the energy spectra of the best models, averaged over the second half of the simulation and over the ensemble members. All three TO-LRS configurations show two small oscillations of the spectrum in $\lVert \boldsymbol{k} \rVert \in [30, 60]$ that are absent from both the reference DNS and the Smagorinsky baseline. The same oscillations show up much more prominently when sharp Fourier filters are used to define the QoIs (\ref{app:fourier-filters}). On top of these oscillations, the $T=14$ spectrum has an additional peak near $\lVert \boldsymbol{k} \rVert = 32$. This peak coincides with the wavenumber at which the fourth replica of that ensemble loses stability.

Taken together, TO-LRS produces stable and statistically accurate online LES when trained on sufficient data, and outperforms the calibrated Smagorinsky baseline on the long-term distributions of every QoI except small-scale energy. Large-scale energy is the hardest QoI to reproduce and benefits most from the largest training set; the other QoIs are already captured well with 28 TU of training data. Regularization improves stability but does not generally improve accuracy. With only 14 TU of training data the model fails to learn the reference dynamics. An equivalent study with sharp-Fourier-filter QoIs (\ref{app:fourier-filters}) reaches the same conclusions on the largest training set but is more fragile with less data, requiring regularization already at 28 TU; evidence that the compact kernel filters do not degrade, and in the small-data regime even improve, the robustness of the closure.

% =============================================================
% 3D channel-flow results section.
% =============================================================

\section{Three-dimensional channel flow}\label{sec:3d-channel}

In this section we apply the TO-LRS model to a three-dimensional turbulent plane channel at $Re_\tau=180$, a canonical wall-bounded flow with strong inhomogeneity in the wall-normal direction. This test exercises every component introduced in Section~\ref{sec:tau-orthogonal-method-reduced-sgs-modeling}: the three-dimensional sensitivities, the compact three-dimensional kernel filters~\eqref{eq:gaussian} and~\eqref{eq:laplace}, and the wall-anchored coarse-graining operator of Section~\ref{sec:coarse-graining-lattice-boltzmann-equation}. Moreover, because the test case is widely adopted, it lets us compare against reference data from the literature~\cite{vreman_comparison_2014}.

\subsection{Setup}\label{sec:3d-setup}

We consider an incompressible turbulent channel between two parallel no-slip walls at friction Reynolds number $Re_\tau=180$. The flow is driven by a constant body force in the streamwise direction. The computational domain is $(L_x, L_y, L_z) = (4\pi H, 2H, 4\pi H /3)$, where $H$ denotes the channel half-height, matching the box used by \cite{vreman_comparison_2014}. The streamwise ($x$) and spanwise ($z$) directions are periodic; the wall-normal ($y$) direction carries no-slip walls at $y \in \{0, 2H\}$, implemented via the LBM halfway bounce-back rule \cite{kruger_lattice_2017}.

All simulations use the D3Q19 velocity stencil with single-precision arithmetic. The high-fidelity grid has $N_{\text{HF}} = 1475 \times 237 \times 490$ and uses a plain BGK collision operator. The low-fidelity grid has $N_{\text{LF}} = 295 \times 49 \times 98$, obtained by subsampling with a factor of $5$ via the wall-anchored projection of Section~\ref{sec:coarse-graining-lattice-boltzmann-equation}. This anchoring ensures that the HF and LF walls coincide. In wall units the HF spacing is $\Delta x^+_{HF} = \Delta y^+_{HF} = \Delta z^+_{HF} \approx 1.53$, placing the first fluid node at $y^+ \approx 0.77$, while the LF spacing is $\Delta^+_{LF} \approx 7.7$ with the first fluid node at $y^+ \approx 3.8$.

The conversion to lattice units is fixed by the grid resolution and a velocity scaling, chosen so that the centerline velocity, $U^+_c \approx 18.3$ in wall units, maps to $\approx 0.1$ in lattice units:
\begin{equation}
    C_l = \frac{2H}{N_y - 2}, \quad C_u=\frac{U^+_c}{0.1}, \quad C_\rho = \rho = 1.
\end{equation}
We report time in characteristic time units (TU), $H / u_\tau$, where $u_\tau$ is the friction velocity; one TU corresponds to $21\,445$ HF steps or $4289$ LF steps.

\begin{figure}[t]
    \centering
    \includegraphics[width=0.85\linewidth]{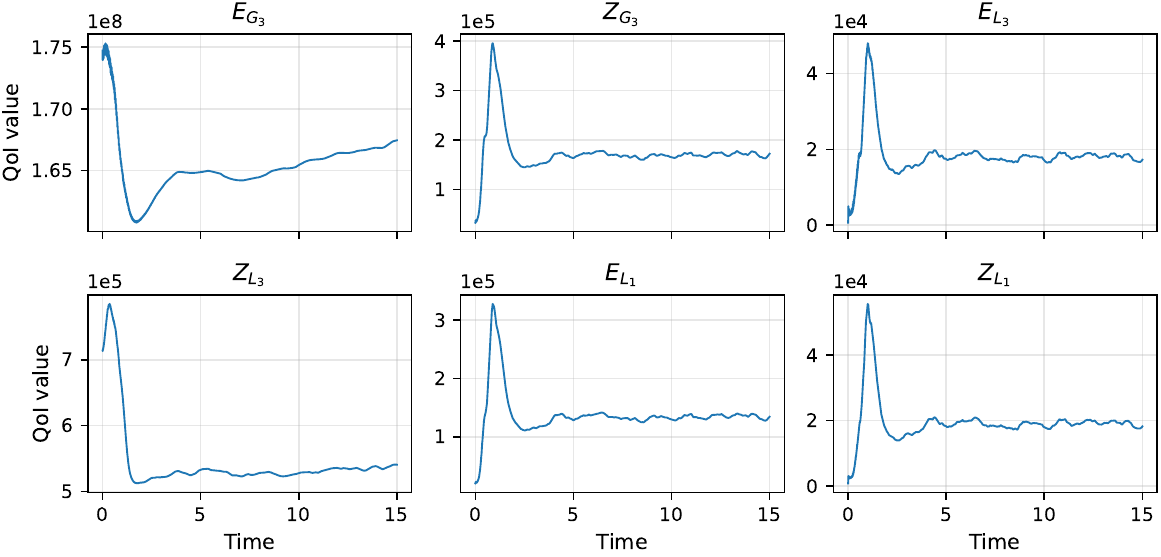}
    \caption{Burn-in transient of the six scale-aware QoIs $\{E_{G_3}, Z_{G_3}, E_{L_3}, Z_{L_3}, E_{L_1}, Z_{L_1}\}$ for the HF $Re_\tau=180$ channel-flow simulation.}
    \label{fig:3d-burnin-settling}
\end{figure}

All simulations start from the same fully developed turbulent field, obtained from a $15$~TU burn-in on the HF grid that is seeded by the initial condition of \cite{moin_numerical_1980}. Figure~\ref{fig:3d-burnin-settling} shows the six scale-aware QoIs over this burn-in. After the initial transient the QoIs settle onto statistically stationary levels within $5$~TU, except for the large-scale energy, which is still increasing slightly.

\subsection{QoIs for the channel}\label{sec:3d-channel-qois}

In the channel flow simulations we use $N_Q = 12$ QoIs. The first six are the scale-aware QoIs, now built with the three-dimensional kernels. The remaining six are mean-profile QoIs:
\begin{equation}
    \{q_i\}_{i=1}^{12}
    = \{ E_{G_3}, Z_{G_3}, E_{L_3}, Z_{L_3}, E_{L_1}, Z_{L_1},
         Q_1^{\text{mp}}, Q_2^{\text{mp}}, \dots, Q_6^{\text{mp}} \}.
\end{equation}
\begin{figure}[htbp]
    \centering
    \includegraphics[width=0.6\linewidth]{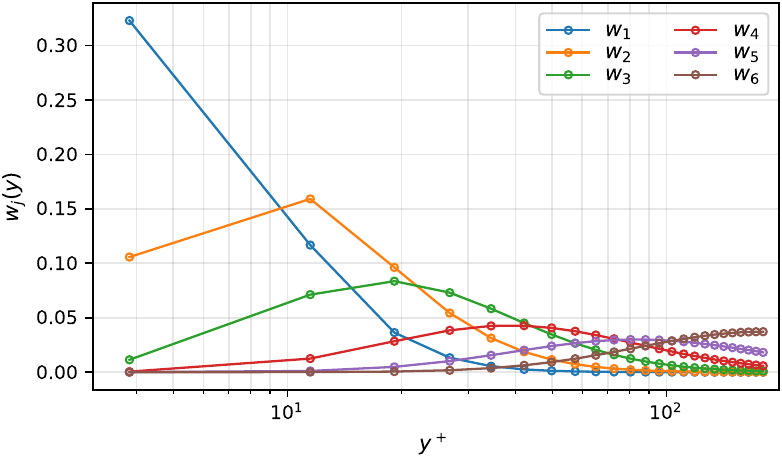}
    \caption{The six wall-normal weight bands $w_j(y)$ of the mean-profile QoIs, on the LF grid. The lower half-channel is shown.}
    \label{fig:3d-mp-band-weights}
\end{figure}

The six mean-profile QoIs $\{Q_j^{\text{mp}}\}_{j=1}^{6}$ use the general definition \eqref{eq:mean-profile-qoi} with band weights
\begin{equation}\label{eq:3d-mp-band-weights}
    w_j(y) = \frac{1}{A_j}\exp\!\left( -\frac{1}{2}\left(\frac{\ln d_w(y) - \ln d_{w,j}}{\sigma}\right)^2 \right),
\end{equation}
where $A_j$ normalizes the integral to $\int_0^{2H} w_j(y)\,\D y = 1$, with $d_w(y) = \min(y, 2H-y)$ the distance to the nearest wall. The six centers $d_{w,j}$ are log-spaced in $y^+$, with bandwidth $\sigma$ equal to the center spacing so adjacent bands overlap smoothly (Figure~\ref{fig:3d-mp-band-weights}). Log-spacing in $y^+$ concentrates the bands where the profile varies most steeply. Including these bands as TO targets tracks the mean streamwise velocity profile, which the scale-aware energies and enstrophies leave free. \ref{app:3d-no-meanprof} shows that this is needed: an LF tracking run constrained by the scale-aware QoIs alone drifts off the HF mean profile.

\subsection{High-fidelity reference}\label{sec:3d-hf-ref}

The HF reference simulation covers $20$~TU, yielding reference trajectories for the $12$ QoIs. The first half of these trajectories provides the training data for TO-LRS, whereas the trajectories after the $5000$-step spin-up serve as the long-term reference distributions against which the online simulations are scored.

\begin{figure}[htbp]
    \centering
    \includegraphics[width=0.6\linewidth]{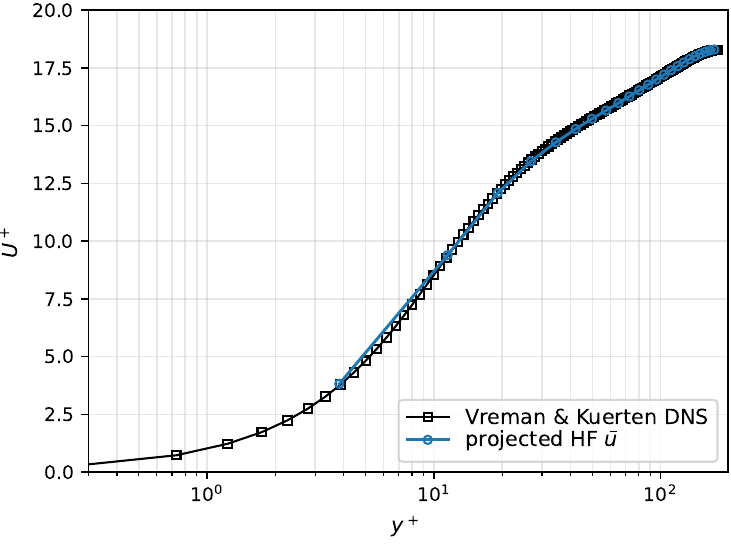}
    \caption{Mean streamwise velocity profile for the $Re_\tau=180$ channel flow: the projected HF reference averaged over 6 snapshots between $t = 10$ and $t = 20$~TU, and the DNS of Vreman and Kuerten~\cite{vreman_comparison_2014}.}
    \label{fig:3d-hf-mean-profile}
\end{figure}

Figure~\ref{fig:3d-hf-mean-profile} validates this HF simulation against the DNS of Vreman and Kuerten~\cite{vreman_comparison_2014}. It shows the mean streamwise velocity profile $U^+(y) = \langle \bar{u}_x \rangle(y)$, obtained from six snapshots of the downsampled HF solution $\bar{\boldsymbol{u}}$ in $t \in [10, 20]$; it closely follows the reference DNS data.

\subsection{Low-fidelity baselines}\label{sec:3d-lf-baselines}

The LF grid leaves the near-wall layer severely under-resolved: plain BGK simulations with no SGS model are unstable, and we will later find that adding TO tracking alone does not resolve the instability. The eddy-viscosity models therefore play two roles. Run standalone at their calibrated coefficients, they provide the \emph{baselines} against which TO accuracy is judged. Secondly, used as a \emph{substrate}, they provide a more dissipative collision operator that stabilizes the TO tracking simulations. In the substrate role the coefficient is left uncalibrated, so that the eddy viscosity only stabilizes the simulation while the TO correction carries the closure. We use the Smagorinsky model and the WALE model in both roles.

\begin{figure}[htbp]
    \centering
    \includegraphics[width=0.85\linewidth]{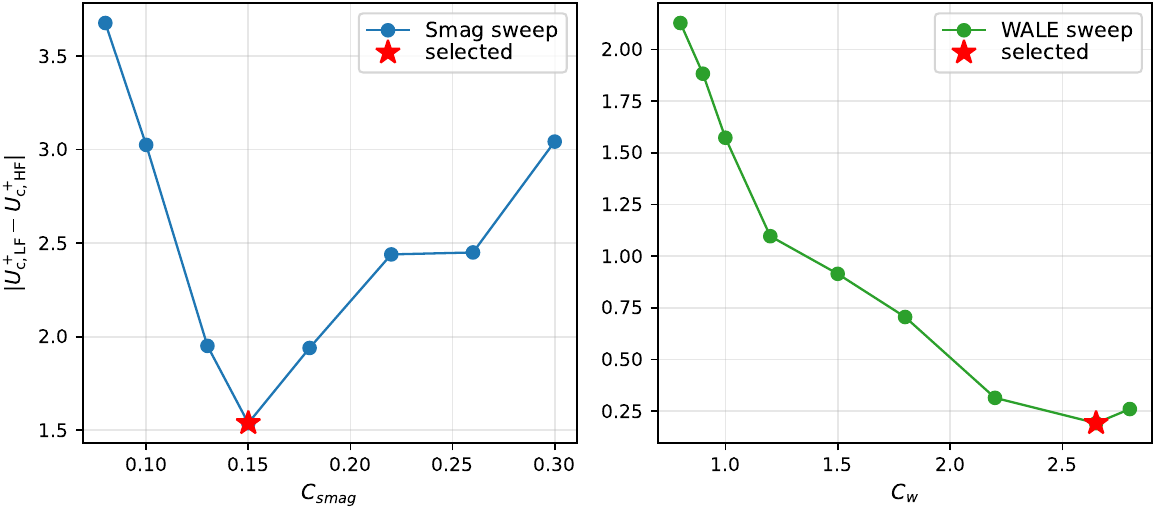}
    \caption{LF calibration of the Smagorinsky and WALE coefficients for the channel flow against the HF centerline mean streamwise velocity at $t = 10$~TU. Each marker is one calibration run; the star marks the selected value.}
    \label{fig:3d-lf-calibration}
\end{figure}

We calibrate both models with a coefficient sweep, selecting for each the coefficient that minimizes the discrepancy with the HF centerline mean streamwise velocity at $t = 10$~TU (Figure~\ref{fig:3d-lf-calibration}): $C_{smag} = 0.15$ for Smagorinsky and $C_w = 2.65$ for WALE. This WALE coefficient is roughly five times the canonical value $C_w = 0.5$, at which our LF simulations are unstable. At this very coarse resolution the first fluid node needs extra dissipation to keep the simulation stable.

\subsection{Tracking}\label{sec:3d-tracking}

\begin{figure}[htbp]
    \centering
    \includegraphics[width=\linewidth]{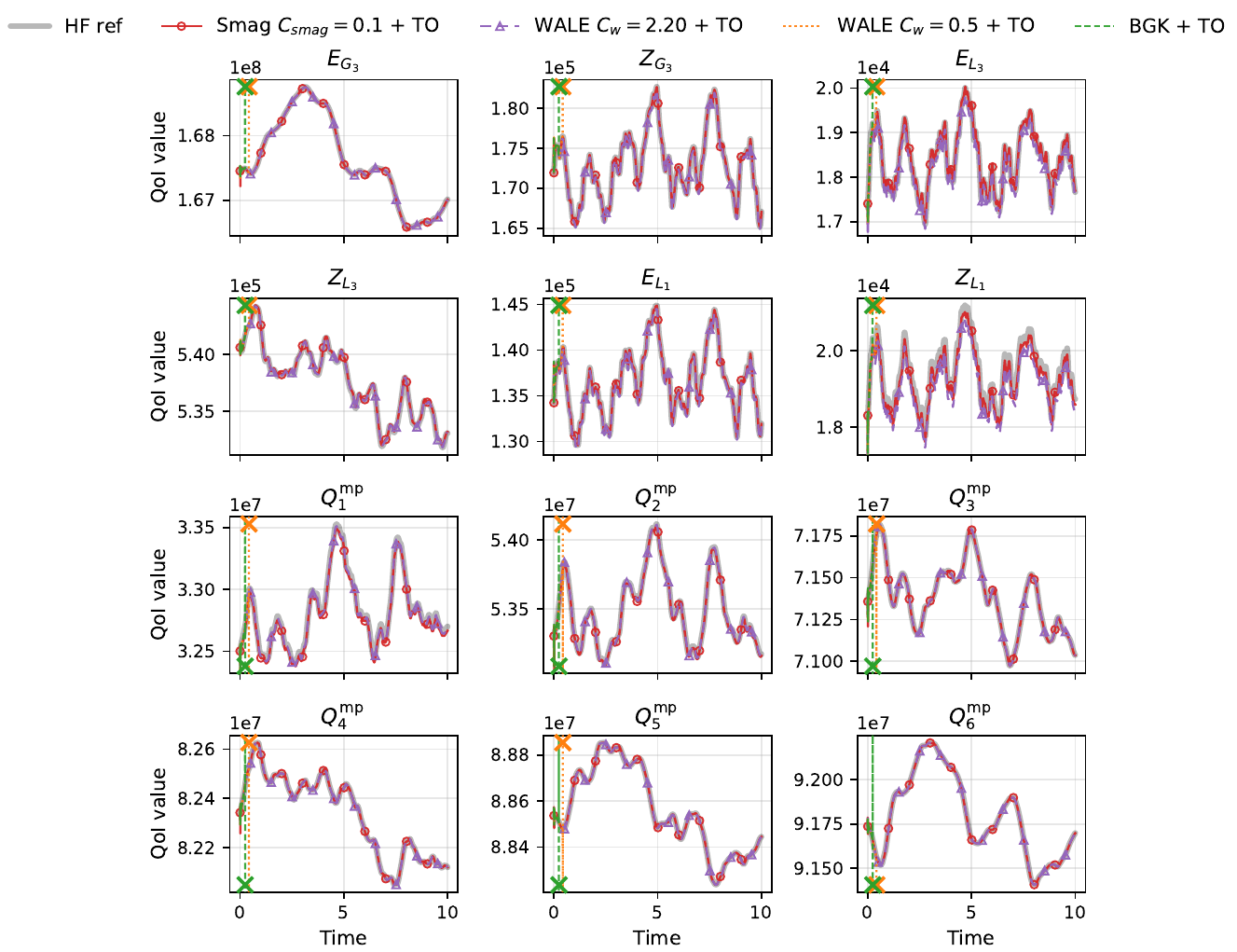}
    \caption{Post-streaming QoI trajectories from LF tracking of the channel flow on the four candidate substrates, over the $10$ TU tracking window. Canonical WALE ($C_w = 0.5$) and plain BGK diverge almost immediately, marked by an `X'.}
    \label{fig:3d-tracking-qois}
\end{figure}

We track the HF reference trajectories over $10$~TU with the TO method on four substrates. Figure~\ref{fig:3d-tracking-qois} shows the resulting QoI trajectories for all of them. On the Smagorinsky substrate at the canonical coefficient $C_{smag} = 0.1$, the TO method tracks all $12$ QoIs on the HF reference for the full $10$~TU without loss of stability. The WALE substrate tracks stably only at the inflated coefficient $C_w = 2.20$. At the canonical $C_w = 0.5$ the tracked simulation diverges at $t \approx 0.43$~TU. The plain BGK substrate likewise diverges almost immediately at $t \approx 0.25$~TU.

\begin{figure}[htbp]
    \centering
    \includegraphics[width=\linewidth]{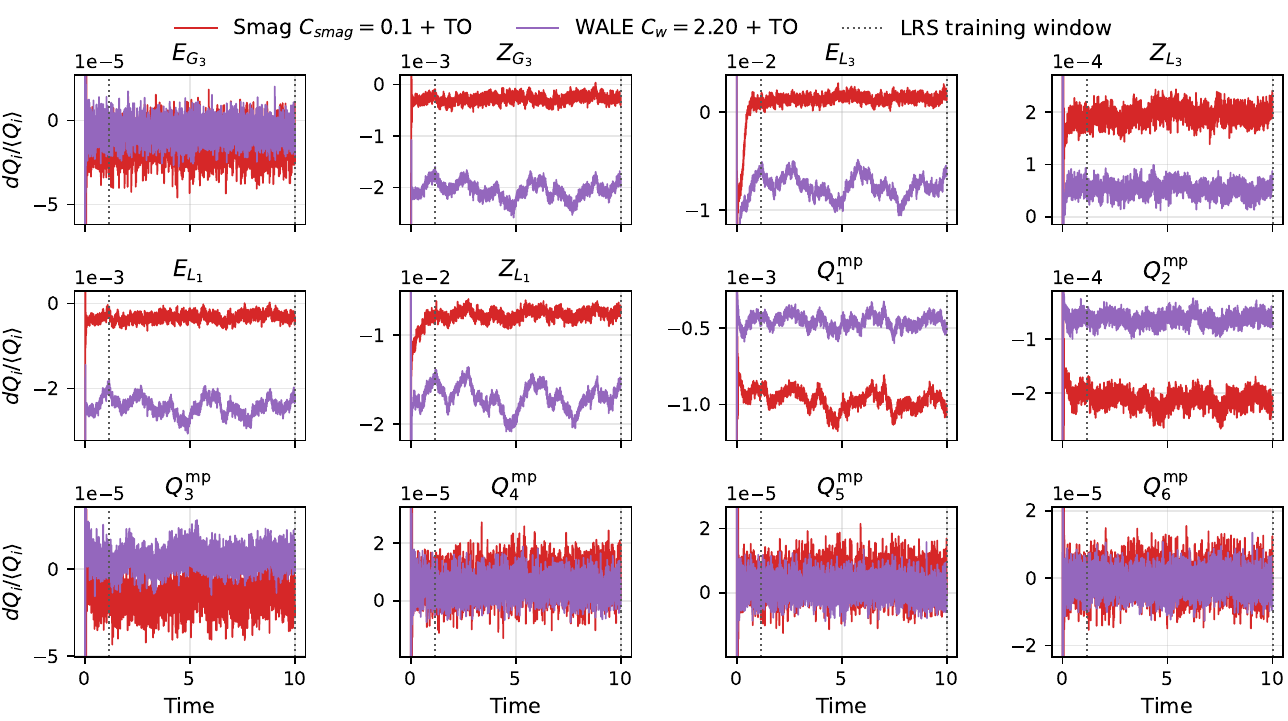}
    \caption{SGS corrections $dQ_i^n$ applied during tracking of the channel flow, normalized by the time-averaged value of the corresponding QoI. The dotted vertical lines mark the LRS training window.}
    \label{fig:3d-tracking-dQ}
\end{figure}

Figure~\ref{fig:3d-tracking-dQ} reports the applied corrections for both stable substrates. The corrections are clearly biased on the small-scale QoIs and on the mean-profile bands close to the wall. The two substrates need them in different places: the WALE substrate requires larger corrections to the scale-aware energy and enstrophy, whereas the Smagorinsky substrate requires larger corrections to the mean-profile QoIs.

As in the two-dimensional Kolmogorov flow, we regard the first $5000$ steps ($\approx 1.17$~TU) of the tracking as a spin-up transient following the resolution change, and exclude them from the training data.


\subsection{TO-LRS hyperparameter sweep}\label{sec:3d-sweep}

We sweep the LRS history length $h \in \{400, 500\}$ and the regularization strength $\lambda \in \{0, 1, 5\}$ on both substrates, running five online replicas over 20 TU per ensemble configuration. We nudge to the HF reference trajectories during the first 5000 steps and then use the fitted LRS model. Each ensemble is scored by the summed KS-distance over the $12$ QoIs against the HF reference.

Regularization proves necessary for stability: all $\lambda = 0$ replicas are unstable, whereas every replica with $\lambda = 1$ or $\lambda = 5$ remains stable. This behavior differs from the two-dimensional Kolmogorov-flow tests, where the best-performing models are obtained at $\lambda = 0$. We found the same contrast with homogeneous isotropic turbulence (HIT) in earlier work on the TO method in three-dimensional turbulence~\cite{hoekstra_reduced_2026}.

\begin{figure}[htbp]
    \centering
    \includegraphics[width=0.85\linewidth]{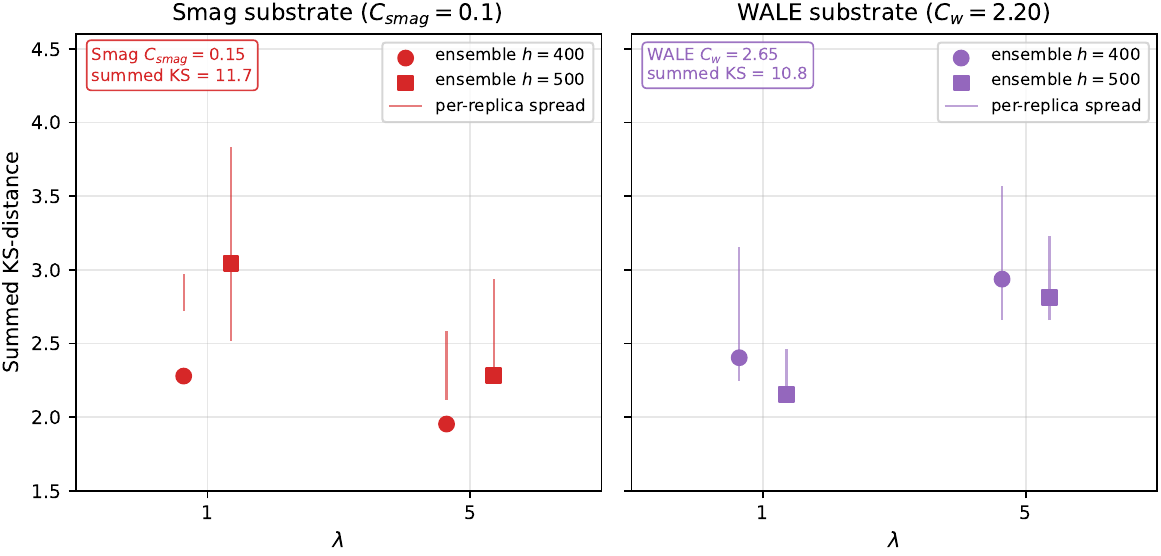}
    \caption{Online performance of TO models against the HF reference over 20 TU as a function of the regularization $\lambda$ and history length $h$, on the channel flow. Results are based on five-replica ensembles; markers show the pooled summed KS-distance, and vertical bars show the replica spread.}
    \label{fig:3d-predict-sweep-KS}
\end{figure}

Figure~\ref{fig:3d-predict-sweep-KS} shows the online performance against $\lambda$ for both substrates. The calibrated eddy-viscosity baselines sit far above the TO-LRS ensembles, at summed KS-distances of $10.8$ (WALE) and $11.7$ (Smagorinsky). Since the KS-distance of a single QoI is at most one, these baselines lie essentially at saturation: their QoI distributions barely overlap the reference. The TO-LRS ensembles, by contrast, all fall in the band $1.95 \le \mathrm{KS} \le 3.05$.

The SGS model does not depend strongly on per-case tuning of $h$ and $\lambda$ over the tested range. The differences between $h = 400$ and $h = 500$, or between $\lambda = 1$ and $\lambda = 5$, are of the same order as the replica spread and small compared with the gap to the calibrated baselines. 

\subsection{Analysis of best models}\label{sec:3d-ensemble}

\begin{figure}[h]
    \centering
    \includegraphics[width=\linewidth]{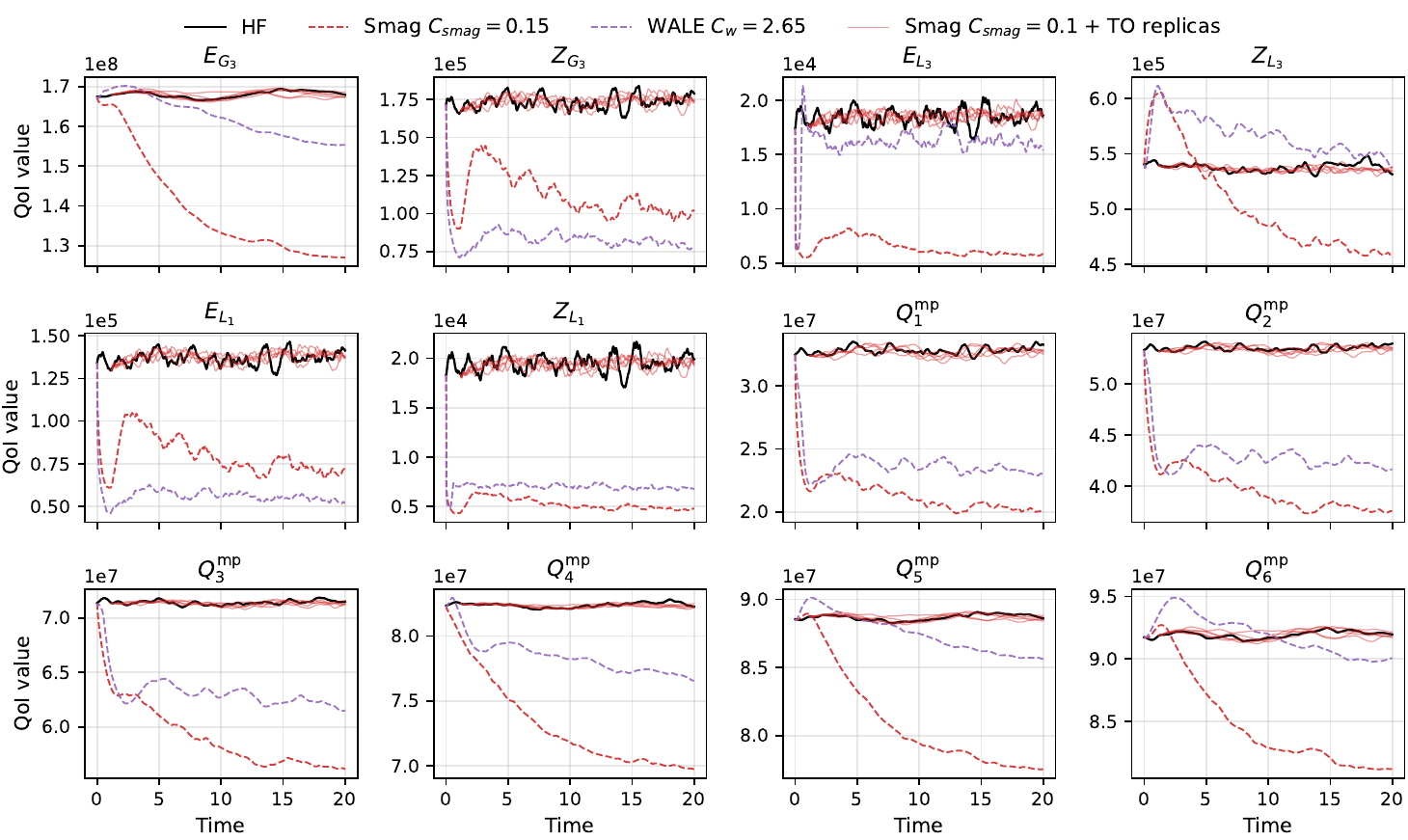}
    \caption{QoI trajectories from the online TO-LRS ensemble for the channel flow on the Smagorinsky substrate ($C_{smag} = 0.1$), at its best-performing configuration $h = 400$, $\lambda = 5$. The five replica trajectories are shown together with the HF reference and the calibrated eddy-viscosity baselines.}
    \label{fig:3d-ensemble-qoi-trajectory}
\end{figure}

\begin{figure}[H]
    \centering
    \includegraphics[width=\linewidth]{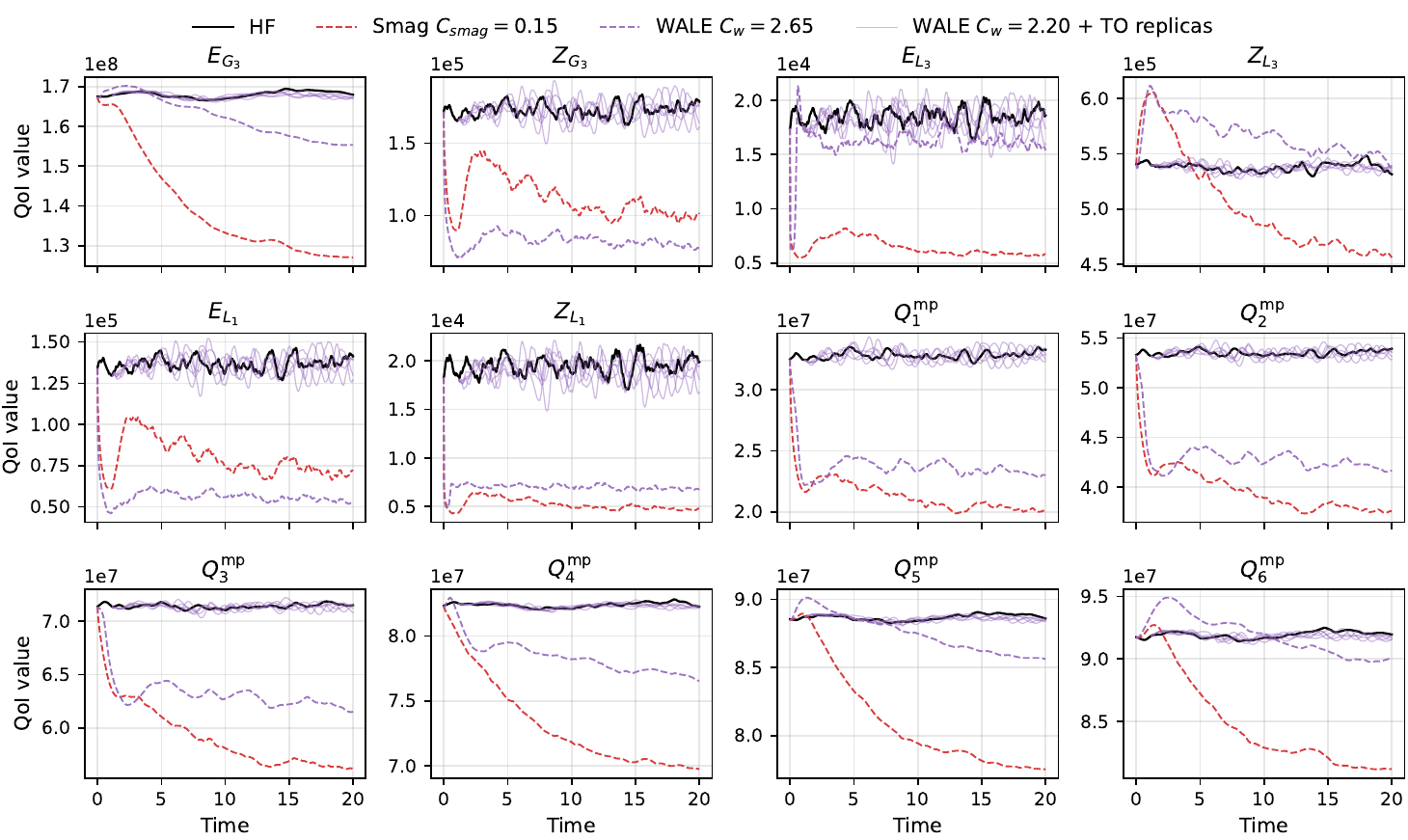}
    \caption{QoI trajectories from the online TO-LRS ensemble for the channel flow on the WALE substrate ($C_w = 2.20$), at its best-performing configuration $h = 500$, $\lambda = 1$. The five replica trajectories are shown together with the HF reference and the calibrated eddy-viscosity baselines.}
    \label{fig:3d-ensemble-qoi-trajectory-wale}
\end{figure}

We now examine the best TO-LRS ensembles in detail, selecting for each substrate its best-performing configuration: $h=400$, $\lambda = 5$ on the Smagorinsky substrate and $h=500$, $\lambda = 1$ on the WALE substrate. Figures~\ref{fig:3d-ensemble-qoi-trajectory} and~\ref{fig:3d-ensemble-qoi-trajectory-wale} show their QoI trajectories. As expected for a chaotic flow, the individual replica trajectories diverge from the reference, yet the TO-LRS replicas stay close to it in a statistical sense. The calibrated baselines, included in the same figures, do not: they settle onto displaced levels.

\begin{figure}[H]
    \centering
    \includegraphics[width=\linewidth]{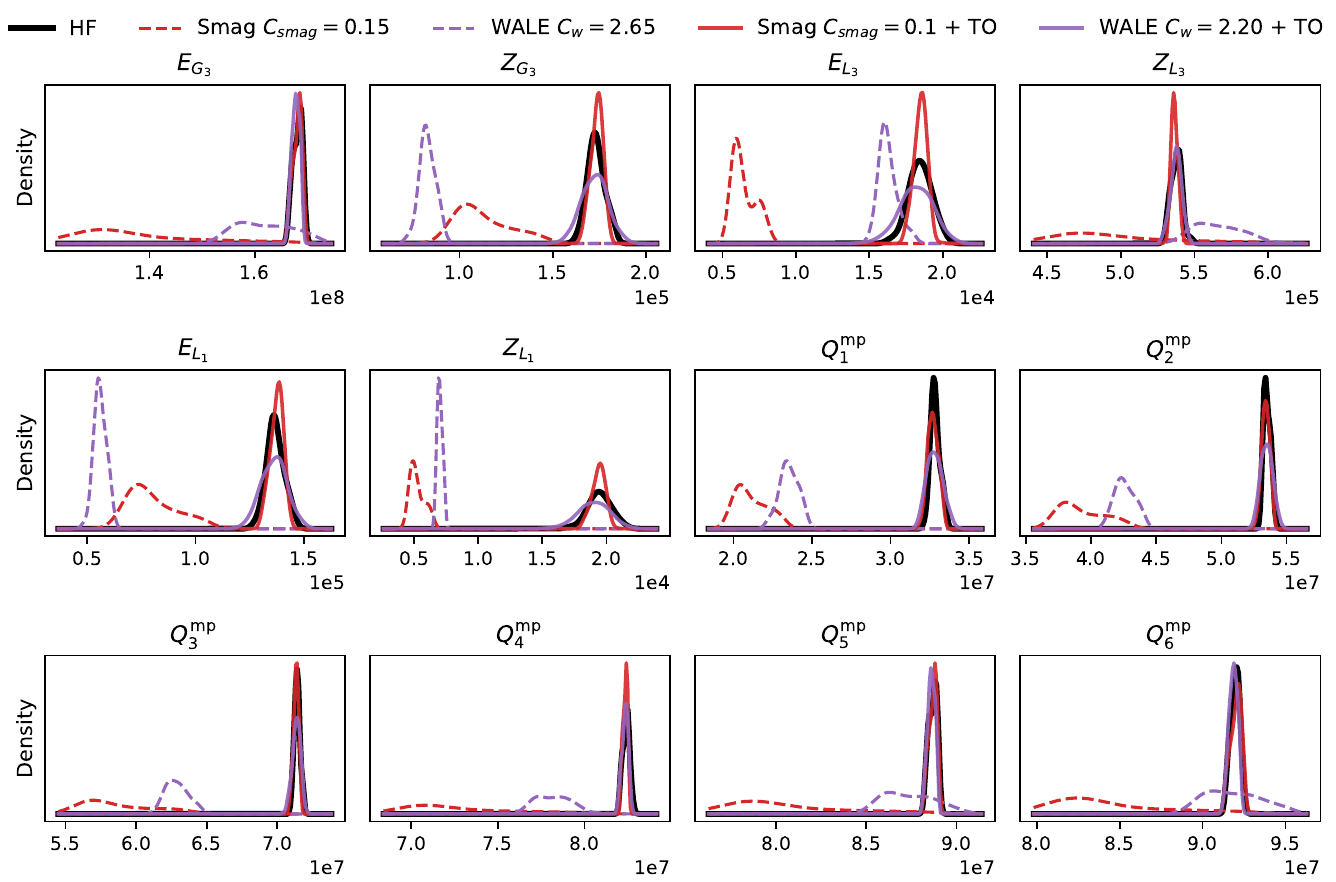}
    \caption{Long-term distributions of the $12$ QoIs for the channel flow, for the online Smagorinsky and WALE TO-LRS ensembles, each at its best-performing configuration, compared with the calibrated Smagorinsky and calibrated WALE baselines and with the HF reference. Each TO-LRS distribution is pooled over the five replicas of the ensemble.}
    \label{fig:3d-ensemble-qoi-distributions}
\end{figure}

Figure~\ref{fig:3d-ensemble-qoi-distributions} shows the long-term distributions of the $12$ QoIs, pooled over the five replicas of each ensemble and compared against the HF reference and the two calibrated baselines. The TO-LRS ensembles track the HF distributions across all $12$ QoIs, whereas the calibrated baselines are displaced on nearly every QoI, most strongly on the small-scale energies and enstrophies, which the over-dissipative eddy-viscosity models suppress heavily.

\begin{figure}[htbp]
    \centering
    \includegraphics[width=0.7\linewidth]{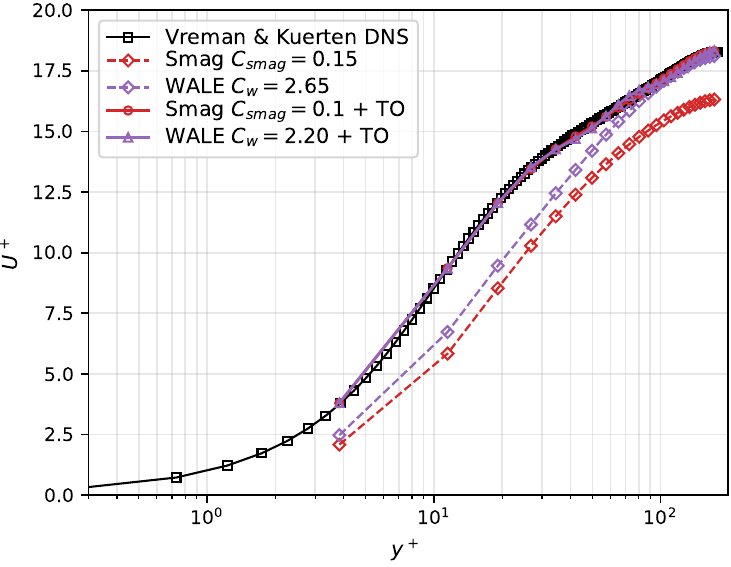}
    \caption{Mean streamwise velocity profile of the channel flow for the Smagorinsky and WALE TO-LRS ensembles (each at its best-performing configuration), compared with the calibrated baselines and the Vreman--Kuerten DNS~\cite{vreman_comparison_2014}. Profiles are averaged over 6 snapshots between $t = 10$ and $t = 20$~TU; TO-LRS profiles are additionally averaged over the five replicas.}
    \label{fig:3d-ensemble-mean-profile}
\end{figure}

Figure~\ref{fig:3d-ensemble-mean-profile} shows the mean streamwise velocity profiles of the online simulations. The six mean-profile QoIs constrain this profile, so the agreement partly reflects the training target; the non-trivial result is that the SGS model maintains the correct profile in online prediction mode as well. The calibrated baselines cannot reproduce the mean profile at this grid resolution.

\begin{figure}[htbp]
    \centering
    \includegraphics[width=\linewidth]{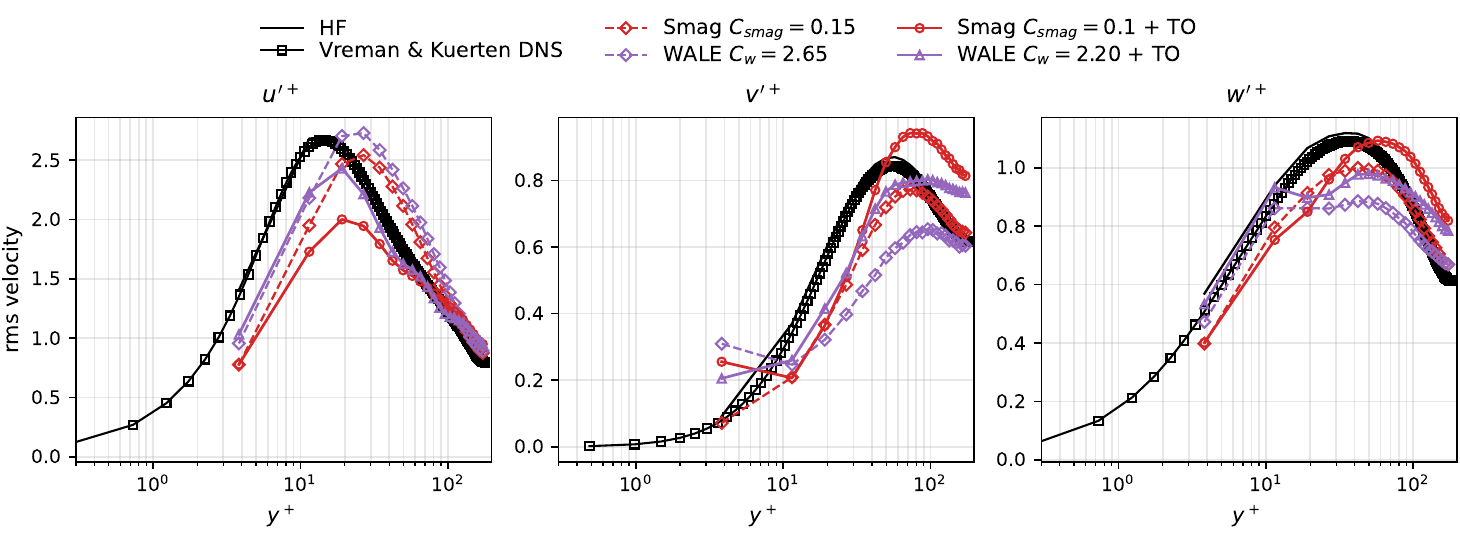}
    \caption{Root-mean-square value of velocity fluctuations of the channel flow for the two TO-LRS ensembles, the calibrated baselines, the HF reference, all based on 6 snapshots between $t = 10$ and $t = 20$~TU, compared against the Vreman--Kuerten DNS~\cite{vreman_comparison_2014}.}
    \label{fig:3d-ensemble-rms-profiles}
\end{figure}

The velocity fluctuations (Figure~\ref{fig:3d-ensemble-rms-profiles}), which no QoI constrains, are less favorable: the TO-LRS ensembles under-predict the streamwise $u'^+$ peak (most so on the Smagorinsky substrate), and the TO-LRS Smagorinsky ensemble overshoots the wall-normal $v'^+$ peak.

\begin{figure}[htbp]
    \centering
    \begin{subfigure}[b]{0.5\linewidth}
        \centering
        \includegraphics[width=\linewidth]{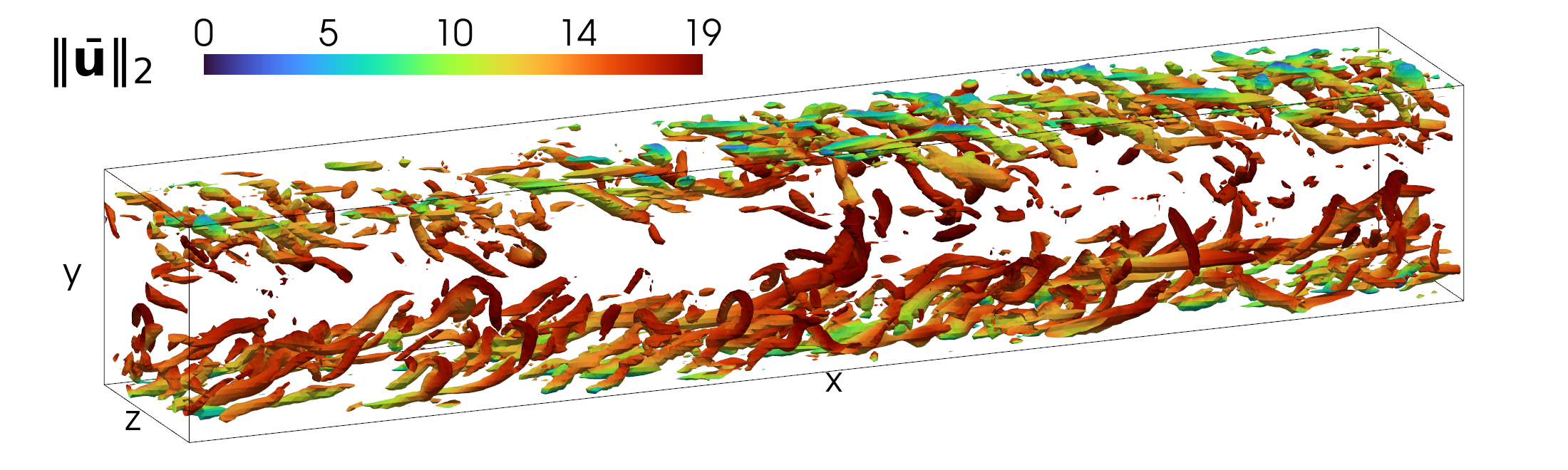}
        \caption{Projected HF reference}
    \end{subfigure}

    \begin{subfigure}[b]{0.49\linewidth}
        \centering
        \includegraphics[width=\linewidth]{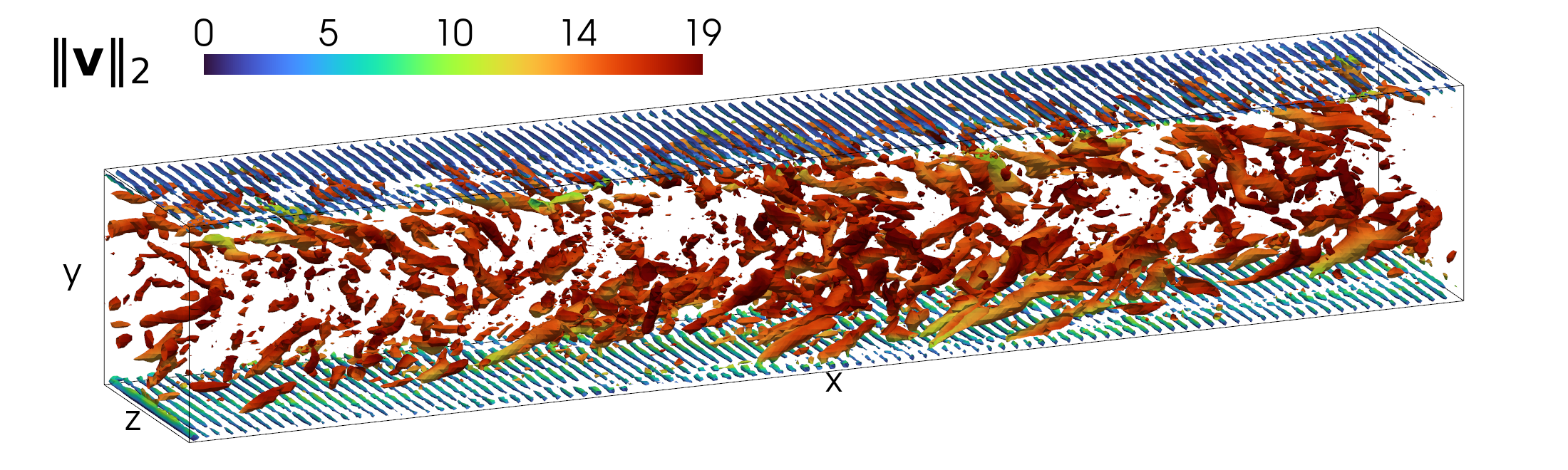}
        \caption{Smagorinsky + TO-LRS ($h = 400$, $\lambda = 5$)}
    \end{subfigure}
    \hfill
    \begin{subfigure}[b]{0.49\linewidth}
        \centering
        \includegraphics[width=\linewidth]{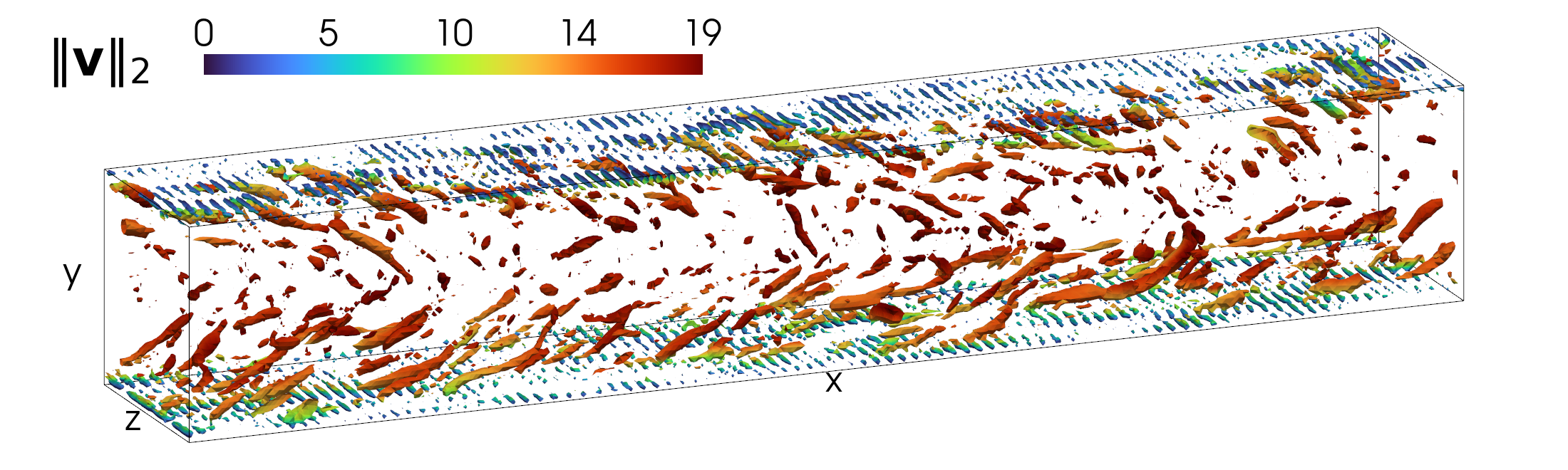}
        \caption{WALE + TO-LRS ($h = 500$, $\lambda = 1$)}
    \end{subfigure}

    \begin{subfigure}[b]{0.49\linewidth}
        \centering
        \includegraphics[width=\linewidth]{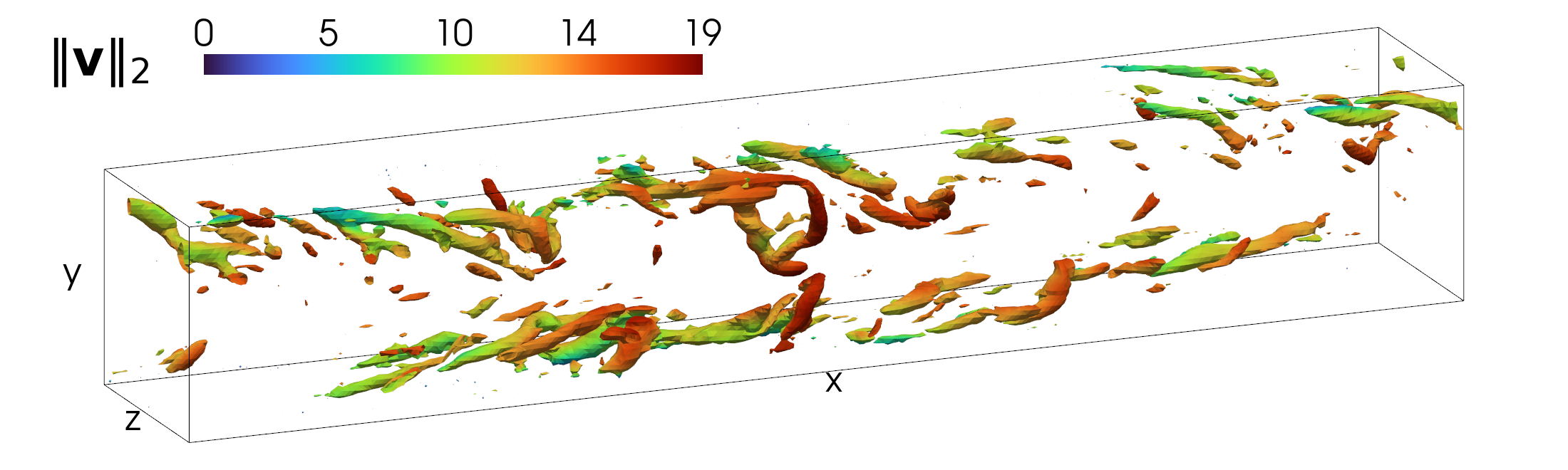}
        \caption{Calibrated Smagorinsky ($C_{smag} = 0.15$)}
    \end{subfigure}
    \hfill
    \begin{subfigure}[b]{0.49\linewidth}
        \centering
        \includegraphics[width=\linewidth]{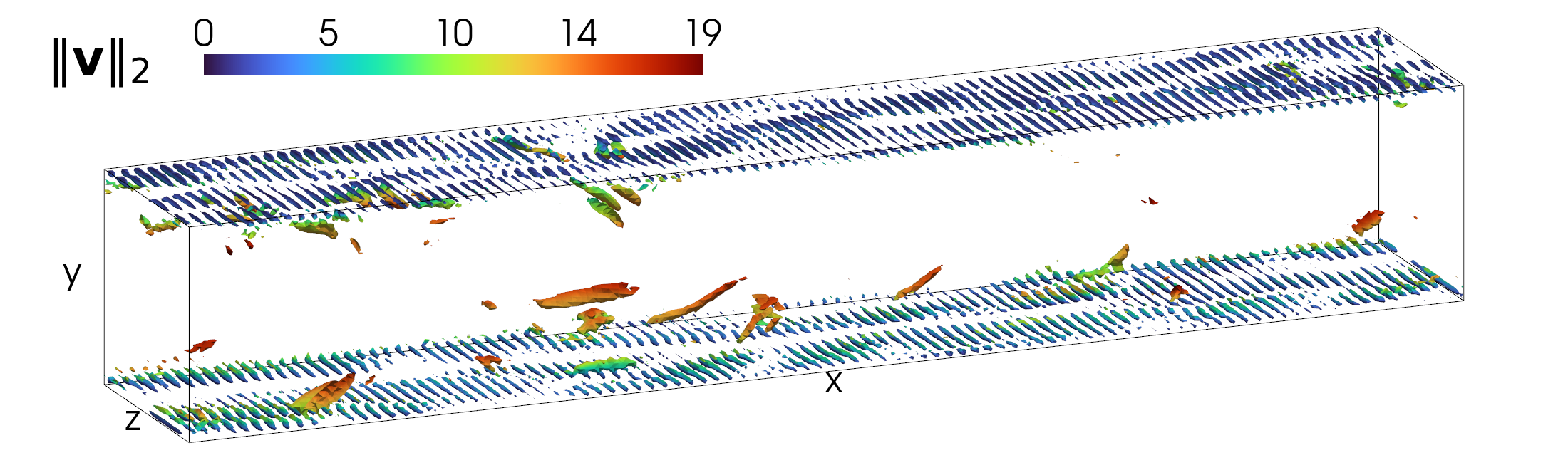}
        \caption{Calibrated WALE ($C_w = 2.65$)}
    \end{subfigure}
    \caption{Vortical structures at the end of the channel-flow simulations ($t = 20$ TU), visualized by isocontours $Q = 300$ of the second invariant of the velocity-gradient tensor, colored by velocity magnitude. The front half of the spanwise domain is shown.}
    \label{fig:3d-Q-criterion}
\end{figure}

Finally, Figure~\ref{fig:3d-Q-criterion} compares the vortical structures at the end of the simulations. They are visualized by isocontours of the second invariant of the velocity-gradient tensor, $Q = \tfrac{1}{2}\left(\Omega_{ij}\Omega_{ij} - S_{ij}S_{ij}\right)$, where $\Omega_{ij}$ is the rotation-rate tensor and $S_{ij}$ the strain-rate tensor. The projected HF reference shows a dense population of streamwise-aligned vortices concentrated in the buffer layers at the walls; we even observe a typical hairpin vortex. The TO-LRS runs retain a comparably dense population of near-wall vortices, but the structures are noticeably noisier and more fragmented than in the projected reference. The stochastic SGS corrections roughen the smallest resolved scales, yet the turbulence does not break down under this forcing, consistent with our finite-volume study~\cite{hoekstra_reduced_2026}.
Both TO-LRS runs, as well as the calibrated WALE baseline, show grid-scale oscillations in the first fluid layers. These oscillations mark the region where the simulations rely heavily on SGS dissipation for stability: in the TO-LRS runs the stochastic corrections re-excite these modes, and the substrate's dissipation keeps them bounded. The calibrated Smagorinsky run is free of such oscillations, since its eddy viscosity remains large at the wall. Both calibrated baselines, however, clearly contain too few vortices, confirming at the structural level the over-dissipation already seen in the QoI distributions.

\subsection{Computational cost}\label{sec:3d-cost}

Table~\ref{tab:timing-3d} reports the computational efficiency of the single-precision three-dimensional channel solvers, implemented in Python using JAX. All simulations ran on a single NVIDIA~H100-SXM5-80\,GB GPU of the Dutch national supercomputer \emph{Snellius}. Timings were obtained with an in-loop profiler that brackets each solver section with device synchronizations and accumulates wall time over a fixed window of steps, after warm-up and JIT compilation are complete. Each number is the mean of three independent runs of $200$ profiled steps; the run-to-run spread was below $1\%$. We separate the bare LBM \emph{step} (collision, streaming, boundary conditions, and the SGS or TO correction) from the per-step diagnostic output.

\begin{table}[h]
    \centering
    \begin{tabular}{l r r r}
        \hline
        Simulation & ms/step & MLUPS & s\,/\,TU \\ \hline
        HF reference & $91.0$ & $1883$ & $1950.8$  \\ \hline
        LF (no model)      & $0.81$ & $1757$ & $3.5$  \\
        LF Smagorinsky     & $1.07$ & $1329$ & $4.6$   \\
        LF WALE            & $1.04$ & $1357$ & $4.5$   \\
        LF + TO track      & $11.45$ & $124$ & $49.1$  \\
        LF + TO-LRS online & $11.38$ & $124$ & $48.8$  \\ \hline
    \end{tabular}
    \caption{Per-step cost, throughput, and wall time per simulated TU for the three-dimensional channel solvers, all measured on a single NVIDIA~H100-SXM5.}
    \label{tab:timing-3d}
\end{table}

On the HF grid the plain solver attains a throughput of $1883$~MLUPS (million lattice updates per second). Our solvers are built on top of the ``XLB'' solver~\cite{ataei2024xlb}, for which a comparable throughput is reported at this precision on an A100 GPU. Saving the $12$-QoI block at the LF cadence (one evaluation every five HF steps) adds under $1\%$ to the bare HF step ($+0.8$~ms on $91.0$~ms), so collecting training data adds little cost.

On the LF grid the plain solver runs at $1757$~MLUPS, and the Smagorinsky and WALE eddy-viscosity substrates add a modest $\sim0.25$~ms per step. Adding the TO method on top of the Smagorinsky substrate raises the per-step time to around $11.4$~ms, for both tracking and online simulations. We attribute this large overhead to the non-local nature of the QoIs, which has two facets. First, the QoIs are global integrals, so the code performs several global reductions that are absent from the otherwise fully local LBM scheme. Second, the scale-aware QoIs use kernel convolutions, which are expensive in the non-periodic direction. Evaluating the QoIs and their derivatives currently takes $7.3$~ms per LF step. Despite this overhead, the LF simulations with the TO method are still $40\times$ faster than the HF simulation.

Once the high-fidelity reference trajectories are available, training the TO method amounts to tracking them with an LF solver and then fitting the LRS model to the corrections. Fitting the linear regression and the multivariate Gaussian takes seconds, a cost negligible against running the simulations. This contrasts sharply with the large computational effort of training a neural network as an SGS model, as done in \cite{ortali_kinetic_2025} for example.

\section{Conclusion}\label{sec:conclusion}
We have adapted the Tau-orthogonal (TO) framework to lattice Boltzmann LES, replacing the prediction of a high-dimensional SGS field with a low-dimensional data-driven model for the dynamics of QoIs. The core of the SGS model is inherited from earlier work on classical Navier--Stokes solvers \cite{edeling_reducing_2020, hoekstra2024Reduced_data-driven, hoekstra_reduced_2026}: a spatial basis derived analytically from QoI sensitivities, a predictor-corrector tracking scheme that extracts SGS corrections without a differentiable solver, and lightweight linear regression models with stochastic residuals (LRS) that drive the LF simulation autonomously at inference time. This paper adds three new ingredients: the adaptation to the kinetic LBM solver via macroscopic-velocity forcing, compact convolution-based filters that remove the dependence on sharp Fourier filtering and on fully periodic domains, and a new set of mean-profile QoIs that track the mean streamwise velocity profile.

On two-dimensional Kolmogorov flow at $Re=10\,000$, TO-LRS produced stable and statistically accurate online LF simulations when trained on sufficient data. The model trained on 57 TU reproduced the long-term QoI distributions and energy spectra most accurately, outperforming the calibrated Smagorinsky baseline on every QoI except small-scale energy. Training on 28 TU already produced good results for most QoIs, while 14 TU was insufficient. Regularization improved stability but did not consistently improve accuracy. The TO-LRS solutions carried diagonal high-frequency oscillations, attributed to a lattice imprint of the direction-dependent dispersion of the D2Q9 stencil. Because the scale-aware QoIs do not constrain the spectrum pointwise, they leave this imprint free.

Beyond the two-dimensional benchmark, we applied the SGS model to three-dimensional channel flow at $Re_\tau=180$, a test of the kernel-based QoIs on a non-periodic geometry. On the low-fidelity grid, the plain BGK solver could not be stabilized by TO tracking alone, so we added dissipation through eddy-viscosity substrates in the collision step and applied the TO correction on top of them. Online TO-LRS ensembles on two uncalibrated eddy-viscosity substrates reproduced the $12$-QoI distributions, comprising six scale-aware QoIs augmented by six mean-profile QoIs, together with the mean streamwise velocity profile of the high-fidelity reference. The calibrated Smagorinsky and WALE baselines reproduced none of this; their long-term QoI distributions lay far from the reference. The velocity fluctuation peaks, for which no QoI was defined, did display notable deviations from the reference solutions.

\paragraph{Extrapolation and generalization}
The principal strength of the TO-LRS model is temporal extrapolation on statistically stationary turbulence: from a short high-fidelity reference trajectory it drives long, autonomous low-fidelity runs that recover the correct long-term QoI distributions. The QoI machinery is costly compared to the low-fidelity LBM solver. Its global reductions and kernel convolutions raise the cost of a coarse step by roughly $10\times$ over the substrate solver. Even so, the low-fidelity TO-LRS run produces correct long-term statistics at $40\times$ lower cost per simulated time unit than the high-fidelity simulation.

Generalization across flow configurations is limited. The model does not transfer between flows: a new geometry, Reynolds number, or boundary condition calls for fresh high-fidelity reference trajectories and a refit of the LRS model. The SGS model is also solver-specific, because the predictor-corrector relation depends on the discretization. Each eddy-viscosity substrate therefore needs its own tracking and LRS fit, as the separately trained Smagorinsky and WALE substrates illustrate. However, re-tracking reuses the existing high-fidelity data, so adapting to a new solver or substrate stays cheap. The successful adaptation of the framework from classical Navier--Stokes solvers to the LBM setting demonstrates the versatility of the TO framework.

\paragraph{Future directions}
This project was motivated by large-scale cardiovascular blood-flow simulation, for which lattice Boltzmann solvers such as HemeLB~\cite{mazzeo2008hemelb} are widely used. Patient-specific simulations of pulsatile flow in vascular networks are very expensive at physiologically relevant resolutions. The cardiac cycle gives the flow a natural periodicity favorable for the data-driven closure: a relatively short reference simulation (say, a few cardiac cycles at high resolution) should provide training data that already cover the most relevant phases of the flow, while the autonomous TO-LRS model can then drive long, low-cost simulations over many subsequent cycles. The natural next steps are therefore to apply the SGS model to the complex non-periodic geometries of vascular networks, to quantify the speed-up against existing high-resolution runs, and to confirm that the stochastic SGS forcing leaves clinically relevant quantities such as wall shear stress intact.

Overall, the TO-LRS method offers an interpretable and computationally attractive alternative to high-dimensional neural SGS closures, by coupling turbulence-relevant QoIs to parsimonious time-series corrections.

%%%%%%%%%%%%%%%%%%%%%%%%%%%%%%%%%%%%%%%%
\section*{Software \& data availability}
%%%%%%%%%%%%%%%%%%%%%%%%%%%%%%%%%%%%%%%%

The implementation of the methods which are presented in this paper is publicly available in the repository ``TO\_for\_LBM'' \cite{TO_for_LBM}. This repository contains scripts and data to reproduce all plots in this paper.

%%%%%%%%%%%%%%%%%%%%%%%%%%%
\section*{Funding}
%%%%%%%%%%%%%%%%%%%%%%%%%%%

R.H. and W.E. acknowledge funding from the Netherlands Organization for Scientific Research (NWO) through the ENW-M1 project ``Learning small closure models for large multiscale problems'' (OCENW.M.21.053). This work used the Dutch national e-infrastructure with the support of the SURF Cooperative using grant no.\ EINF-15461. P.V.C. and X.X. acknowledge funding from the European Commission through the CompBioMed Centre of Excellence (Grant Nos.\ 675451 and 823712). 

%%%%%%%%%%%%%%%%%%%%%%%%%%%%%%%%%%
\section*{CRediT author statement}
%%%%%%%%%%%%%%%%%%%%%%%%%%%%%%%%%%

{\bf Rik Hoekstra}: Conceptualization, Data curation, Formal analysis, Investigation, Methodology, Resources, Software, Validation, Visualization, Writing -- original draft, Writing -- review \& editing.
{\bf Xiao Xue}: Conceptualization, Data curation, Formal analysis, Investigation, Methodology, Resources, Software, Validation, Writing -- original draft, Writing -- review \& editing.
{\bf Peter V. Coveney}: Conceptualization, Formal analysis, Methodology, Resources, Supervision, Writing -- review \& editing.
{\bf Wouter Edeling}: Conceptualization, Formal analysis, Funding acquisition, Investigation, Methodology, Project administration, Supervision, Validation, Writing -- review \& editing.

%%%%%%%%%%%%%%%%%%%%%%%%%%%%%%%%%%%%%%%%%%%%
\section*{Declaration of competing interest}
%%%%%%%%%%%%%%%%%%%%%%%%%%%%%%%%%%%%%%%%%%%%

The authors declare that they have no known competing financial interests or personal relationships that could have appeared to influence the work reported in this paper.

%%%%%%%%%%%%%%%%%%%%%%%%%%%%%%%%%%%%%%%%%%%%%%%%%%%%%%%%
\section*{Declaration of Generative AI and AI-assisted technologies in the writing process}
%%%%%%%%%%%%%%%%%%%%%%%%%%%%%%%%%%%%%%%%%%%%%%%%%%%%%%%%%

During the preparation of this work the authors used Claude Opus in order to (re)write sections of the manuscript and code base. After using this tool/service, the authors reviewed and edited the content as needed and take full responsibility for the content of the published article.

\bibliographystyle{plainnat}
\bibliography{references}

@article{ataei2024xlb,
  title = {{XLB}: A differentiable massively parallel lattice {B}oltzmann library in {P}ython},
  author = {Ataei, M. and Salehipour, H.},
  journal = {Computer Physics Communications},
  volume = {300},
  pages = {109187},
  year = {2024},
  publisher = {Elsevier},
  doi = {https://doi.org/10.1016/j.cpc.2024.109187}
}

@article{xue2026fast,
  title = {Fast-forward prediction of lattice {B}oltzmann dynamics with physics-informed neural operators},
  author = {Xue, Xiao and ten Eikelder, Marco F. P. and Gao, Mingyang and Cheng, Xiaoyuan and Yang, Yiming and He, Yi and Wang, Shuo and Cheng, Sibo and Hu, Yukun and Coveney, Peter V.},
  journal = {Nature Communications},
  volume = {17},
  pages = {9243},
  year = {2026},
  publisher = {Nature Publishing Group},
  doi = {https://doi.org/10.1038/s41467-026-75730-1}
}

@article{chikatamarla2013entropic,
  title = {Entropic lattice {B}oltzmann method for turbulent flow simulations: Boundary conditions},
  author = {Chikatamarla, S.S. and Karlin, I.V.},
  journal = {Physica A: Statistical Mechanics and its Applications},
  volume = {392},
  number = {9},
  pages = {1925--1930},
  year = {2013},
  publisher = {Elsevier},
  doi = {https://doi.org/10.1016/j.physa.2012.12.034}
}

@article{edeling_reducing_2020,
  title = {Reducing data-driven dynamical subgrid scale models by physical constraints},
  author = {Edeling, W. and Crommelin, D.},
  journal = {Computers \& Fluids},
  volume = {201},
  pages = {104470},
  year = {2020},
  publisher = {Elsevier},
  doi = {https://doi.org/10.1016/j.compfluid.2020.104470}
}

@article{ephrati_continuous_2025,
  title = {Continuous data assimilation closure for modeling statistically steady turbulence in large-eddy simulation},
  author = {Ephrati, S.R. and Franken, A. and Luesink, E. and Cifani, P. and Geurts, B.J.},
  journal = {Physical Review Fluids},
  volume = {10},
  number = {1},
  pages = {013801},
  year = {2025},
  publisher = {American Physical Society},
  doi = {https://doi.org/10.1103/PhysRevFluids.10.013801}
}

@article{fischer_optimal_2025,
  author = {Fischer, P. and Kaltenbach, S. and Litvinov, S. and Succi, S. and Koumoutsakos, P.},
title = {Optimal Lattice {B}oltzmann Closures through Multi-Agent Reinforcement Learning},
year = {2026},
publisher = {Association for Computing Machinery},
address = {New York, NY, USA},
doi = {https://doi.org/10.1145/3822504},
journal = {ACM Trans. AI Sci.},
month = jul
}

@article{germano_dynamic_1991,
  title = {A dynamic subgrid-scale eddy viscosity model},
  author = {Germano, M. and Piomelli, U. and Moin, P. and Cabot, W.H.},
  journal = {Physics of Fluids A: Fluid Dynamics},
  volume = {3},
  number = {7},
  pages = {1760--1765},
  year = {1991},
  publisher = {American Institute of Physics},
  doi = {https://doi.org/10.1063/1.857955}
}

@phdthesis{gkoudesnes_implementation_nodate,
  title = {Implementation and Verification of {LES} models for {SRT} Lattice {B}oltzmann Methods},
  author = {Gkoudesnes, C.},
  school = {University of Southampton},
  type = {{PhD} thesis},
  year = {2021},
  url = {https://eprints.soton.ac.uk/455268/}
}

@article{grad1949kinetic,
  title = {On the kinetic theory of rarefied gases},
  author = {Grad, H.},
  journal = {Communications on Pure and Applied Mathematics},
  volume = {2},
  number = {4},
  pages = {331--407},
  year = {1949},
  publisher = {Wiley},
  doi = {https://doi.org/10.1002/cpa.3160020403}
}

@incollection{grad1958principles,
	title = {Principles of the {Kinetic} {Theory} of {Gases}},
	volume = {3 / 12},
	isbn = {978-3-642-45894-1 978-3-642-45892-7},
	doi = {10.1007/978-3-642-45892-7_3},
	booktitle = {Thermodynamik der {Gase} / {Thermodynamics} of {Gases}},
	publisher = {Springer Berlin Heidelberg},
	author = {Grad, H.},
	editor = {Flügge, S.},
	year = {1958},
	pages = {205--294},
}

@article{he1997theory,
  title = {Theory of the lattice {B}oltzmann method: From the {B}oltzmann equation to the lattice {B}oltzmann equation},
  author = {He, X. and Luo, L.S.},
  journal = {Physical Review E},
  volume = {56},
  number = {6},
  pages = {6811--6817},
  year = {1997},
  publisher = {American Physical Society},
  doi = {https://doi.org/10.1103/PhysRevE.56.6811}
}

@article{hoekstra2024Reduced_data-driven,
  title = {Reduced data-driven turbulence closure for capturing long-term statistics},
  author = {Hoekstra, R. and Crommelin, D. and Edeling, W.},
  journal = {Computers \& Fluids},
  volume = {285},
  pages = {106469},
  year = {2024},
  publisher = {Elsevier},
  doi = {https://doi.org/10.1016/j.compfluid.2024.106469}
}

@article{hoekstra_reduced_2026,
  title = {Reduced subgrid scale terms in three-dimensional turbulence},
  author = {Hoekstra, R. and Edeling, W.},
  journal = {Computer Methods in Applied Mechanics and Engineering},
  volume = {449},
  pages = {118506},
  year = {2026},
  publisher = {Elsevier},
  doi = {https://doi.org/10.1016/j.cma.2025.118506}
}

@article{khan_physics-constrained_2026,
  title = {Physics-Constrained Neural Closure for Lattice {B}oltzmann Large-Eddy Simulation},
  author = {Khan, M.I. and Succi, S. and Yao, H.D. and Falcucci, G.},
  journal = {arXiv preprint arXiv:2603.15992},
  year = {2026},
  doi = {https://doi.org/10.48550/arXiv.2603.15992}
}

@article{kochkov_machine_2021,
  title = {Machine learning--accelerated computational fluid dynamics},
  author = {Kochkov, D. and Smith, J.A. and Alieva, A. and Wang, Q. and Brenner, M.P. and Hoyer, S.},
  journal = {Proceedings of the National Academy of Sciences},
  volume = {118},
  number = {21},
  pages = {e2101784118},
  year = {2021},
  publisher = {National Academy of Sciences},
  doi = {https://doi.org/10.1073/pnas.2101784118}
}

@book{kruger_lattice_2017,
  title = {The Lattice {B}oltzmann Method: Principles and Practice},
  author = {Kr{\"u}ger, T. and Kusumaatmaja, H. and Kuzmin, A. and Shardt, O. and Silva, G. and Viggen, E.M.},
  series = {Graduate Texts in Physics},
  publisher = {Springer International Publishing},
  address = {Cham},
  year = {2017},
  doi = {https://doi.org/10.1007/978-3-319-44649-3}
}

@article{kupershtokh_equations_2009,
  title = {On equations of state in a lattice {B}oltzmann method},
  author = {Kupershtokh, A.L. and Medvedev, D.A. and Karpov, D.I.},
  journal = {Computers \& Mathematics with Applications},
  volume = {58},
  number = {5},
  pages = {965--974},
  year = {2009},
  publisher = {Elsevier},
  doi = {https://doi.org/10.1016/j.camwa.2009.02.024}
}

@article{ladd1993short,
  title = {Short-time motion of colloidal particles: Numerical simulation via a fluctuating lattice-{B}oltzmann equation},
  author = {Ladd, A.J.C.},
  journal = {Physical Review Letters},
  volume = {70},
  number = {9},
  pages = {1339--1342},
  year = {1993},
  publisher = {American Physical Society},
  doi = {https://doi.org/10.1103/PhysRevLett.70.1339}
}

@article{lallemand_theory_2000,
  title = {Theory of the lattice {B}oltzmann method: Dispersion, dissipation, isotropy, {G}alilean invariance, and stability},
  author = {Lallemand, P. and Luo, L.S.},
  journal = {Physical Review E},
  volume = {61},
  number = {6},
  pages = {6546--6562},
  year = {2000},
  publisher = {American Physical Society},
  doi = {https://doi.org/10.1103/PhysRevE.61.6546}
}

@article{lallemand2021lattice,
  title = {The lattice {B}oltzmann method for nearly incompressible flows},
  author = {Lallemand, P. and Luo, L.S. and Krafczyk, M. and Yong, W.A.},
  journal = {Journal of Computational Physics},
  volume = {431},
  pages = {109713},
  year = {2021},
  publisher = {Elsevier},
  doi = {https://doi.org/10.1016/j.jcp.2020.109713}
}

@article{lo2024uncertainty,
  title = {Uncertainty quantification of the impact of peripheral arterial disease on abdominal aortic aneurysms in blood flow simulations},
  author = {Lo, S.C.Y. and McCullough, J.W.S. and Xue, X. and Coveney, P.V.},
  journal = {Journal of the Royal Society Interface},
  volume = {21},
  number = {213},
  pages = {20230656},
  year = {2024},
  publisher = {The Royal Society},
  doi = {https://doi.org/10.1098/rsif.2023.0656}
}

@article{lo2025multi,
  title = {A multi-component, multi-physics computational model for solving coupled cardiac electromechanics and vascular haemodynamics},
  author = {Lo, S.C.Y. and Zingaro, A. and McCullough, J.W.S. and Xue, X. and Gonzalez-Martin, P. and Joo, B. and V{\'a}zquez, M. and Coveney, P.V.},
  journal = {Computer Methods in Applied Mechanics and Engineering},
  volume = {446},
  pages = {118185},
  year = {2025},
  publisher = {Elsevier},
  doi = {https://doi.org/10.1016/j.cma.2025.118185}
}

@article{malaspinas_consistent_2012,
  title = {Consistent subgrid scale modelling for lattice {B}oltzmann methods},
  author = {Malaspinas, O. and Sagaut, P.},
  journal = {Journal of Fluid Mechanics},
  volume = {700},
  pages = {514--542},
  year = {2012},
  publisher = {Cambridge University Press},
  doi = {https://doi.org/10.1017/jfm.2012.155}
}

@article{mazzeo2008hemelb,
  title = {{HemeLB}: A high performance parallel lattice-{B}oltzmann code for large scale fluid flow in complex geometries},
  author = {Mazzeo, M.D. and Coveney, P.V.},
  journal = {Computer Physics Communications},
  volume = {178},
  number = {12},
  pages = {894--914},
  year = {2008},
  publisher = {Elsevier},
  doi = {https://doi.org/10.1016/j.cpc.2008.02.013}
}

@article{moin_numerical_1980,
  title = {On the numerical solution of time-dependent viscous incompressible fluid flows involving solid boundaries},
  author = {Moin, P. and Kim, J.},
  journal = {Journal of Computational Physics},
  volume = {35},
  number = {3},
  pages = {381--392},
  year = {1980},
  publisher = {Elsevier},
  doi = {https://doi.org/10.1016/0021-9991(80)90076-5}
}

@article{nathen_adaptive_2018,
  title = {Adaptive filtering for the simulation of turbulent flows with lattice {B}oltzmann methods},
  author = {Nathen, P. and Haussmann, M. and Krause, M.J. and Adams, N.A.},
  journal = {Computers \& Fluids},
  volume = {172},
  pages = {510--523},
  year = {2018},
  publisher = {Elsevier},
  doi = {https://doi.org/10.1016/j.compfluid.2018.03.042}
}

@article{nicoud_ducros_1999,
  title = {Subgrid-scale stress modelling based on the square of the velocity gradient tensor},
  author = {Nicoud, F. and Ducros, F.},
  journal = {Flow, Turbulence and Combustion},
  volume = {62},
  number = {3},
  pages = {183--200},
  year = {1999},
  publisher = {Springer},
  doi = {https://doi.org/10.1023/A:1009995426001}
}

@article{ortali_kinetic_2025,
  title = {Kinetic data-driven approach to turbulence subgrid modeling},
  author = {Ortali, G. and Gabbana, A. and Demo, N. and Rozza, G. and Toschi, F.},
  journal = {Physical Review Research},
  volume = {7},
  number = {1},
  pages = {013202},
  year = {2025},
  publisher = {American Physical Society},
  doi = {https://doi.org/10.1103/PhysRevResearch.7.013202}
}

@article{premnath_dynamic_2009,
  title = {Dynamic subgrid scale modeling of turbulent flows using lattice-{B}oltzmann method},
  author = {Premnath, K.N. and Pattison, M.J. and Banerjee, S.},
  journal = {Physica A: Statistical Mechanics and its Applications},
  volume = {388},
  number = {13},
  pages = {2640--2658},
  year = {2009},
  publisher = {Elsevier},
  doi = {https://doi.org/10.1016/j.physa.2009.02.041}
}

@article{shan1993lattice,
  title = {Lattice {B}oltzmann model for simulating flows with multiple phases and components},
  author = {Shan, X. and Chen, H.},
  journal = {Physical Review E},
  volume = {47},
  number = {3},
  pages = {1815--1819},
  year = {1993},
  publisher = {American Physical Society},
  doi = {https://doi.org/10.1103/PhysRevE.47.1815}
}

@article{shan2006kinetic,
  title = {Kinetic theory representation of hydrodynamics: a way beyond the {N}avier--{S}tokes equation},
  author = {Shan, X. and Yuan, X.F. and Chen, H.},
  journal = {Journal of Fluid Mechanics},
  volume = {550},
  pages = {413--441},
  year = {2006},
  publisher = {Cambridge University Press},
  doi = {https://doi.org/10.1017/S0022112005008153}
}

@article{simeoni2024lattice,
  title = {Lattice {B}oltzmann method for warm fluid simulations of plasma wakefield acceleration},
  author = {Simeoni, D. and Parise, G. and Guglietta, F. and Rossi, A.R. and Rosenzweig, J. and Cianchi, A. and Sbragaglia, M.},
  journal = {Physics of Plasmas},
  volume = {31},
  number = {1},
  pages = {013904},
  year = {2024},
  publisher = {AIP Publishing},
  doi = {https://doi.org/10.1063/5.0175910}
}

@article{smagorinsky_general_1963,
  title = {General circulation experiments with the primitive equations: {I}. {T}he basic experiment},
  author = {Smagorinsky, J.},
  journal = {Monthly Weather Review},
  volume = {91},
  number = {3},
  pages = {99--164},
  year = {1963},
  publisher = {American Meteorological Society},
  doi = {https://doi.org/10.1175/1520-0493(1963)091%3C0099:GCEWTP%3E2.3.CO;2}
}

@article{stolz_approximate_1999,
  title = {An approximate deconvolution procedure for large-eddy simulation},
  author = {Stolz, S. and Adams, N.A.},
  journal = {Physics of Fluids},
  volume = {11},
  number = {7},
  pages = {1699--1701},
  year = {1999},
  publisher = {AIP Publishing},
  doi = {https://doi.org/10.1063/1.869867}
}

@book{succi2001lattice,
  title = {The Lattice {B}oltzmann Equation for Fluid Dynamics and Beyond},
  author = {Succi, S.},
  publisher = {Oxford University Press},
  address = {Oxford},
  year = {2001},
  doi = {https://doi.org/10.1093/oso/9780198503989.001.0001}
}

@article{vreman_comparison_2014,
  title = {Comparison of direct numerical simulation databases of turbulent channel flow at {$Re_{\tau}$}=180},
  author = {Vreman, A.W. and Kuerten, J.G.M.},
  journal = {Physics of Fluids},
  volume = {26},
  number = {1},
  pages = {015102},
  year = {2014},
  publisher = {AIP Publishing},
  doi = {https://doi.org/10.1063/1.4861064}
}

@article{xue2018effects,
  title = {Effects of thermal fluctuations in the fragmentation of a nanoligament},
  author = {Xue, X. and Sbragaglia, M. and Biferale, L. and Toschi, F.},
  journal = {Physical Review E},
  volume = {98},
  number = {1},
  pages = {012802},
  year = {2018},
  publisher = {American Physical Society},
  doi = {https://doi.org/10.1103/PhysRevE.98.012802}
}

@article{xue2022synthetic,
  title = {Synthetic turbulence generator for lattice {B}oltzmann method at the interface between {RANS} and {LES}},
  author = {Xue, X. and Yao, H.D. and Davidson, L.},
  journal = {Physics of Fluids},
  volume = {34},
  number = {5},
  pages = {055118},
  year = {2022},
  publisher = {AIP Publishing},
  doi = {https://doi.org/10.1063/5.0090641}
}

@article{xue_physics_2024,
  title = {Physics informed data-driven near-wall modelling for lattice {B}oltzmann simulation of high {R}eynolds number turbulent flows},
  author = {Xue, X. and Wang, S. and Yao, H.D. and Davidson, L. and Coveney, P.V.},
  journal = {Communications Physics},
  volume = {7},
  number = {1},
  pages = {338},
  year = {2024},
  publisher = {Nature Publishing Group},
  doi = {https://doi.org/10.1038/s42005-024-01832-1}
}

@book{TO_for_LBM,
	title = {{TO\_for\_LBM} (v1.0) [Software]},
	author = {Hoekstra, R.},
        year = {2026},
        month = sep,
        doi = {https://doi.org/10.5281/zenodo.22299377},
        publisher = {Zenodo}
}

\newpage
%%%%%%%%%%%%%%%%%%%%%%%%%%%%%%%%%%%%%%%%%%%%%%%%%%%%%%%%%%%
\appendix
%%%%%%%%%%%%%%%%%%%%%%%%%%%%%%%%%%%%%%%%%%%%%%%%%%%%%%%%%%%%
\section{Functional derivatives of QoIs}\label{app:derivatives of QoIs}
This appendix derives the functional derivatives (sensitivities) $\delta Q_i/\delta\boldsymbol{v}$ of the scale-aware energy and enstrophy QoIs and the mean-profile QoIs. We derive closed-form expressions for these sensitivities under a general kernel filter
\begin{equation}
    R_i \boldsymbol{v} = K_i * \boldsymbol{v}.
\end{equation}
Note that the sharp Fourier filters also fit this category, albeit with non-compact kernels $K_i$.

For the scale-aware energy $E_{R_i} = \frac{1}{2}\int_\Omega \|R_i \boldsymbol{v}\|^2 \, \D\boldsymbol{x}$, the sensitivity with respect to the velocity field $\boldsymbol{v}$ is found by considering the first order terms in the variation:
\begin{align}
    E_{R_i} + \delta E_{R_i} &= \int_\Omega \frac{1}{2} R_{i} (\boldsymbol{v} + \delta \boldsymbol{v}) \cdot R_{i} (\boldsymbol{v} + \delta \boldsymbol{v}) \, \D \boldsymbol{x} \nonumber\\
    \delta E_{R_i} &= \int_\Omega R_{i} \boldsymbol{v} \cdot R_{i}  \delta \boldsymbol{v} \, \D \boldsymbol{x} \nonumber \\
    &= \int_\Omega (R_{i}^* R_{i} \boldsymbol{v}) \cdot \delta \boldsymbol{v} \, \D \boldsymbol{x},
\end{align}
where $R_{i}^*$ denotes the adjoint of the filter operator. Since both the sharp Fourier filters and the symmetric convolution kernels ($G_s$ and $L_s$) are self-adjoint, we have $R_{i}^* = R_{i}$, which gives
\begin{equation}
    \frac{\delta E_{R_i}}{\delta \boldsymbol{v}} = R_{i} R_i \boldsymbol{v}.
\end{equation}

For the scale-aware enstrophy $Z_{R_i} = \frac{1}{2}\int_\Omega \|R_i\,\mathrm{curl}(\boldsymbol{v})\|^2 \, \D\boldsymbol{x}$, the derivation is less straightforward due to the curl operator. The variation is:
\begin{align}
    Z_{R_i} + \delta Z_{R_i} &= \int_\Omega \frac{1}{2}R_i \ \text{curl}(\boldsymbol{v}+\delta \boldsymbol{v}) \cdot R_i \ \text{curl}(\boldsymbol{v}+\delta \boldsymbol{v}) \, \D \boldsymbol{x} \nonumber\\
    \delta Z_{R_i} &=\int_\Omega R_i \ \text{curl}(\boldsymbol{v}) \cdot R_i \, \text{curl}(\delta \boldsymbol{v}) \, \D \boldsymbol{x} \nonumber \\
    &=\int_\Omega (R_i^* R_i \ \text{curl}(\boldsymbol{v})) \cdot \text{curl}(\delta \boldsymbol{v}) \, \D \boldsymbol{x}.
\end{align}
Applying the adjoint of the curl operator with vanishing boundary terms (we assume periodicity or $\delta \boldsymbol{v} = 0$ at no-slip boundaries) transfers the differential operator onto the filtered vorticity field:
\begin{align}
    \delta Z_{R_i} &= \int_\Omega \text{curl}^*(R_i^* R_i \ \text{curl}(\boldsymbol{v})) \cdot \delta \boldsymbol{v} \, \D \boldsymbol{x}.
\end{align}
In two dimensions the vorticity is a scalar and $\text{curl}^* \omega = (\partial_y \omega, -\partial_x \omega)^T$, so that
\begin{equation}
    \frac{\delta Z_{R_i}}{\delta \boldsymbol{v}} = \begin{pmatrix} \partial_y (R_{i} R_i{\omega}) \\ -\partial_x (R_{i}R_i \omega) \end{pmatrix}.
    \label{eq:enstrophy-sensitivity-2d}
\end{equation}
In three dimensions the vorticity $\boldsymbol{\omega} = \mathrm{curl}(\boldsymbol{v})$ is a vector and the curl is self-adjoint, which gives
\begin{equation}\label{eq:enstrophy-sensitivity-3d}
    \frac{\delta Z_{R_i}}{\delta \boldsymbol{v}} = \mathrm{curl}\!\left(R_i R_i \boldsymbol{\omega}\right).
\end{equation}

The mean-profile QoIs \eqref{eq:mean-profile-qoi} are linear in $\boldsymbol{v}$, so their functional derivative is the fixed field
\begin{equation}\label{eq:mean-profile-sensitivity}
    \frac{\delta Q_j^{\text{mp}}}{\delta \boldsymbol{v}} = w_j(y)\,\boldsymbol{e}_x,
\end{equation}
independent of $\boldsymbol{v}$.

\section{Individual QoIs in Smagorinsky constant optimization on Kolmogorov flow} \label{appendix:KS-distances for individual QOIs}
Figure \ref{fig:app-smag-optimization-individual} shows, for each QoI, the KS-distances between the trajectories from the Smagorinsky calibration runs and the reference simulation. Different QoIs are optimized by different values of the Smagorinsky constant.

\begin{figure}[htbp]
    \centering
    \includegraphics[width=0.8\linewidth]{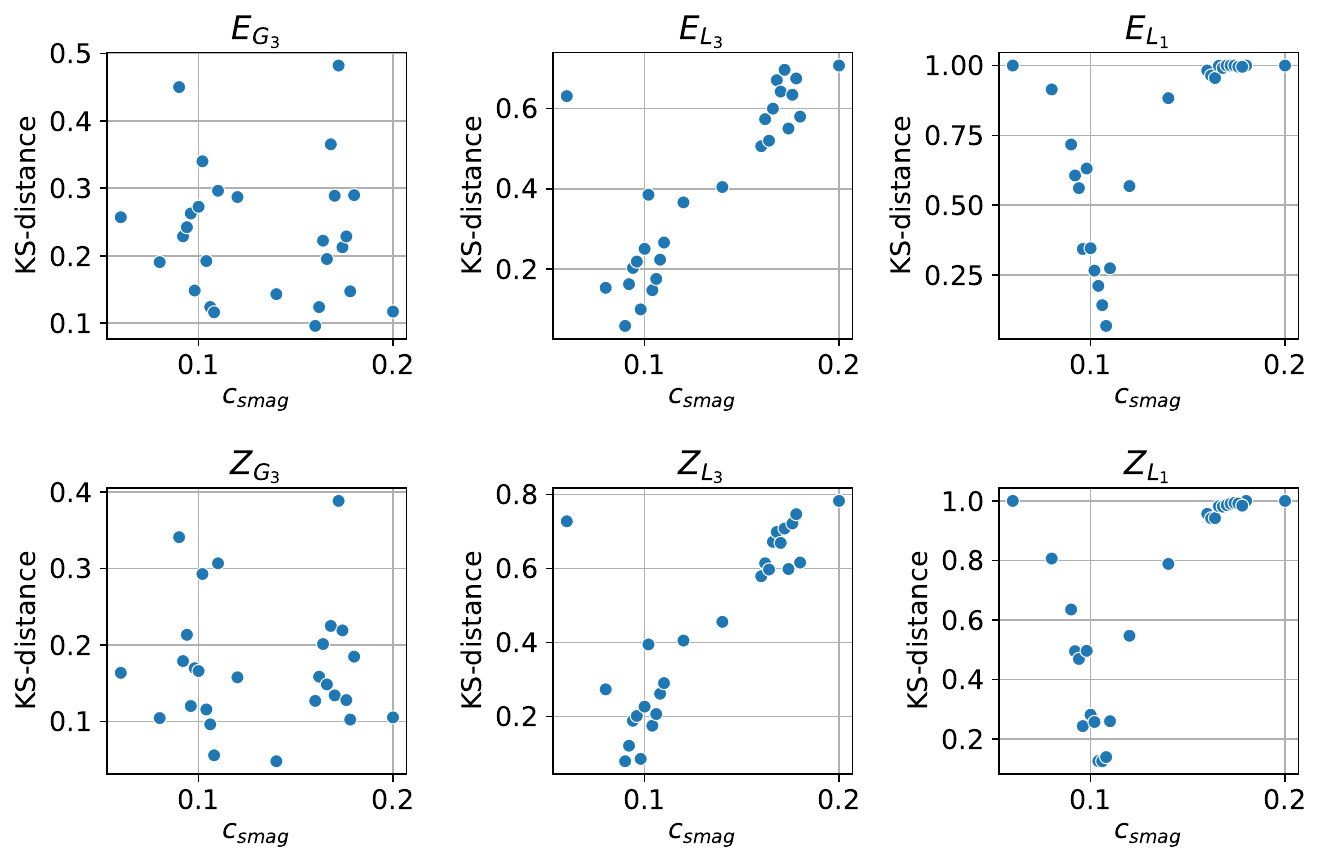}
    \caption{Optimization of the Smagorinsky constant for the Kolmogorov flow: KS-distance between calibration runs and reference trajectories for each QoI.}
    \label{fig:app-smag-optimization-individual}
\end{figure}

\section{Tracking QoIs with the TO method on Kolmogorov flow}\label{app:tracking}
Figures~\ref{fig:qois-in-lf-track-precollision} and~\ref{fig:qois-in-lf-track-postcollision} show QoI trajectories from the same tracking simulation, evaluated on the post-streaming and post-collision fields respectively. Because nudging acts on the post-collision state, its trajectories follow the reference more closely; the post-streaming state is corrected only indirectly, giving slightly larger deviations, most notably an overestimation of small-scale energy.

\begin{figure}[H]
    \centering
    \includegraphics[width=0.9\linewidth]{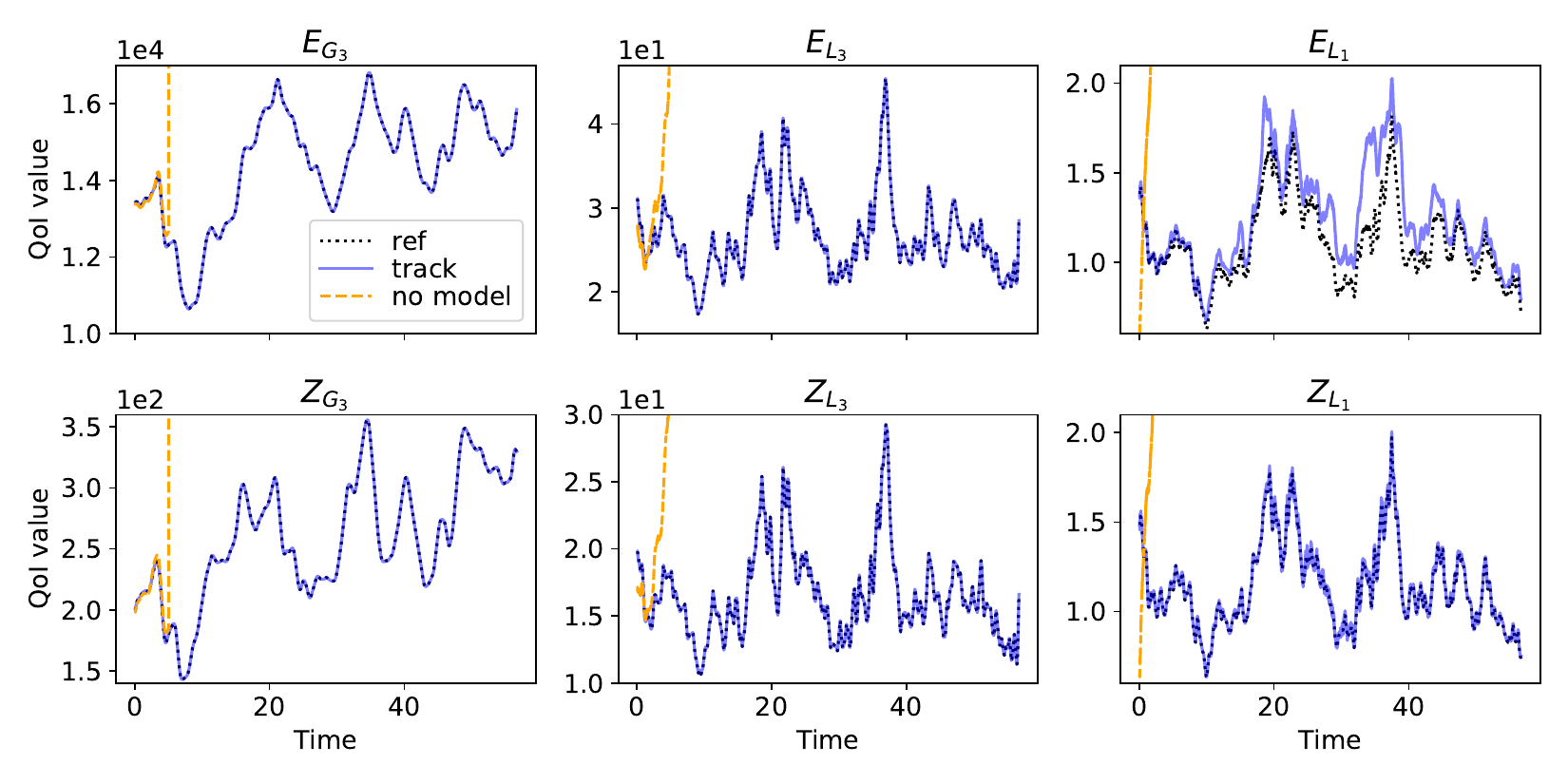}
    \caption{QoI trajectories from the post-streaming state in a low-fidelity tracking simulation of the Kolmogorov flow.}
    \label{fig:qois-in-lf-track-precollision}
\end{figure}

\begin{figure}[H]
    \centering
    \includegraphics[width=0.9\linewidth]{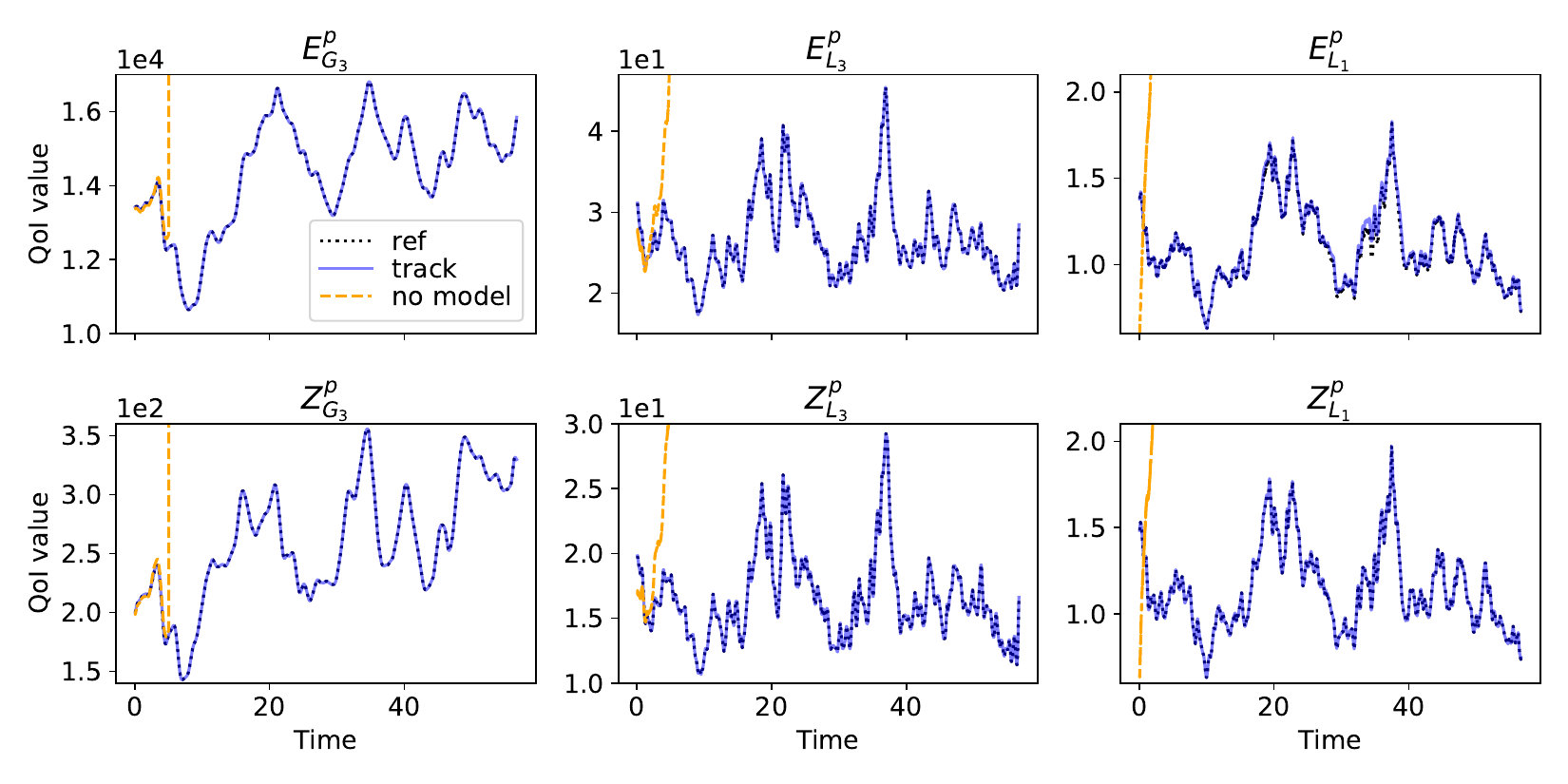}
    \caption{QoI trajectories from the post-collision state in a low-fidelity tracking simulation of the Kolmogorov flow.}
    \label{fig:qois-in-lf-track-postcollision}
\end{figure}

\section{Online trajectories for the best models on Kolmogorov flow}\label{app:online-best-models}
This appendix collects the online trajectories behind the best-model analysis on the two-dimensional Kolmogorov flow of Section~\ref{sec:results}: the QoI trajectories for the two shorter training windows (the $T=57$ case is Figure~\ref{fig:qoi-trajectories-h500-T57} in the main text), and the predicted SGS corrections $dQ$ for all three best models.

\begin{figure}[H]
    \centering
    \includegraphics[width=\linewidth]{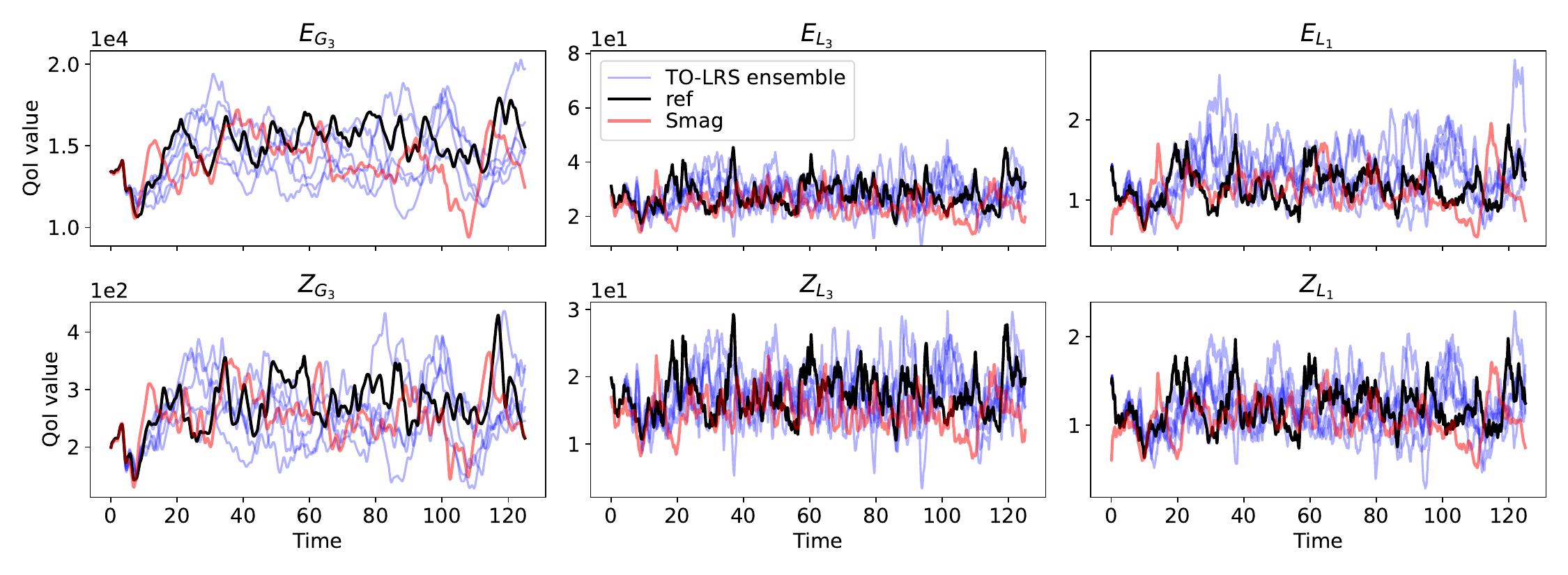}
    \caption{QoI trajectories from the online TO-LRS ensemble with history length $h=400$, $\lambda = 0$, and training data up to $T=28$, for the Kolmogorov flow. The reference DNS trajectory is shown together with the five replica trajectories of the ensemble and the calibrated Smagorinsky baseline.}
    \label{fig:qoi-trajectories-h400-T28}
\end{figure}

\begin{figure}[H]
    \centering
    \includegraphics[width=\linewidth]{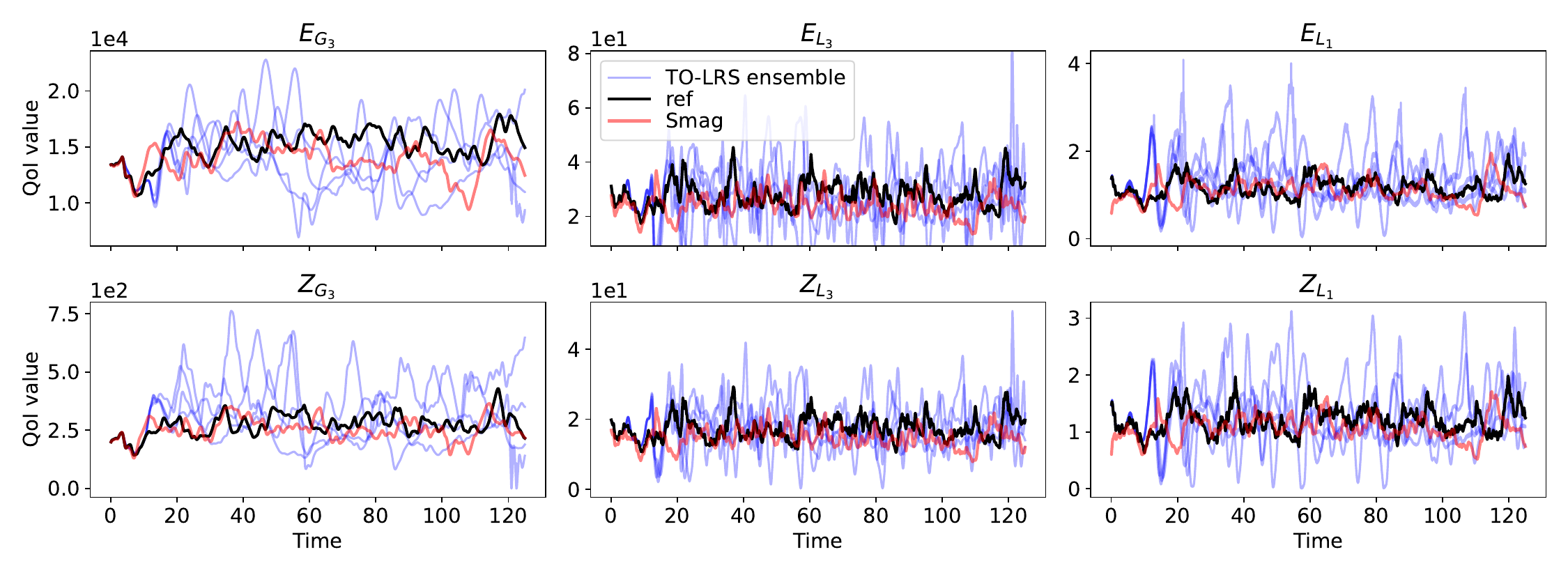}
    \caption{QoI trajectories from the online TO-LRS ensemble with history length $h=400$, $\lambda = 10$, and training data up to $T=14$, for the Kolmogorov flow. The reference DNS trajectory is shown together with the four stable replicas of the ensemble and the calibrated Smagorinsky baseline; one replica became unstable and is excluded.}
    \label{fig:qoi-trajectories-h400-T14}
\end{figure}

\begin{figure}[H]
    \centering
    \includegraphics[width=\linewidth]{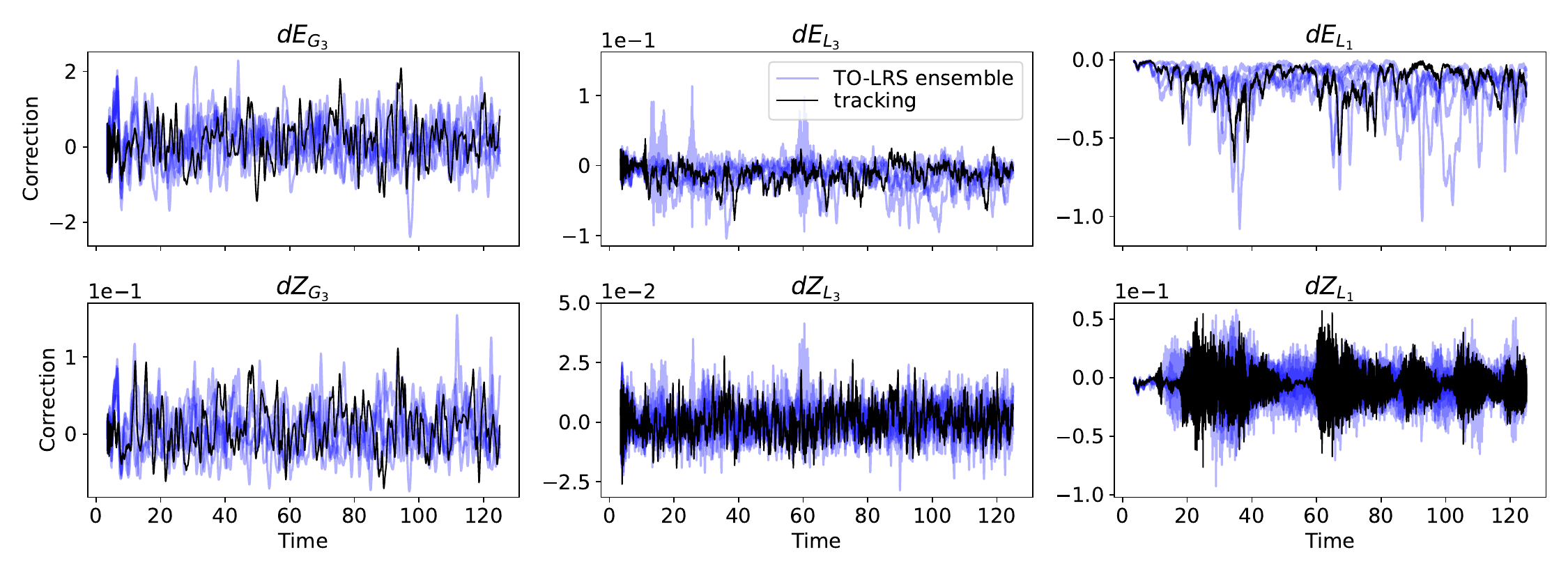}
    \caption{Predicted SGS corrections $dQ_i^n$ from the online TO-LRS ensemble with $h=500$, $\lambda = 0$, and training data up to $T=57$, for the Kolmogorov flow. The corrections in the tracking simulation and the five online replicas are shown together.}
    \label{fig:dQ-h500-T57}
\end{figure}

\begin{figure}[H]
    \centering
    \includegraphics[width=\linewidth]{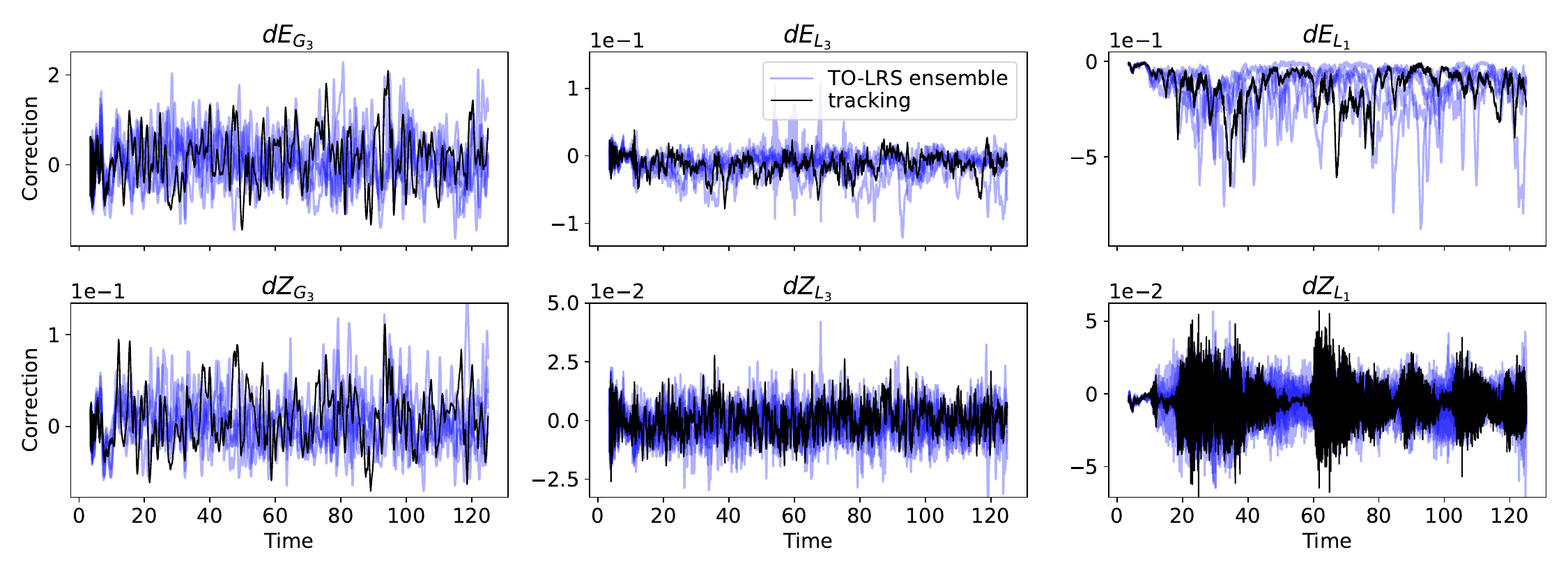}
    \caption{Predicted SGS corrections $dQ_i^n$ from the online TO-LRS ensemble with $h=400$, $\lambda = 0$, and training data up to $T=28$, for the Kolmogorov flow. The corrections in the tracking simulation and the five online replicas are shown together.}
    \label{fig:dQ-h400-T28}
\end{figure}

\begin{figure}[H]
    \centering
    \includegraphics[width=\linewidth]{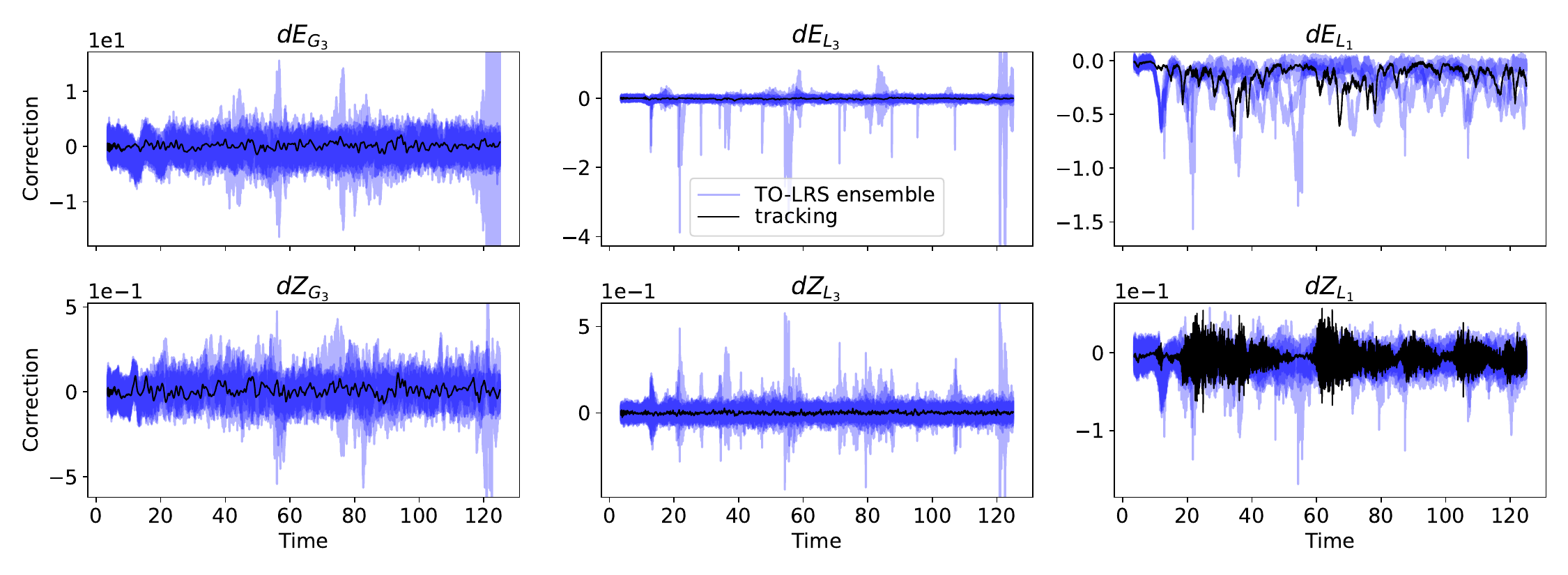}
    \caption{Predicted SGS corrections $dQ_i^n$ from the online TO-LRS ensemble with $h=400$, $\lambda = 10$, and training data up to $T=14$, for the Kolmogorov flow. The corrections in the tracking simulation and the four stable online replicas are shown together.}
    \label{fig:dQ-h400-T14}
\end{figure}

\section{TO-LRS with sharp-Fourier-filter QoIs on Kolmogorov flow}\label{app:fourier-filters}
This appendix repeats the study of Section~\ref{sec:kolmogorov-flow} with scale-aware QoIs based on the sharp Fourier filters \eqref{eq:scale-aware-energy-fourier} in place of the compact kernel filters. The wavenumber bins are as in Figure~\ref{fig:fourier-filters}, namely $\{R_{[0,12]}, R_{[13,32]}, R_{[33,128]}\}$, so that ``large'', ``mid'' and ``small'' scales refer to comparable spectral ranges across the two setups. Only the shape of the filter changes.

The filter shape does, however, affect the numerical QoI values. Figure~\ref{fig:qois-hf-fourier-vs-kernel} compares the trajectories of both families on the same high-fidelity simulation. The scale-aware QoIs agree in shape and order of magnitude, but the two trajectories do not coincide. Absolute QoI values, summed KS-distances, and predictions are therefore not directly comparable between the families; we compare trends rather than raw numbers. The Smagorinsky baseline keeps the constant calibrated in Section~\ref{sec:kolmogorov-flow}.

\begin{figure}[htbp]
    \centering
    \includegraphics[width=0.8\linewidth]{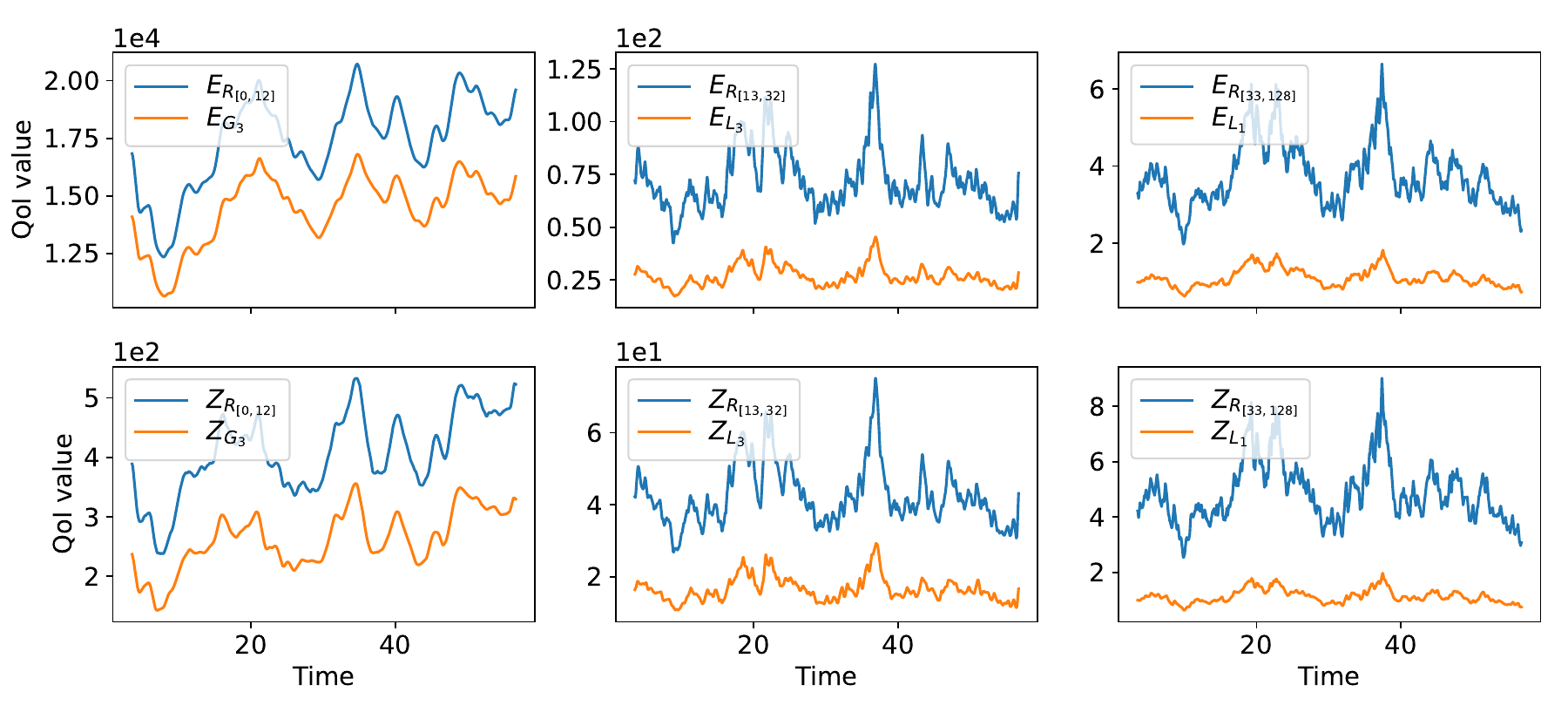}
    \caption{Scale-aware QoI trajectories computed from the same Kolmogorov flow DNS reference for the two filter families: sharp Fourier filters $\{R_{[0,12]}, R_{[13,32]}, R_{[33,128]}\}$ and the compact kernel filters $\{G_3, L_3, L_1\}$ used in the main text.}
    \label{fig:qois-hf-fourier-vs-kernel}
\end{figure}

Figure~\ref{fig:track-qois-fourier-correction} shows the SGS corrections during tracking with Fourier-based QoIs, normalized by each QoI's time-averaged value. They behave much as in the kernel-based tracking (Figure~\ref{fig:dQ-in-tracking}). The main difference is the mid-scale energy correction, which is mostly negative with the kernel filter but more symmetric about zero here.

\begin{figure}[htbp]
    \centering
    \includegraphics[width=0.8\linewidth]{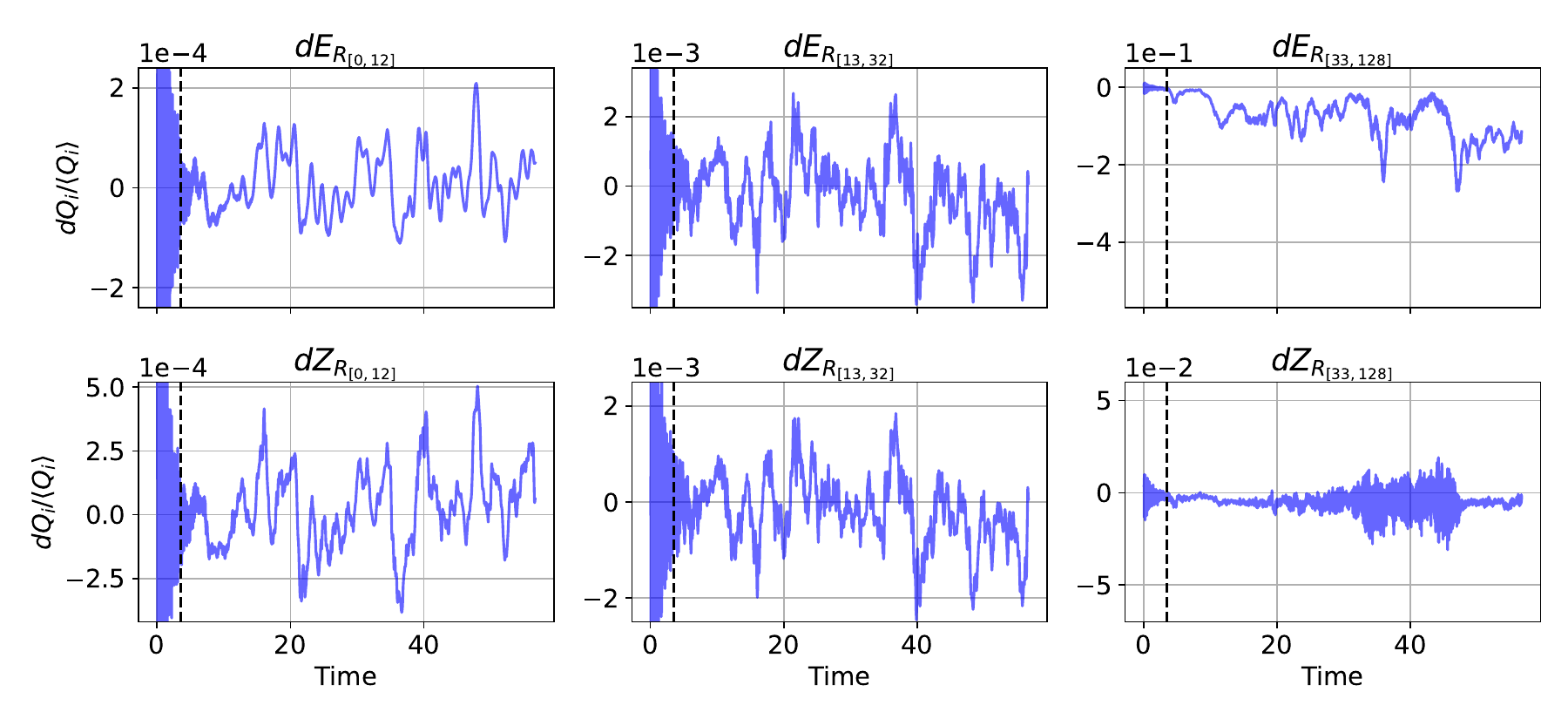}
    \caption{SGS corrections $dQ_i^n$ applied during tracking of the Kolmogorov flow with sharp Fourier-filter QoIs, normalized by the time-averaged value of the corresponding QoI. The dashed vertical line marks the end of the spin-up window.}
    \label{fig:track-qois-fourier-correction}
\end{figure}

We use the same training windows as before. In the unregularized case (Figure~\ref{fig:TO-LRS-fourier-online-performance-lambda0}), the ensembles trained on 57 TU closely mirror the kernel-filter results: most history lengths are stable with summed KS-distance well below the Smagorinsky baseline, best around $h=400$ and $h=800$ (only $h=500$ has an unstable replica). At 28 TU the picture deteriorates relative to the kernel filters; every ensemble has at least one unstable replica and the stable ones are no better than Smagorinsky. At 14 TU all unregularized models are unstable.

\begin{figure}[htbp]
    \centering
    \includegraphics[width=0.7\linewidth]{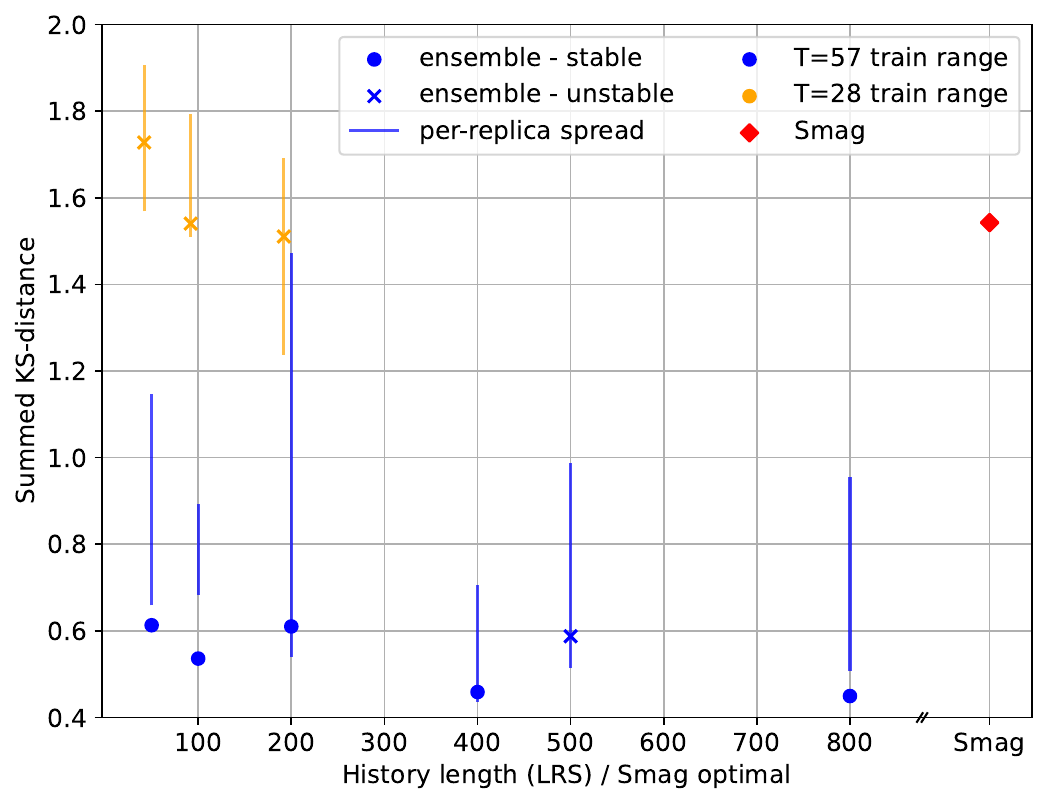}
    \caption{Online performance over 125 TU on the Kolmogorov flow, for Fourier-based TO-LRS models, trained on 57 or 28 TU without regularization. Results are based on five-replica ensembles; dots show the pooled summed KS-distance over stable replicas, vertical bars show the replica spread, and an `X' marks ensembles in which at least one replica became unstable. The Smagorinsky baseline uses the constant calibrated against the kernel-filter QoIs in Section~\ref{sec:kolmogorov-flow}.}
    \label{fig:TO-LRS-fourier-online-performance-lambda0}
\end{figure}

Regularization helps on the two smaller training windows (Figure~\ref{fig:TO_LRS-fourier-online-smaller-training-regularized}). At $T=28$, strong regularization ($\lambda=10$) stabilizes the long histories and brings $h=400$ and $h=800$ noticeably below Smagorinsky. At $T=14$ regularization again rescues several configurations, but accuracy stays weak. Relative to the kernel-filter case, regularization is thus needed earlier---already at $T=28$, where the kernel filters still tolerated $\lambda=0$---and the small-data regime is more fragile.

\begin{figure}[htbp]
    \begin{subfigure}[b]{0.49\textwidth}
        \centering
        \includegraphics[width = \linewidth]{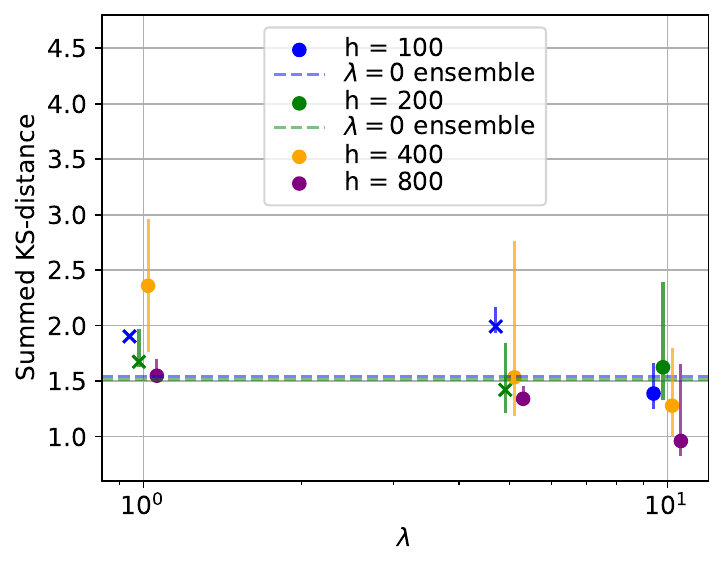}
        \caption{Training data up to $T=28$.}
    \end{subfigure}
    \begin{subfigure}[b]{0.49\textwidth}
        \centering
        \includegraphics[width = \linewidth]{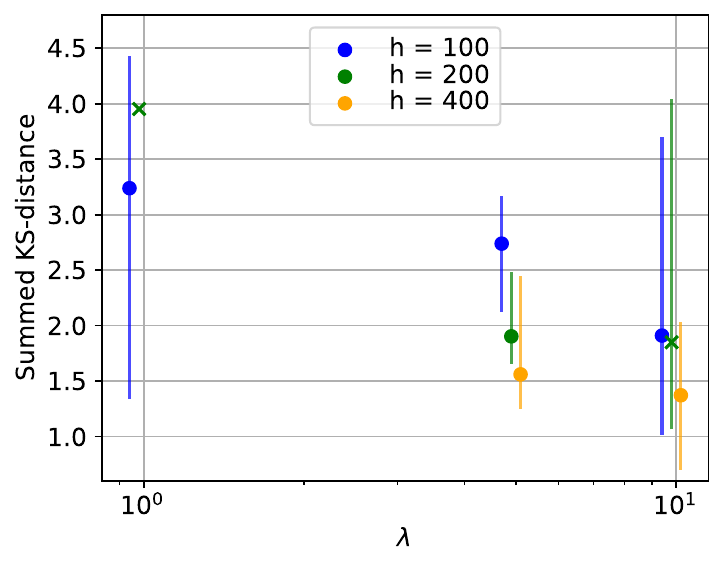}
        \caption{Training data up to $T=14$.}
    \end{subfigure}
    \caption{Influence of regularization on Fourier-based TO-LRS models trained on smaller datasets, for the Kolmogorov flow. Results are based on five-replica ensembles over 125 TU. Dots show the pooled summed KS-distance over stable replicas; an `X' marks ensembles in which at least one replica became unstable. Vertical bars show the replica spread, and dashed lines show the pooled summed KS-distance at $\lambda = 0$ when those ensembles were stable.}
    \label{fig:TO_LRS-fourier-online-smaller-training-regularized}
\end{figure}

We now examine the three best-performing Fourier-filter TO-LRS configurations: $h=400$ trained on $T=57$ with $\lambda=0$, $h=800$ trained on $T=28$ with $\lambda=10$, and $h=400$ trained on $T=14$ with $\lambda=10$. Figure~\ref{fig:energy-spectra-fourier} shows their time-averaged energy spectra. All three configurations show the same two oscillations in $\|\boldsymbol{k}\| \in [30, 60]$ that were already present in the kernel-filter case (Figure~\ref{fig:energy-spectra}), but here they are markedly more pronounced.

\begin{figure}[htbp]
    \centering
    \includegraphics[width=0.7\linewidth]{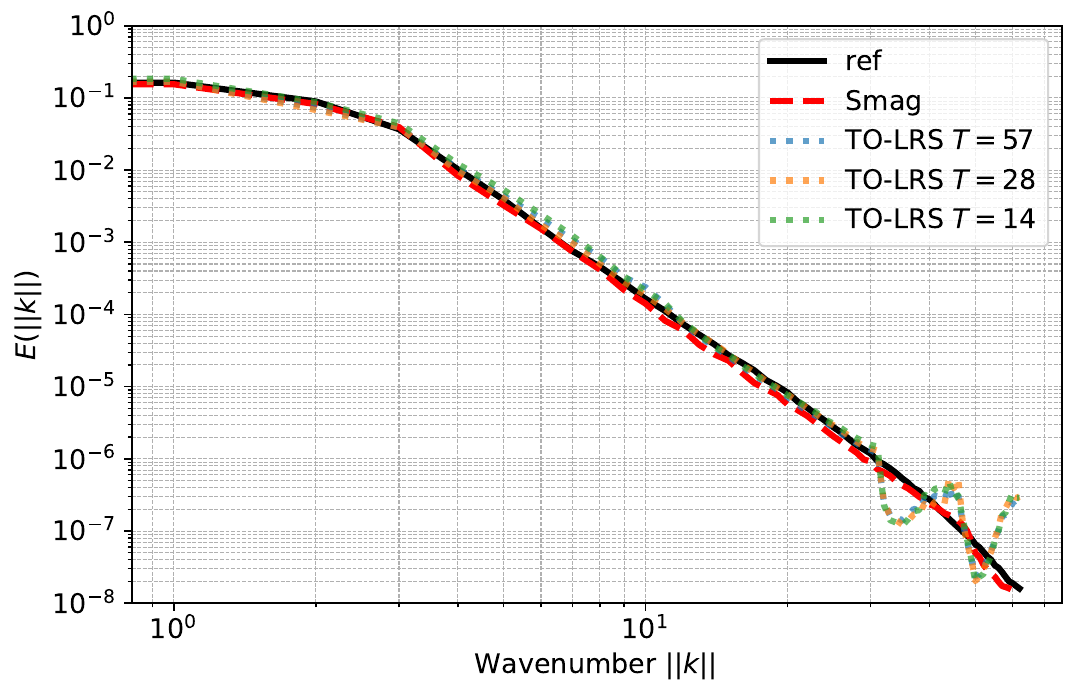}
    \caption{Energy spectra of the best-performing Fourier-based TO-LRS models, the calibrated Smagorinsky baseline, and the filtered-DNS reference, for the Kolmogorov flow. Each spectrum is averaged over 11 snapshots in $t \in [62.5, 125]$ and over stable ensemble members.}
    \label{fig:energy-spectra-fourier}
\end{figure}

Finally, Figure~\ref{fig:final-fields-fourier} shows the vorticity fields at the end of the simulations. The diagonal high-frequency pattern flagged in Section~\ref{sec:results} is also clearly present in the Fourier-filter tracking simulation and persists in all three online TO-LRS configurations. Because the sharp Fourier filters are exactly isotropic in Fourier space, this rules out filter discretization as the origin of the artifact and supports the lattice-anisotropy hypothesis. 

\begin{figure}[htbp]
    \centering
    \includegraphics[width=0.9\linewidth]{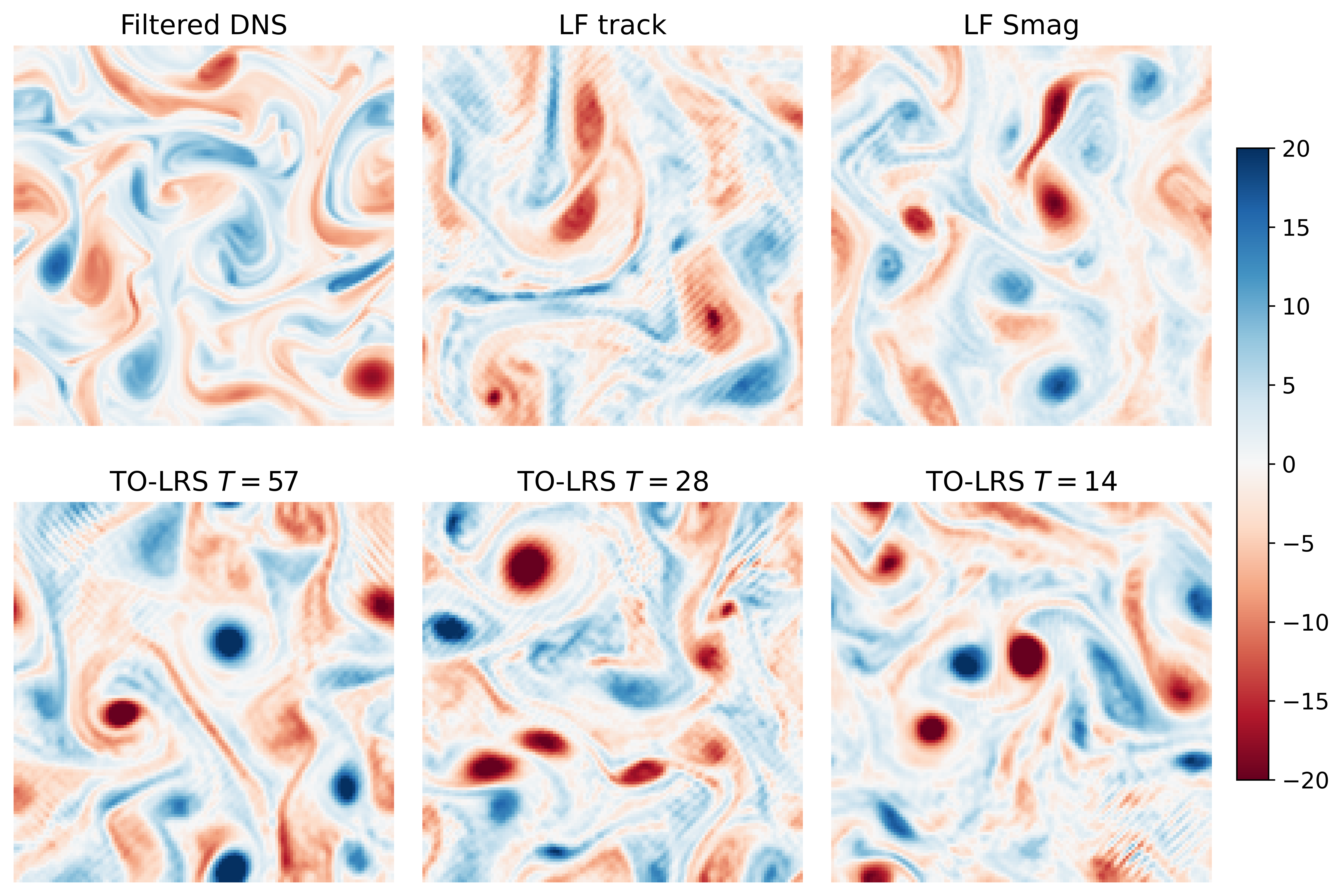}
    \caption{Vorticity fields at the end of the long-term Kolmogorov flow simulations ($t=125$ TU) for the filtered DNS, the low-fidelity tracking simulation, the calibrated Smagorinsky baseline, and the three best-performing Fourier-based TO-LRS models. For each TO-LRS ensemble the field from the first replica is shown.}
    \label{fig:final-fields-fourier}
\end{figure}

\section{Channel-flow tracking without the mean-profile QoIs}\label{app:3d-no-meanprof}
The scale-aware energies and enstrophies do not constrain the mean streamwise velocity profile. To demonstrate that the six mean-profile QoIs $Q_j^{\text{mp}}$ are necessary, we repeat the tracking simulation on the Smagorinsky substrate with just the six scale-aware QoIs. Figure~\ref{fig:3d-app-no-meanprof} shows the resulting velocity profile. The scale-aware-only run under-predicts $U^+$ throughout the buffer and lower logarithmic layers, much like the calibrated Smagorinsky baseline, and overshoots the reference at the channel center. The full $12$-QoI tracking run, by contrast, lies on the DNS profile across the whole channel.

\begin{figure}[htbp]
    \centering
    \includegraphics[width=0.7\linewidth]{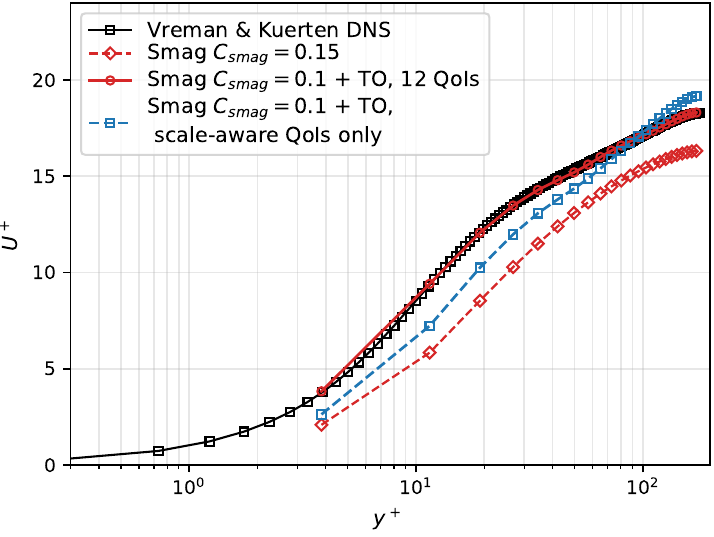}
    \caption{Mean streamwise velocity profile of the channel-flow tracking simulations on the Smagorinsky substrate constrained by the six scale-aware QoIs, compared with the full $12$-QoI tracking run, the calibrated Smagorinsky baseline, and the Vreman--Kuerten DNS~\cite{vreman_comparison_2014}.}
    \label{fig:3d-app-no-meanprof}
\end{figure}

\end{document}